\documentclass[draftcls,journal,onecolumn]{IEEEtran}
\usepackage[latin1]{inputenc}
\usepackage[T1]{fontenc}
\usepackage{float}
\usepackage{hyperref,graphicx,multirow}
\usepackage[cmex10]{amsmath}
\usepackage{amsthm,amsfonts,amssymb,stackrel}
\DeclareMathOperator{\erfc}{erfc}

\DeclareMathOperator{\res}{res}
\numberwithin{equation}{section}
\newtheorem{thm}{Theorem}[section]

\newtheorem{cor}[thm]{Corollary}

\begin{document}
\title{Coherent Fading Channels Driven by Arbitrary Inputs: Asymptotic Characterization of the Constrained Capacity and Related Information- and Estimation-Theoretic Quantities}
\author{Alberto~Gil~C.~P.~Ramos,~\IEEEmembership{Student Member,~IEEE,}
        and~Miguel~R.~D.~Rodrigues,~\IEEEmembership{Member,~IEEE}
\thanks{A. G. C. P. Ramos was with Instituto de Telecomunica\c{c}\~oes--Porto, Portugal. He is now with the Cambridge Centre for Analysis, University of Cambridge, United Kingdom [e-mail: a.g.c.p.ramos@maths.cam.ac.uk]. The work of A. G. C. P. Ramos was supported by Fundação para a Ciência e a Tecnologia, Portugal through the research project PTDC/EEA-TEL/100854/2008 and through the fellowship SFRH/BD/71692/2010.}
\thanks{M. R. D. Rodrigues was with Instituto de Telecomunica\c{c}\~oes--Porto and Departamento de Ciência de Computadores, Universidade do Porto, Portugal. He is now with the Department of Electronic and Electrical Engineering, University College London, United Kingdom [e-mail: m.rodrigues@ucl.ac.uk]. The work of M. R. D. Rodrigues was supported by Fundação para a Ciência e a Tecnologia, Portugal through the research project PTDC/EEA-TEL/100854/2008.}
}

\maketitle



\begin{abstract}
\addcontentsline{toc}{section}{Abstract}
We consider the characterization of the asymptotic behavior of the average minimum mean-squared error (MMSE) and the average mutual information in scalar and vector fading coherent channels, where the receiver knows the exact fading channel state but the transmitter knows only the fading channel distribution, driven by a range of inputs. 
We construct low-$snr$ and -- at the heart of the novelty of the contribution -- high-$snr$ asymptotic expansions for the average MMSE and the average mutual information for coherent channels subject to Rayleigh fading, Ricean fading or Nakagami fading and driven by discrete inputs (with finite support) or various continuous inputs.  We reveal the role that the so-called \emph{canonical} MMSE in a standard additive white Gaussian noise (AWGN) channel plays in the characterization of the asymptotic behavior of the average MMSE and the average mutual information in a fading coherent channel: in the regime of low-$snr$, the derivatives of the \emph{canonical} MMSE define the expansions of the estimation- and information-theoretic quantities; in contrast, in the regime of high-$snr$, the Mellin transform of the \emph{canonical} MMSE define the expansions of the quantities. We thus also provide numerically and -- whenever possible -- analytically the Mellin transform of the \emph{canonical} MMSE for the most common input distributions. We also reveal connections to and generalizations of the MMSE dimension.  The most relevant element that enables the construction of these non-trivial expansions is the realization that the integral representation of the estimation- and information-theoretic quantities can be seen as an $h$-transform of a kernel with a monotonic argument: this enables the use of a novel asymptotic expansion of integrals technique -- the Mellin transform method -- that leads immediately to not only the high-$snr$ but also the low-$snr$ expansions of the average MMSE and -- via the I-MMSE relationship -- to expansions of the average mutual information. We conclude with applications of the results to the characterization and optimization of the constrained capacity of a bank of parallel independent coherent fading channels driven by arbitrary discrete inputs.
\end{abstract}



\section{Introduction}
\IEEEPARstart{T}{he} characterization of the constrained capacity of fading coherent channels, where the receiver knows the exact fading channel realization but the transmitter knows only the fading channel distribution, driven by arbitrary inputs is a problem with significant practical interest. The importance relates to the fact that this quantity defines the highest information transmission rate between the transmitter and the receiver with a message error probability that approaches zero in systems that employ arbitrary and practical signalling schemes, such as $m$-phase shift keying (PSK), $m$-pulse amplitude modulation (PAM) or $m$-quadrature amplitude modulation (QAM) constellations, rather than the capacity-achieving Gaussian one, thereby establishing a benchmark to the design and optimization of practical communications systems.

Unfortunately, the characterization of the constrained capacity, which -- assuming that the fading channel variation over time is stationary and ergodic -- is given by the average over the fading channel gain distribution of the mutual information between the input and the output of the channel conditioned on the channel gain, is complex due to the absence of closed-form expressions for the mutual information as well as the average mutual information.

An innovative approach that also leads to a characterization -- in asymptotic regimes -- of the constrained capacity of key communications channels was put forth in~\cite{Lozano06optimumpower,Lozano_optimumpower}. The approach capitalizes on connections between mutual information and minimum mean-squared error (MMSE)~\cite{Guo05mutualinformation,Palomar_Verdu_Gradient_of_Mutual_Information}, key quantities in information theory and estimation theory, in order to express the mutual information or bounds to the mutual information in terms of the MMSE or bounds to the MMSE, respectively. Other applications of the relation between mutual information and MMSE can be found in~\cite{Verdu2006,Monotonic_Decrease_Tulino_Verdu,Estimation_in_Gaussian_Noise_Properties_of_the_Minimum_Mean-Square_Error_Guo_Wu_Shamai_and_Verdu,Cruz_Rodrigues_Verdu_MIMO_Gaussian_Channels_With_Arbitrary_Inputs,Parayo09a,Parayo09b,Lamarca09,GloballyXiao,LinearZeng}.

In this paper, we leverage connections between the average value of the mutual information and the average value of the MMSE in a coherent fading channel to obtain characterizations of the average mutual information from characterizations of the average MMSE. We pursue the characterizations exclusively in the asymptotic regime of low signal-to-noise ratio (SNR) and, at the heart of the novelty of the contribution, in the asymptotic regime of high SNR. The asymptotic analysis, which bypasses the difficulty associated with the construction of general non-asymptotic results, often applies to a variety of practical scenarios leading to considerable insight.

We also use, in addition to the connections between the average mutual information and the average MMSE, key techniques that lead to the asymptotic expansions of the quantities not only in the regime of high SNR but also low SNR. In particular, by recognizing that the quantities can be seen as an $h$-transform with a kernel of monotonic argument~\cite{asymptotic_expansion_of_integrals_Bleistein_and_Handelsman}, we capitalize on expansions of integrals techniques, such as Mellin transform based methods or integration by part methods, in order to construct the asymptotic expansions of the average MMSE and -- via the connections between estimation and information theory -- the average mutual information.

The original contributions of the article include:
\begin{itemize}
	\item
The identification of a comprehensive framework to construct asymptotic expansions of the average MMSE and the average mutual information in a single-input--single-output fading coherent channel driven by arbitrary inputs. The framework enables the construction of low- and -- more importantly -- high-SNR asymptotic expansions of the quantities in a unified way.
	\item
The construction of novel asymptotic expansions for the average MMSE and the average mutual information in a single-input--single-output fading coherent channel driven by arbitrary inputs. We conceive expansions applicable to arbitrary fading channels, which are specialized to Rayleigh fading channels, Ricean fading channels as well as Nakagami fading channels. We also conceive expansions applicable to discrete inputs with finite support and continuous inputs distributed according to $\infty$-PSK, $\infty$-PAM, $\infty$-QAM and standard complex Gaussian distributions.
	\item
The generalization of the asymptotic results from single-input--single-output to single-input--multiple-output fading coherent channels driven by arbitrary inputs.
	\item
The application of the results in key communications problems. In particular, inspired by the ubiquitous use of Orthogonal Frequency-Division Multiplexing (OFDM) and multi-user OFDM based systems, we consider the determination of the power allocation policy that maximizes the constrained capacity of a bank of parallel independent fading coherent channels driven by arbitrary discrete inputs in the asymptotic regime of high SNR.
\end{itemize}

This paper is organized as follows: Section \ref{section: Model and Definitions} describes the channel model and the main estimation-theoretic and information-theoretic quantities. Sections \ref{section: High-snr Regime} and \ref{section: Low-snr Regime}, which contain the main contributions, describe in detail the approach used to construct the asymptotic expansions of the estimation-theoretic and the information-theoretic measures. Section \ref{section: Generalizations} describes generalizations of the approach. Section \ref{section: Some MMSE Mellin Transform Results} includes a series of analytic and numerical Mellin transform results useful for the construction of the asymptotic expansions. Section \ref{section: Numerical Results} presents a series of simulations that confirm the accuracy of the asymptotic expansions. Applications of the approach are covered in Section \ref{section: Applications: Optimal Power Allocation in a Bank of Parallel Independent Fading Coherent Channels driven by Arbitrary Inputs}. Finally, we conclude with various remarks in Section \ref{section: Conclusions}.

\subsection{Notation}
We use the following notation: Italic lower case letters denote scalars, boldface lower case letters denote column vectors and boldface upper case letters denote matrices, e.g., $c$, $\boldsymbol{c}$ and $\boldsymbol{C}$, respectively. $\left(\cdot\right)^*$, $\left(\cdot\right)^T$ and $\left(\cdot\right)^{\dagger}$ denote the complex conjugate, transpose, and complex conjugate transpose operators, respectively. The norm of a vector $\boldsymbol{v}$ is defined as $||\boldsymbol{v}||:=\sqrt{\boldsymbol{v}^{\dagger}\boldsymbol{v}}$. $\mathcal{R}\left(\cdot\right)$ and $\mathcal{I}\left(\cdot\right)$ denote the real and imaginary part operators, respectively. $\boldsymbol{I}_k$ denotes the $k\times k$ identity matrix. We denote the probability mass/density function of a discrete/continuous random variable $x$ by $f_x\left(\cdot\right)$ and the expectation of a function $g\left(\cdot\right)$ with respect to the random variable $x$ by $E_x\left\{g\left(x\right)\right\}$. Given $\sigma>0$, we say that a complex random variable $\boldsymbol{x}$ is distributed according to $\mathcal{CN}\left(\boldsymbol{\mu},2\sigma^2\boldsymbol{I}_k\right)$ if $[\mathcal{R}(\boldsymbol{x})^T,\mathcal{I}(\boldsymbol{x})^T]^T\sim\mathcal{N}\left([\mathcal{R}(\boldsymbol{\mu})^T,\mathcal{I}(\boldsymbol{\mu})^T]^T,\sigma^2\boldsymbol{I}_{2k}\right)$. $\log{\left(\cdot\right)}$ denotes the natural logarithm throughout the paper.

We also use the following asymptotic notation as in~\cite{asymptotic_expansion_of_integrals_Bleistein_and_Handelsman}: We write
\begin{equation*}
f\left(x\right)=O\left(g\left(x\right)\right), \qquad x\rightarrow x_0
\end{equation*}
if there exists a positive real number $k$, and a neighborhood of $x_0$, $N_{x_0}$, such that, $\forall x\in N_{x_0}$, $\left|f\left(x\right)\right|\leq k\left|g\left(x\right)\right|$. We also write
\begin{equation*}
f\left(x\right)=o\left(g\left(x\right)\right), \qquad x\rightarrow x_0
\end{equation*}
if for any positive real number $\epsilon$, there exists a neighborhood of $x_0$, $N_{x_0}$, such that, $\forall x\in N_{x_0}$, $\left|f\left(x\right)\right|\leq\epsilon\left|g\left(x\right)\right|$. In addition, if $\left\{\phi_n\left(x\right)\right\}, n=0,1,2,\ldots$, is a sequence of continuous functions such that, $\forall n\in\mathbb{Z}_0^+$,
\begin{equation*}
\phi_{n+1}\left(x\right)=o\left(\phi_n\left(x\right)\right), \qquad x\rightarrow x_0
\end{equation*}
then we write
\begin{equation*}
f\left(x\right)\sim\sum_{n=0}^{+\infty}a_n\phi_n\left(x\right), \qquad x\rightarrow x_0
\end{equation*}
where the formal series in the right-hand-side may or may not converge, if
\begin{equation*}
f\left(x\right)=\sum_{n=0}^ma_n\phi_n\left(x\right)+O\left(\phi_{m+1}\left(x\right)\right), \qquad x\rightarrow x_0
\end{equation*}
holds $\forall m\in\mathbb{Z}_0^+$ and we write
\begin{equation*}
f\left(x\right)\sim\sum_{n=0}^{N-1}a_n\phi_n\left(x\right), \qquad x\rightarrow x_0
\end{equation*}
if
\begin{equation*}
f\left(x\right)=\sum_{n=0}^ma_n\phi_n\left(x\right)+O\left(\phi_{m+1}\left(x\right)\right), \qquad x\rightarrow x_0
\end{equation*}
holds only for $m\in\left\{0,1,\ldots,N-1\right\}$.

\section{Model and Definitions}
\label{section: Model and Definitions}
We consider a standard frequency-flat fading channel, which for a single time instant, can be modeled as follows:
\begin{equation}
y=\sqrt{snr}hx+n
\label{eq: Additive Gaussian Noise channel}
\end{equation}
where $y\in\mathbb{C}$ represents the channel output, $x\in\mathbb{C}$ represents the channel input, $h$ is a complex scalar random variable (with support $\mathbb{C}$ or $\mathbb{C}\setminus\{0\}$) such that $E_{h}\left\{|h|^2\right\}<+\infty$ which represents the random channel fading gain between the input and the output of the channel, and $n\in\mathbb{C}$ is a circularly symmetric complex scalar Gaussian random variable with zero mean and unit variance which represents standard noise. The scaling factor $snr\in\mathbb{R}^+$ relates to the signal-to-noise ratio. We assume that $x$, $h$ and $n$ are independent random variables. We also assume that the receiver knows the exact realization of the channel gain but the transmitter knows only the distribution of the channel gain.

In particular, we consider three conventional fading models: \emph{i}) the Rayleigh fading model; \emph{ii}) the Ricean fading model; and \emph{iii}) the Nakagami fading model. In the Rayleigh fading model, the channel gain $h\sim\mathcal{CN}\left(0,2\sigma^2\right)$, where $\sigma>0$, and hence $|h|\sim\text{Rayleigh}\left(\sigma\right)$~\cite{Digital_Communications_Proakis}, i.e.,
\begin{equation}
f_{|h|}\left(r\right)=\frac{r}{\sigma^2}\exp{\left(-\frac{r^2}{2\sigma^2}\right)}.
\label{eq: Rayleigh fading model}
\end{equation}
In the Ricean fading model, the channel gain $h\sim\mathcal{CN}\left(\mu,2\sigma^2\right)$, where $\mu\in\mathbb{C}\setminus\{0\}$ and $\sigma>0$, and hence $|h|\sim\text{Rice}\left(|\mu|, \sigma\right)$~\cite{Digital_Communications_Proakis}, i.e.,
\begin{equation}
f_{|h|}\left(r\right)=\frac{r}{\sigma^2}\exp{\left(-\frac{r^2+|\mu|^2}{2\sigma^2}\right)}I_0\left(\frac{r|\mu|}{\sigma^2}\right)
\label{eq: Rice fading model}
\end{equation}
where $I_0\left(\cdot\right)$ is the modified Bessel function of the first kind with order zero~\cite[Equation 10.25.2]{abramowitz_stegun}. In the Nakagami fading model, $|h|\sim\text{Nakagami}\left(\mu ,w\right)$, where $\mu\geq\frac{1}{2}$ and $w>0$~\cite{Digital_Communications_Proakis}, i.e.,
\begin{equation}
f_{|h|}\left(r\right)=\frac{2\mu^{\mu}}{\Gamma\left(\mu\right)w^{\mu}}r^{2\mu-1}\exp{\left(-\frac{\mu}{w}r^2\right)}.
\label{eq: Nakagami fading model}
\end{equation}
where $\Gamma\left(\cdot\right)$ is the Gamma function~\cite[Equation 5.2.1]{abramowitz_stegun}.

We now introduce a series of quantities which will be used throughout the paper. We define the conditional MMSE given that the channel gain $h=h_0$ associated with the estimation of the noiseless output given the noisy output of the channel model in \eqref{eq: Additive Gaussian Noise channel} as
\begin{equation*}
mmse_{h_0}\left(snr\right):=E_{x,y|h}\left\{\left|h_0x-E_{x|y,h}\left\{h_0x|y,h_0\right\}\right|^2\Bigg|h=h_0\right\}
\end{equation*}
and the conditional mutual information given that the channel gain $h=h_0$ between the input and the output of the channel model in \eqref{eq: Additive Gaussian Noise channel} as
\begin{equation*}
I_{h_0}\left(snr\right):=E_{x,y|h}\left\{\log{\left(\frac{f_{x,y|h}\left(x,y|h_0\right)}{f_{x|h}\left(x|h_0\right)f_{y|h}\left(y|h_0\right)}\right)}\Bigg|h=h_0\right\}
\end{equation*}
We also define the average value of the MMSE and the average value of the mutual information as follows:
\begin{align}
\overline{mmse}\left(snr\right)&:=E_h\left\{mmse_h\left(snr\right)\right\}\label{eq: definition average mmse}\\
\overline{I}\left(snr\right)&:=E_h\left\{I_h\left(snr\right)\right\}\label{eq: definition average mi}
\end{align}

It will also be relevant to define the MMSE and the mutual information associated with the \emph{canonical} additive white Gaussian noise (AWGN) channel model given by:
\begin{equation}
y=\sqrt{snr}x+n
\label{eq: Canonical Additive Gaussian Noise channel}
\end{equation}
where $y\in\mathbb{C}$ represents the channel output, $x\in\mathbb{C}$ represents the channel input and $n\in\mathbb{C}$ is a circularly symmetric complex scalar Gaussian random variable with zero mean and unit variance which represents standard noise. The scaling factor $snr\in\mathbb{R}^+$ also relates to the signal-to-noise ratio. We assume that $x$ and $n$ are independent random variables.

Now, the MMSE associated with the estimation of the input given the output of the \emph{canonical} AWGN channel model in \eqref{eq: Canonical Additive Gaussian Noise channel} is defined as:
\begin{equation}
mmse\left(snr\right):=E_{x,y}\left\{\left|x-E_{x|y}\left\{x|y\right\}\right|^2\right\}=mmse_1\left(snr\right)
\label{eq: relation between cammse and commse}
\end{equation}
where $mmse_1\left(snr\right)$ denotes the conditional MMSE given that the channel gain $h=1$ associated with the estimation of the noiseless output given the noisy output of the channel model in \eqref{eq: Additive Gaussian Noise channel}. The mutual information between the input and the output of the \emph{canonical} AWGN channel model in \eqref{eq: Canonical Additive Gaussian Noise channel} is defined as:
\begin{equation}
I\left(snr\right):=E_{x,y}\left\{\log{\left(\frac{f_{x,y}\left(x,y\right)}{f_{x}\left(x\right)f_{y}\left(y\right)}\right)}\right\}=I_1\left(snr\right)
\label{eq: relation between cami and comi}
\end{equation}
where $I_1\left(snr\right)$ denotes the conditional mutual information given that the channel gain $h=1$ between the input and the output of the channel model in \eqref{eq: Additive Gaussian Noise channel}. Therefore, it is very simple to express the average MMSE in \eqref{eq: definition average mmse} in terms of the \emph{canonical} MMSE in \eqref{eq: relation between cammse and commse} as:
\begin{equation}
\overline{mmse}\left(snr\right)=E_{|h|}\left\{|h|^2mmse\left(snr|h|^2\right)\right\}
\label{eq: average mmse-mmse relation}
\end{equation}
and the average mutual information in \eqref{eq: definition average mi} in terms of the \emph{canonical} mutual information in \eqref{eq: relation between cami and comi} as:
\begin{equation}
\overline{I}\left(snr\right)=E_{|h|}\left\{I\left(snr|h|^2\right)\right\}
\label{eq: average mi-mi relation}
\end{equation}

The objective is to characterize the asymptotic behavior, as $snr\rightarrow\infty$ or as $snr\rightarrow 0^+$, of the average MMSE and the average mutual information, which, when the channel variation over time is stationary and ergodic, leads to the constrained capacity of a fading coherent channel driven by a specific input distribution. We adopt a two-step procedure: \emph{i}) We first obtain, via Mellin transform expansion techniques or via integration by parts expansion techniques, a characterization of the asymptotic behavior of the average MMSE in \eqref{eq: definition average mmse}; \emph{ii}) We then obtain a characterization of the asymptotic behavior of the average mutual information in \eqref{eq: definition average mi} by capitalizing on the now well-known relation between average MMSE and average mutual information given by~\cite{Palomar_Verdu_Gradient_of_Mutual_Information}:
\begin{equation}
\frac{d\overline{I}\left(snr\right)}{dsnr}=\overline{mmse}\left(snr\right)
\label{eq: average mi-average mmse relation}
\end{equation}
We note that the Mellin transform of a function $f(\cdot)$, which is defined as follows~\cite{asymptotic_expansion_of_integrals_Bleistein_and_Handelsman}:
\begin{equation*}
M\left[f;1+z\right]:=\int_0^{+\infty}t^{z}f\left(t\right)dt,
\end{equation*}
plays a key role in the definition of the asymptotic expansions.

\section{High-snr Regime}
\label{section: High-snr Regime}
We now consider the construction of high-$snr$ asymptotic expansions of the average MMSE and the average mutual information in a fading coherent channel driven by inputs that conform to either arbitrary discrete distributions (with finite support) or $\infty$-PSK, $\infty$-PAM, $\infty$-QAM, and standard complex Gaussian continuous distributions. The study of the first three continuous distributions, which represent the limit as $m\rightarrow+\infty$ of the well-known equiprobable $m$-PSK, $m$-PAM or $m$-QAM discrete distributions, is justified by the fact that it is often analytically convenient to approximate discrete constellations using continuous distributions over a suitable region on the complex plane, in order to define meaningful asymptotic notions such as the shaping gain~\cite{Forney1989}.

We proceed by obtaining a high-$snr$ asymptotic expansion of the average MMSE and then -- via the connection between average MMSE and average mutual information in \eqref{eq: average mi-average mmse relation} -- a high-$snr$ asymptotic expansion of the average mutual information. The construction of the high-$snr$ expansions capitalizes on the fact that the average MMSE, which can be written as
\begin{equation}
\overline{mmse}\left(snr\right)=E_{|h|}\left\{|h|^2mmse\left(snr|h|^2\right)\right\}
=\int_0^{+\infty}\frac{\sqrt{t}f_{|h|}\left(\sqrt{t}\right)}{2}mmse\left(snr\cdot t\right)dt,
\label{eq: average mmse integral representation h transform}
\end{equation}
can be seen as an $h$-transform with a kernel of monotonic argument~\cite[Chapter 4]{asymptotic_expansion_of_integrals_Bleistein_and_Handelsman}. This enables the use of a key asymptotic expansion of integrals technique, which exploits Mellin transforms~\cite[Section 4.4]{asymptotic_expansion_of_integrals_Bleistein_and_Handelsman}, that leads to a dissection of the asymptotic behavior of the quantities in a large variety of scenarios. In fact, the construction of the high-$snr$ expansions of the integral representation in \eqref{eq: average mmse integral representation h transform} can be effected by exploiting a range of techniques, such as the Mellin transform method~\cite[Section 4.4]{asymptotic_expansion_of_integrals_Bleistein_and_Handelsman} or the integration by parts methods~\cite[Chapter 3]{asymptotic_expansion_of_integrals_Bleistein_and_Handelsman}. It is important to emphasize though that the Mellin transform technique -- when compared to the integration by parts technique -- is able to produce expansions for a wider range of fading and input distributions.

\subsection{Discrete Inputs}
\label{subsection: High-snr Discrete Inputs}
We consider arbitrary discrete inputs with finite support $\{x_1,\ldots,x_m\}$ with $m\in\mathbb{Z}^+$ and $x_1,\ldots,x_m\in\mathbb{C}$. The construction of the high-$snr$ expansions of the average MMSE associated with arbitrary discrete inputs with finite support capitalizes on the characterization of the decay of the \emph{canonical} MMSE in the regime of high-$snr$ in~\cite{Lozano06optimumpower}.

The following Theorem defines the asymptotic expansion of the average MMSE in fading coherent channels driven by arbitrary (but fixed) discrete inputs with finite support, by capitalizing on the Mellin transform method~\cite[Section 4.4]{asymptotic_expansion_of_integrals_Bleistein_and_Handelsman}.

\begin{thm}
Consider the fading coherent channel in \eqref{eq: Additive Gaussian Noise channel} driven by an arbitrary discrete input with finite support and define
\begin{equation*}
f\left(t\right):=\frac{\sqrt{t}f_{|h|}\left(\sqrt{t}\right)}{2}
\end{equation*}
which relates to the distribution of the fading process and
\begin{gather*}
\gamma:=\inf\left\{\gamma^*: f\left(t\right)=O\left(t^{-\gamma^*}\right), t\rightarrow 0^+\right\},\\
\delta:=\sup\left\{\delta^*: f\left(t\right)=O\left(t^{-\delta^*}\right), t\rightarrow +\infty\right\},\\
C:=]0,+\infty[\cap]1-\delta,1-\gamma[.
\end{gather*}
If
\begin{equation}
f\left(t\right)\sim\exp{\left(-qt^{-\mu}\right)}\sum_{m=0}^{+\infty}\sum_{n=0}^{\overline{N}\left(m\right)}p_{mn}t^{a_m}\left(\log{\left(t\right)}\right)^n, \qquad t\rightarrow 0^+
\label{eq: asymptotic behavior in Additive Gaussian Noise average mmse general h}
\end{equation}
where $\mathcal{R}\left(q\right)\geq 0$, $\mu>0$, $\overline{N}\left(m\right)$ is finite for each $m$, $p_{mn}\in\mathbb{C}$ and $\mathcal{R}\left(a_m\right)\uparrow +\infty$,
\begin{gather}
\gamma<\delta\label{eq: gamma less delta in Additive Gaussian Noise average mmse general h Mellin transform}\\
C\neq\emptyset\label{eq: C not empty in Additive Gaussian Noise average mmse general h Mellin transform}\\
\exists c\in C: \forall x\in[c,+\infty[, M\left[f;1-x-iy\right]=O\left(|y|^{-2}\right), \qquad |y|\rightarrow +\infty\label{eq: asymptotic behavior in Additive Gaussian Noise average mmse general h Mellin transform}
\end{gather}
where $M\left[f;1-x-iy\right]$ is to be understood as the analytic continuation of $M\left[f;1-x-iy\right]$ from $\{x+iy: 1-\delta<x<1-\gamma\}$ to $\{x+iy: x>1-\delta\}$~\cite{introduction_to_complex_analysis_second_edition_Priestley}, then
\begin{enumerate}
	\item
If $q=0$ it follows that
\begin{align}
&\overline{mmse}\left(snr\right)\sim\nonumber\\
&\sim\sum_{m=0}^{+\infty}snr^{-1-a_m}\sum_{n=0}^{\overline{N}\left(m\right)}p_{mn}\sum_{j=0}^n\binom{n}{j}\left(-\log{\left(snr\right)}\right)^jM^{\left(n-j\right)}\left[mmse;z\right]\Bigg|_{z=1+a_m}, \qquad snr\rightarrow +\infty
\label{eq: Additive Gaussian Noise average mmse general h part 1}
\end{align}
	\item
If $q\neq 0$ it follows that
\begin{equation}
\forall R\in\mathbb{R}^+, \overline{mmse}\left(snr\right)=o\left(snr^{-R}\right), \qquad snr\rightarrow +\infty
\label{eq: Additive Gaussian Noise average mmse general h part 2}
\end{equation}
\end{enumerate}
\label{thm: Additive Gaussian Noise average mmse general h}
\end{thm}

\begin{IEEEproof}
See Appendix \ref{section: Proof of Theorem Additive Gaussian Noise average mmse general h}.
\end{IEEEproof}

\vspace{0.25cm}

Theorem \ref{thm: Additive Gaussian Noise average mmse general h} provides a dissection of the asymptotic behavior as $snr\rightarrow +\infty$ of the average MMSE in a fading coherent channel driven by arbitrary discrete inputs (with finite support) under very general fading conditions. In particular, the Theorem requires that the asymptotic expansion as $t\rightarrow 0^+$ of $f\left(t\right)$, which relates to the probability density function of the fading random variable, behaves as in \eqref{eq: asymptotic behavior in Additive Gaussian Noise average mmse general h} and that the Mellin transform of the function $f\left(\cdot\right)$ exists, is holomorphic in a certain vertical strip in the complex plane and decays along vertical lines as in \eqref{eq: asymptotic behavior in Additive Gaussian Noise average mmse general h Mellin transform}. If the fading model does not accommodate for exponential decay of $f\left(t\right)$ as $t\rightarrow 0^+$, i.e., $q=0$, then the average MMSE behaves as in \eqref{eq: Additive Gaussian Noise average mmse general h part 1}; otherwise, if the fading model accommodates for exponential decay of $f\left(t\right)$ as $t\rightarrow 0^+$, i.e., $q\neq 0$, then the average MMSE behaves as in \eqref{eq: Additive Gaussian Noise average mmse general h part 2}. The Theorem also requires the existence of the Mellin transform of the \emph{canonical} MMSE and their higher-order derivatives, which is guaranteed by the fact that the input distribution is discrete without accumulation points.

Theorem \ref{thm: Additive Gaussian Noise average mmse general h} reveals that in the important scenario which accommodates for the most common fading models (i.e., $q=0$) the asymptotic behavior as $snr\rightarrow +\infty$ of the average MMSE depends mainly on the asymptotic behavior as $t\rightarrow 0^+$ of $f\left(t\right)$. Interestingly, the fact that the behavior of some quantities around zero determine the behavior of other quantities in the infinity has already been pointed out in e.g.~\cite{Zheng2003,RodriguesOn_the_constrained,Rodrigues_Characterization}. Theorem \ref{thm: Additive Gaussian Noise average mmse general h} also reveals that the asymptotic behavior as $snr\rightarrow +\infty$ of the average MMSE depends on the input distribution via the Mellin transform of the \emph{canonical} MMSE.

Theorem \ref{thm: Additive Gaussian Noise average mmse general h} can also now be immediately specialized to the most common fading distributions, including: \emph{i}) Rayleigh fading; \emph{ii}) Ricean fading; and \emph{iii}) Nakagami fading.
\begin{cor}
Consider a Rayleigh or Ricean fading coherent channel as in \eqref{eq: Additive Gaussian Noise channel} driven by an arbitrary discrete input with finite support, where \eqref{eq: Rayleigh fading model} or \eqref{eq: Rice fading model} hold, respectively. Then, in the regime of high-$snr$ the average MMSE behaves as follows:
\begin{align*}
&\overline{mmse}\left(snr\right)\sim\\
&\sim\frac{\exp{\left(-\frac{|\mu|^2}{2\sigma^2}\right)}}{snr^2}\sum_{m=0}^{+\infty}\Bigg(\sum_{n=0}^m\left(\frac{\left(-1\right)^{m-n}}{\left(m-n\right)!\left(2\sigma^2\right)^{m+n+1}}\frac{|\mu|^{2n}}{\left(n!\right)^2}\right)M\left[mmse;2+m\right]\Bigg)\frac{1}{snr^m}, \qquad snr\rightarrow +\infty
\end{align*}
\label{cor: Additive Gaussian Noise average mmse h Rayleigh and Rice}
\end{cor}

\begin{IEEEproof}
See Appendix \ref{section: Proof of Corollary Additive Gaussian Noise average mmse h Rayleigh and Rice}.
\end{IEEEproof}

\vspace{0.25cm}

\begin{cor}
Consider a Nakagami fading coherent channel as in \eqref{eq: Additive Gaussian Noise channel} driven by an arbitrary discrete input with finite support, where \eqref{eq: Nakagami fading model} holds. Then, in the regime of high-$snr$ the average MMSE behaves as follows:
\begin{equation*}
\overline{mmse}\left(snr\right)\sim\frac{\mu^{\mu}}{snr^{1+\mu}\Gamma\left(\mu\right)w^{\mu}}\sum_{m=0}^{+\infty}\Bigg(\frac{\left(-1\right)^m\mu^m}{m!w^m}M\left[mmse;1+\mu+m\right]\Bigg)\frac{1}{snr^m}, \qquad snr\rightarrow +\infty
\end{equation*}
\label{cor: Additive Gaussian Noise average mmse h Nakagami}
\end{cor}

\begin{IEEEproof}
See Appendix \ref{section: Proof of Corollary Additive Gaussian Noise average mmse h Nakagami}.
\end{IEEEproof}

\vspace{0.25cm}

The high-$snr$ asymptotic expansions of the average MMSE embodied in Corollaries \ref{cor: Additive Gaussian Noise average mmse h Rayleigh and Rice} and \ref{cor: Additive Gaussian Noise average mmse h Nakagami} lead immediately to a high-$snr$ asymptotic expansion of the average mutual information, by leveraging the following integral representation of the connection between average MMSE and average mutual information in \eqref{eq: average mi-average mmse relation} given by:
\begin{equation}
\overline{I}\left(snr\right)=\log{\left(m\right)}-\int_{snr}^{\infty}\overline{mmse}\left(\epsilon\right)d\epsilon\label{eq: average mi-average mmse relation integral representation}
\end{equation}
\begin{cor}
Consider a Rayleigh or Ricean fading coherent channel as in \eqref{eq: Additive Gaussian Noise channel} driven by an arbitrary discrete input with finite support, where \eqref{eq: Rayleigh fading model} or \eqref{eq: Rice fading model} hold, respectively. Then, in the regime of high-$snr$ the average mutual information obeys the expansion:
\begin{align*}
&\overline{I}\left(snr\right)\sim\log{\left(m\right)}-\\
&-\frac{\exp{\left(-\frac{|\mu|^2}{2\sigma^2}\right)}}{snr}\sum_{m=0}^{+\infty}\Bigg(\sum_{n=0}^m\left(\frac{\left(-1\right)^{m-n}}{\left(m+1\right)\left(m-n\right)!\left(2\sigma^2\right)^{m+n+1}}\frac{|\mu|^{2n}}{\left(n!\right)^2}\right)M\left[mmse;2+m\right]\Bigg)\frac{1}{snr^m}, \qquad snr\rightarrow +\infty
\end{align*}
\label{cor: Additive Gaussian Noise average mi h Rayleigh and Rice}
\end{cor}

\begin{cor}
Consider a Nakagami fading coherent channel as in \eqref{eq: Additive Gaussian Noise channel} driven by an arbitrary discrete input with finite support, where \eqref{eq: Nakagami fading model} holds. Then, in the regime of high-$snr$ the average mutual information obeys the expansion:
\begin{equation*}
\overline{I}\left(snr\right)\sim \log{\left(m\right)}-\frac{\mu^{\mu}}{snr^{\mu}\Gamma\left(\mu\right)w^{\mu}}\sum_{m=0}^{+\infty}\Bigg(\frac{\left(-1\right)^m\mu^m}{\left(\mu+m\right)m!w^m}M\left[mmse;1+\mu+m\right]\Bigg)\frac{1}{snr^m}, \qquad snr\rightarrow +\infty
\end{equation*}
\label{cor: Additive Gaussian Noise average mi h Nakagami}
\end{cor}

\begin{IEEEproof}[Proof of Corollaries \ref{cor: Additive Gaussian Noise average mi h Rayleigh and Rice} and \ref{cor: Additive Gaussian Noise average mi h Nakagami}]
The Corollaries follow immediately from Corollaries \ref{cor: Additive Gaussian Noise average mmse h Rayleigh and Rice} or \ref{cor: Additive Gaussian Noise average mmse h Nakagami} and the relationship in \eqref{eq: average mi-average mmse relation integral representation}, together with the fact that an order relation can be integrated with respect to the independent variable~\cite{asymptotic_expansion_of_integrals_Bleistein_and_Handelsman}.
\end{IEEEproof}

\vspace{0.25cm}

It is now relevant to reflect on the nature of the asymptotic expansions embodied in Corollaries \ref{cor: Additive Gaussian Noise average mmse h Rayleigh and Rice}, \ref{cor: Additive Gaussian Noise average mmse h Nakagami}, \ref{cor: Additive Gaussian Noise average mi h Rayleigh and Rice} and \ref{cor: Additive Gaussian Noise average mi h Nakagami}. The expansions expose a dissection of the asymptotic behavior as $snr\rightarrow +\infty$ of the quantities. For Rayleigh and Ricean fading the $\left(m+1\right)$-th term in the average MMSE expansion is $O\left(snr^{-m-2}\right)$ and the $\left(m+1\right)$-th term in the average mutual information expansion is $O\left(snr^{-m-1}\right)$; in contrast, for Nakagami fading the $\left(m+1\right)$-th terms in the average MMSE and the average mutual information expansions are $O\left(snr^{-m-1-\mu}\right)$ and $O\left(snr^{-m-\mu}\right)$, respectively. In Rayleigh or Ricean fading channels, a first-order high-$snr$ expansion of the quantities can be written as follows:
\begin{equation*}
\overline{mmse}\left(snr\right)=\epsilon\cdot\frac{1}{snr^2}+O\left(\frac{1}{snr^3}\right)
\end{equation*}
and
\begin{equation*}
\overline{I}\left(snr\right)\sim \log{\left(m\right)}-\epsilon\cdot\frac{1}{snr}+O\left(\frac{1}{snr^2}\right)
\end{equation*}
where
\begin{equation*}
\epsilon=\lim_{snr\rightarrow +\infty}snr^2\cdot\overline{mmse}\left(snr\right)= \frac{\exp{\left(-\frac{|\mu|^2}{2\sigma^2}\right)}}{2\sigma^2}M\left[mmse;2\right]
\end{equation*}
In contrast, in Nakagami fading channels, a first-order high-$snr$ expansion of the average MMSE and the average mutual information can be written as follows:
\begin{equation*}
\overline{mmse}\left(snr\right)=\epsilon\cdot\frac{1}{snr^{1+\mu}}+O\left(\frac{1}{snr^{2+\mu}}\right)
\end{equation*}
and
\begin{equation*}
\overline{I}\left(snr\right)\sim \log{\left(m\right)}-\frac{\epsilon}{\mu}\cdot\frac{1}{snr^\mu}+O\left(\frac{1}{snr^{1+\mu}}\right)
\end{equation*}
where
\begin{equation*}
\epsilon=\lim_{snr\rightarrow +\infty}snr^{1+\mu}\cdot\overline{mmse}\left(snr\right)=\frac{\mu^{\mu}}{\Gamma\left(\mu\right)w^{\mu}}M\left[mmse;1+\mu\right]
\end{equation*}
This shows that the rate (on a $\log(snr)$ scale) at which the average MMSE or the average mutual information tend to the infinite-$snr$ value is given by:
\begin{equation}
-\lim_{snr\rightarrow +\infty}\frac{\log{\left(\overline{mmse}\left(snr\right)\right)}}{\log{\left(snr\right)}}=2
\label{rate_mmse}
\end{equation}
and
\begin{equation}
-\lim_{snr\rightarrow +\infty}\frac{\log{\left(\log{\left(m\right)}-\overline{I}\left(snr\right)\right)}}{\log{\left(snr\right)}}=1
\label{rate_i}
\end{equation}
respectively, both for Rayleigh and Ricean fading channels, and is equal to
\begin{equation*}
-\lim_{snr\rightarrow +\infty}\frac{\log{\left(\overline{mmse}\left(snr\right)\right)}}{\log{\left(snr\right)}}=1+\mu
\end{equation*}
and
\begin{equation*}
-\lim_{snr\rightarrow +\infty}\frac{\log{\left(\log{\left(m\right)}-\overline{I}\left(snr\right)\right)}}{\log{\left(snr\right)}}=\mu
\end{equation*}
respectively, for Nakagami fading channels. The parameter $\epsilon$, which is akin to the MMSE dimension put forth in~\cite{mmse_dimension_Wu_Verdu}, provides a more refined representation of the high-$snr$ asymptotics. The incorporation of a higher number of terms in the expansions provides a more accurate approximation of the behavior of the quantities not only at high-$snr$ but also medium-$snr$.

Finally, it is also relevant to note that one can also construct asymptotic expansions by using other asymptotic expansions of integrals techniques. The following Theorem defines the asymptotic expansion of the average MMSE in fading coherent channels driven by arbitrary (but fixed) discrete inputs with finite support, which follows from an integration by parts method~\cite[Chapter 3]{asymptotic_expansion_of_integrals_Bleistein_and_Handelsman} rather than the Mellin transform method~\cite[Section 4.4]{asymptotic_expansion_of_integrals_Bleistein_and_Handelsman}.

\begin{thm}
Consider the fading coherent channel in \eqref{eq: Additive Gaussian Noise channel} driven by an arbitrary discrete input with finite support and define
\begin{equation*}
f\left(t\right):=\frac{\sqrt{t}f_{|h|}\left(\sqrt{t}\right)}{2}
\end{equation*}
which relates to the distribution of the fading process. If
\begin{gather}
f\in C^{\infty}\left(\mathbb{R}^+\right)\label{eq: differentiability in Additive Gaussian Noise average mmse general h via integration by parts}\\
\forall n\in\mathbb{Z}_0^+, \exists\lim_{t\rightarrow 0^+}f^{\left(n\right)}\left(t\right)<+\infty\label{eq: existence and finiteness of limits in Additive Gaussian Noise average mmse general h via integration by parts}\\
\forall n\in\mathbb{Z}_0^+, f^{\left(n\right)}\left(t\right)=O\left(t^{-2}\right), \qquad t\rightarrow +\infty\label{eq: asymptotic behavior in Additive Gaussian Noise average mmse general h via integration by parts}
\end{gather}
then
\begin{equation}
\overline{mmse}\left(snr\right)\sim\sum_{m=0}^{+\infty}\frac{\left(-1\right)^{m+1}}{snr^{m+1}}f^{\left(m\right)}\left(0\right){mmse}^{\left(-m-1\right)}\left(0\right), \qquad snr\rightarrow +\infty
\label{eq: Additive Gaussian Noise average mmse general h via integration by parts}
\end{equation}
where
\begin{equation}
\forall m\in\mathbb{Z}_0^+, mmse^{\left(-m-1\right)}\left(x\right):=\int_{+\infty}^x\int_{+\infty}^{t_1}\cdots\int_{+\infty}^{t_{m-1}}\int_{+\infty}^{t_{m}}mmse\left(t_{m+1}\right)dt_{m+1}dt_m\cdots dt_2dt_1
\label{eq: [m+1]th repeated integral}
\end{equation}
\label{thm: Additive Gaussian Noise average mmse general h via integration by parts}
\end{thm}

\begin{IEEEproof}
See Appendix \ref{section: Proof of Theorem Additive Gaussian Noise average mmse general h via integration by parts}.
\end{IEEEproof}

\vspace{0.25cm}

We note that the asymptotic expansion of the average MMSE provided in Theorem \ref{thm: Additive Gaussian Noise average mmse general h} is defined via the Mellin transform of the \emph{canonical} MMSE whereas the asymptotic expansion of the average MMSE provided in Theorem \ref{thm: Additive Gaussian Noise average mmse general h via integration by parts} is defined in terms of repeated integrals of the \emph{canonical} MMSE. It is possible though to reconcile the asymptotic expansion in Theorem \ref{thm: Additive Gaussian Noise average mmse general h} with the asymptotic expansion in Theorem \ref{thm: Additive Gaussian Noise average mmse general h via integration by parts}, in some key scenarios. In particular, if the requirements of Theorems \ref{thm: Additive Gaussian Noise average mmse general h} and \ref{thm: Additive Gaussian Noise average mmse general h via integration by parts} hold and
\begin{equation}
\left(\forall m\in\mathbb{Z}_0^+, \left(\overline{N}\left(m\right)=0\wedge a_m=m\right)\right)\wedge q=0
\label{eq: correspondence between Mellin transform and integration by parts part 1}
\end{equation}
then the asymptotic expansions \eqref{eq: Additive Gaussian Noise average mmse general h part 1} and \eqref{eq: Additive Gaussian Noise average mmse general h via integration by parts} coincide because
\begin{align*}
&snr^{-1-a_m}\sum_{n=0}^{\overline{N}\left(m\right)}p_{mn}\sum_{j=0}^n\binom{n}{j}\left(-\log{\left(snr\right)}\right)^jM^{\left(n-j\right)}\left[mmse;z\right]\Bigg|_{z=1+a_m}\\
&=snr^{-1-m}p_{m0}M\left[mmse;1+m\right]\\
&=snr^{-1-m}p_{m0}\left(-1\right)^{m+1}m!mmse^{\left(-m-1\right)}\left(0\right)\\
&=\frac{\left(-1\right)^{m+1}}{snr^{m+1}}f^{\left(m\right)}\left(0\right){mmse}^{\left(-m-1\right)}\left(0\right)
\end{align*}
where the first and the last equalities are due to \eqref{eq: correspondence between Mellin transform and integration by parts part 1} and the second equality is due to
\begin{equation*}
mmse^{\left(-m-1\right)}\left(0\right)=\frac{\left(-1\right)^{m+1}}{m!}M[mmse;m+1]
\end{equation*}
in view of \eqref{eq: [m+1]th repeated integral} and~\cite[Equation 1.4.31]{abramowitz_stegun}.

There are various scenarios of relevance in practice where both Theorems produce the same expansions, such as the Rayleigh and Ricean fading coherent channel driven by arbitrary discrete inputs with finite support, because the Theorem requirements hold and \eqref{eq: correspondence between Mellin transform and integration by parts part 1} also holds. However, there are also scenarios where the Mellin transform method yields an expansion but the integration by parts method does not. 
One important case relates to the definition of the asymptotic expansions in a fading coherent channel driven by arbitrary discrete inputs with finite support, subject to Nakagami fading (where $|h|\sim\text{Nakagami}\left(\mu ,w\right)$). The Mellin transform method allows us to define the asymptotic expansions when the fading parameter $\mu\in\left[\frac{1}{2},+\infty\right[$. In contrast, the integration by parts method only allows us to define the asymptotic expansions when the fading parameter $\mu\in\left[\frac{1}{2},+\infty\right[\setminus\mathbb{Z}^+$. This is due to the fact that such a method requires stronger smoothness properties of the integrand functions. The other important case, which is the subject of the next subsection, relates to the definition of the asymptotic expansions in a fading coherent channel driven by continuous inputs. Note that in such a setting the repeated integrals of the \emph{canonical} MMSE -- which are necessary to conceive the asymptotic expansions via the integration by parts method -- do not often exist.

\subsection{Continuous Inputs}
\label{subsection: High-snr Continuous Inputs}
We now consider the continuous inputs: \emph{i}) unit-power $\infty$-PSK, where $x$ is uniformly distributed over the circle $\{z\in\mathbb{C}: |z|=1\}$ on the complex plane; \emph{ii}) unit-power $\infty$-PAM, where $x$ is uniformly distributed over the line $\left[-\sqrt{3},\sqrt{3}\right]$ on the complex plane; \emph{iii}) unit-power $\infty$-QAM, where $x$ is uniformly distributed over the square $\left[-\sqrt{\frac{3}{2}},\sqrt{\frac{3}{2}}\right]\times\left[-\sqrt{\frac{3}{2}},\sqrt{\frac{3}{2}}\right]$ on the complex plane; and \emph{iv}) standard complex Gaussian inputs, where $x$ is distributed according to $\mathcal{CN}\left(0,1\right)$.

The construction of the high-$snr$ expansions of the average MMSE associated with $\infty$-PSK, $\infty$-PAM and $\infty$-QAM uses the high-$snr$ expansions of the \emph{canonical} MMSEs given by~\cite{Lozano06optimumpower}:
\begin{gather}
mmse^{\infty\text{-PSK}}\left(snr\right)=\frac{1}{2snr}+O\left(\frac{1}{snr^2}\right), \qquad snr\rightarrow +\infty\label{eq: decay mmse Infinite PSK intro}\\
mmse^{\infty\text{-PAM}}\left(snr\right)=\frac{1}{2snr}+O\left(\frac{1}{snr^\frac{3}{2}}\right), \qquad snr\rightarrow +\infty\label{eq: decay mmse Infinite PAM intro}\\
mmse^{\infty\text{-QAM}}\left(snr\right)=\frac{1}{snr}+O\left(\frac{1}{snr^\frac{3}{2}}\right), \qquad snr\rightarrow +\infty\label{eq: decay mmse Infinite QAM intro}
\end{gather}
respectively. It is important to note that this one-term expansions of the \emph{canonical} MMSE lead to one-term expansions for the average MMSE; higher-order expansions for the \emph{canonical} MMSE would lead to higher-order expansions for the average MMSE, by exploiting identical techniques. In contrast, the construction of the high-$snr$ expansions of the average MMSE associated with standard complex Gaussian inputs uses the high-$snr$ expansion of the \emph{canonical} MMSE given by~\cite{Lozano06optimumpower}:
\begin{equation}
mmse^{\mathcal{G}}(snr)=\frac{1}{1+snr}=\frac{1}{snr}\frac{1}{\frac{1}{snr}+1}\sim\sum_{m=0}^{+\infty}(-1)^msnr^{-\left(m+1\right)}, \qquad snr\rightarrow +\infty
\label{eq: decay mmse Infinite Gaussian intro}
\end{equation}

The following Theorem defines the asymptotic expansions of the average MMSE in a fading coherent channel driven by $\infty$-PSK, $\infty$-PAM, $\infty$-QAM or standard complex Gaussian continuous input distributions. It is relevant to note that the proof relies on the use of the more powerful Mellin transform asymptotic expansion of integrals technique.

\begin{thm}
Consider the fading coherent channel in \eqref{eq: Additive Gaussian Noise channel} driven by an $\infty$-PSK, $\infty$-PAM, $\infty$-QAM or standard complex Gaussian continuous input and define
\begin{equation*}
f\left(t\right):=\frac{\sqrt{t}f_{|h|}\left(\sqrt{t}\right)}{2}
\end{equation*}
which relates to the distribution of the fading process and
\begin{gather*}
\gamma:=\inf\left\{\gamma^*: f\left(t\right)=O\left(t^{-\gamma^*}\right), t\rightarrow 0^+\right\}\\
\delta:=\sup\left\{\delta^*: f\left(t\right)=O\left(t^{-\delta^*}\right), t\rightarrow +\infty\right\}\\
C:=]0,1[\cap]1-\delta,1-\gamma[
\end{gather*}

If
\begin{equation}
f\left(t\right)\sim\exp{\left(-qt^{-\mu}\right)}\sum_{m=0}^{+\infty}\sum_{n=0}^{\overline{N}\left(m\right)}p_{mn}t^{a_m}\left(\log{\left(t\right)}\right)^n, \qquad t\rightarrow 0^+
\label{eq: asymptotic behavior in Additive Gaussian Noise average mmse general h Continuous Inputs}
\end{equation}
where $\mathcal{R}\left(q\right)\geq 0$, $\mu>0$, $\overline{N}\left(m\right)$ is finite for each $m$ and $\mathcal{R}\left(a_m\right)\uparrow +\infty$,
\begin{gather}
\left(x\not\sim\mathcal{CN}(0,1)\wedge q=0\right)\Rightarrow\mathcal{R}\left(a_0\right)>0\label{eq: real part of a_0 greater than zero in Additive Gaussian Noise average mmse general h Mellin transform Continuous Inputs}\\
\gamma<\delta\label{eq: gamma less delta in Additive Gaussian Noise average mmse general h Mellin transform Continuous Inputs}\\
C\neq\emptyset\label{eq: C not empty in Additive Gaussian Noise average mmse general h Mellin transform Continuous Inputs}\\
\exists c\in C: \forall x\in[c,+\infty[, M\left[f;1-x-iy\right]=O\left(|y|^{-2}\right), \qquad |y|\rightarrow +\infty\label{eq: asymptotic behavior in Additive Gaussian Noise average mmse general h Mellin transform Continuous Inputs}
\end{gather}
where $M\left[f;1-x-iy\right]$ is to be understood as the analytic continuation of $M\left[f;1-x-iy\right]$ from $\{x+iy: 1-\delta<x<1-\gamma\}$ to $\{x+iy: x>1-\delta\}$~\cite{introduction_to_complex_analysis_second_edition_Priestley}, then
\begin{enumerate}
	\item
If $q=0$ then:
\begin{enumerate}
	\item
If $x$ is distributed according to $\infty$-PSK it follows that
\begin{equation*}
\overline{mmse}\left(snr\right)\sim\frac{M\left[f;0\right]}{2snr}+O\left(\frac{1}{snr^R}\right), \qquad snr\rightarrow +\infty, \qquad \forall R\in]1,\min\left\{\mathcal{R}\left(a_0\right)+1,2\right\}[
\end{equation*}
	\item
If $x$ is distributed according to $\infty$-PAM it follows that
\begin{equation*}
\overline{mmse}\left(snr\right)\sim\frac{M\left[f;0\right]}{2snr}+O\left(\frac{1}{snr^R}\right), \qquad snr\rightarrow +\infty, \qquad \forall R\in\left]1,\min\left\{\mathcal{R}\left(a_0\right)+1,\frac{3}{2}\right\}\right[
\end{equation*}
	\item
If $x$ is distributed according to $\infty$-QAM it follows that
\begin{equation*}
\overline{mmse}\left(snr\right)\sim\frac{M\left[f;0\right]}{snr}+O\left(\frac{1}{snr^R}\right), \qquad snr\rightarrow +\infty, \qquad \forall R\in\left]1,\min\left\{\mathcal{R}\left(a_0\right)+1,\frac{3}{2}\right\}\right[
\end{equation*}
	\item
If $x$ is distributed according to standard complex Gaussian it follows that
\begin{align*}
\overline{mmse}\left(snr\right)&\sim\sum_{m\in R}\frac{(-1)^mM\left[f;-m\right]}{snr^{m+1}}+\nonumber\\
&\quad +
\sum_{m\in A}\frac{1}{snr^{1+a_m}}
\sum_{n=0}^{\overline{N}\left(m\right)}p_{mn}
\sum_{j=0}^n\binom{n}{j}\left(-\log{\left(snr\right)}\right)^j
\left(\frac{d}{dz}\right)^{\left(n-j\right)}\left\{\frac{\pi}{\sin{\left(\pi z\right)}}\right\}\Bigg|_{z=1+a_m}+\nonumber\\
&\quad +
\sum_{m\in RA}-\frac{1}{snr^{m+1}}
\sum_{j=0}^{\overline{N}\left(m\right)+1}
\frac{\left(-\log{\left(snr\right)}\right)^j}{j!\left(\overline{N}\left(m\right)+1-j\right)!}\times\nonumber\\
&\quad\quad\times\left(\frac{d}{dz}\right)^{\left(\overline{N}\left(m\right)+1-j\right)}\left\{\left(z-m-1\right)^{\overline{N}\left(m\right)+2}\frac{\pi M\left[f;1-z\right]}{\sin{\left(\pi z\right)}}\right\}\Bigg|_{z=m+1}\\
&\quad,\qquad snr\rightarrow +\infty
\end{align*}
where
\begin{align*}
R&:=\left\{m\in\mathbb{Z}_0^+: \nexists n\in\mathbb{Z}_0^+: m=a_n\right\}\\
A&:=\left\{n\in\mathbb{Z}_0^+: \nexists m\in\mathbb{Z}_0^+: m=a_n\right\}\\
RA&:=\left\{m\in\mathbb{Z}_0^+: m+1\in\left\{z\in\mathbb{C}: \exists\left(p,q\right)\in\mathbb{Z}_0^+\times\mathbb{Z}_0^+: z=p+1=a_q+1\right\}\right\}
\end{align*}
\end{enumerate}
	\item
If $q\neq 0$ then:
\begin{enumerate}
	\item
If $x$ is distributed according to $\infty$-PSK it follows that
\begin{equation*}
\overline{mmse}\left(snr\right)\sim\frac{M\left[f;0\right]}{2snr}+O\left(\frac{1}{snr^R}\right), \qquad snr\rightarrow +\infty, \qquad \forall R\in]1,2[
\end{equation*}
	\item
If $x$ is distributed according to $\infty$-PAM it follows that
\begin{equation*}
\overline{mmse}\left(snr\right)\sim\frac{M\left[f;0\right]}{2snr}+O\left(\frac{1}{snr^R}\right), \qquad snr\rightarrow +\infty, \qquad \forall R\in\left]1,\frac{3}{2}\right[
\end{equation*}
	\item
If $x$ is distributed according to $\infty$-QAM it follows that
\begin{equation*}
\overline{mmse}\left(snr\right)\sim\frac{M\left[f;0\right]}{snr}+O\left(\frac{1}{snr^R}\right), \qquad snr\rightarrow +\infty, \qquad \forall R\in\left]1,\frac{3}{2}\right[
\end{equation*}
	\item
If $x$ is distributed according to standard complex Gaussian it follows that
\begin{equation*}
\overline{mmse}\left(snr\right)\sim\sum_{m=0}^{+\infty}\frac{(-1)^mM\left[f;-m\right]}{snr^{m+1}}, \qquad snr\rightarrow +\infty
\end{equation*}

\end{enumerate}
\end{enumerate}
\label{thm: Additive Gaussian Noise average mmse general h Continuous Inputs}
\end{thm}

\begin{IEEEproof}
See Appendix \ref{section: Proof of Theorem Additive Gaussian Noise average mmse general h Continuous Inputs}.
\end{IEEEproof}

\vspace{0.25cm}

Theorem \ref{thm: Additive Gaussian Noise average mmse general h Continuous Inputs} provides a simple asymptotic expansion as $snr\rightarrow +\infty$ of the average MMSE in a fading coherent channel driven by $\infty$-PSK, $\infty$-PAM, $\infty$-QAM or standard complex Gaussian continuous inputs under very general fading conditions. In particular, this Theorem -- akin to Theorem \ref{thm: Additive Gaussian Noise average mmse general h} -- only requires that the asymptotic expansion as $t\rightarrow 0^+$ of $f\left(t\right)$, which relates to the probability density function of the fading random variable, behaves as in \eqref{eq: asymptotic behavior in Additive Gaussian Noise average mmse general h Continuous Inputs} and that the Mellin transform of the function $f\left(\cdot\right)$ exists, is holomorphic in a certain vertical strip in the complex plane and decays along vertical lines as in \eqref{eq: asymptotic behavior in Additive Gaussian Noise average mmse general h Mellin transform Continuous Inputs}. The error term in the asymptotic expansion is more or less refined depending on whether the fading model does not ($q=0$) or does ($q\neq 0$) accommodate for exponential decay of $f\left(t\right)$ as $t\rightarrow 0^+$.

This Theorem -- akin to Theorem \ref{thm: Additive Gaussian Noise average mmse general h} -- reveals that the asymptotic behavior as $snr\rightarrow +\infty$ of the average MMSE also depends on the asymptotic behavior as $t\rightarrow 0^+$ of $f\left(t\right)$. The proof of the Theorem also reveals that the \emph{canonical} MMSE -- via its Mellin transform -- plays an important role in the construction of the asymptotic behavior of the average MMSE.

Theorem \ref{thm: Additive Gaussian Noise average mmse general h Continuous Inputs} can also be immediately specialized to the most common fading distributions.
\begin{cor}
Consider a Rayleigh or Ricean fading coherent channel as in \eqref{eq: Additive Gaussian Noise channel} driven by an $\infty$-PSK, $\infty$-PAM, $\infty$-QAM or standard complex Gaussian continuous input, where \eqref{eq: Rayleigh fading model} or \eqref{eq: Rice fading model} hold, respectively. Then, in the regime of high-$snr$ the average MMSE behaves as follows:
\begin{enumerate}
	\item
If $x$ is distributed according to $\infty$-PSK then
\begin{equation*}
\overline{mmse}\left(snr\right)\sim\frac{1}{2snr}+O\left(\frac{1}{snr^R}\right), \qquad snr\rightarrow +\infty, \qquad \forall R\in]1,2[
\end{equation*}
	\item
If $x$ is distributed according to $\infty$-PAM then
\begin{equation*}
\overline{mmse}\left(snr\right)\sim\frac{1}{2snr}+O\left(\frac{1}{snr^R}\right), \qquad snr\rightarrow +\infty, \qquad \forall R\in\left]1,\frac{3}{2}\right[
\end{equation*}
	\item
If $x$ is distributed according to $\infty$-QAM then
\begin{equation*}
\overline{mmse}\left(snr\right)\sim\frac{1}{snr}+O\left(\frac{1}{snr^R}\right), \qquad snr\rightarrow +\infty, \qquad \forall R\in\left]1,\frac{3}{2}\right[
\end{equation*}
	\item
If $x$ is distributed according to standard complex Gaussian it follows that
\begin{align*}
\overline{mmse}\left(snr\right)&\sim\frac{1}{snr}+
\frac{1}{snr^2}\sum_{m=0}^{+\infty}-\frac{1}{snr^m}
\sum_{j=0}^1
\frac{\left(-\log{\left(snr\right)}\right)^j}{j!\left(1-j\right)!}\times\nonumber\\
&\quad\times\left(\frac{d}{dz}\right)^{\left(1-j\right)}\left\{\left(z-m-2\right)^2\frac{\pi \exp{\left(-\frac{|\mu|^2}{2\sigma^2}\right)}\left(2\sigma^2\right)^{1-z}\Gamma\left(2-z\right){_1}F_1\left(2-z;1;\frac{|\mu|^2}{2\sigma^2}\right)}{\sin{\left(\pi z\right)}}\right\}\Bigg|_{z=m+2}\\
&\quad,\qquad snr\rightarrow +\infty
\end{align*}
where ${_1}F_1\left(a;b;c\right)$ is the confluent hypergeometric series~\cite[Equation 13.2.2]{abramowitz_stegun}.
\end{enumerate}
\label{cor: Additive Gaussian Noise average mmse h Continuous Inputs Rayleigh and Rice}
\end{cor}

\begin{IEEEproof}
The conditions required for this specialization have already been established in Subappendices \ref{subsection: Proof of Corollary Additive Gaussian Noise average mmse h Rayleigh and Rice case mu=0} and \ref{subsection: Proof of Corollary Additive Gaussian Noise average mmse h Rayleigh and Rice case mu not 0}. Note also that $R=\left\{0\right\}$, $A=\emptyset$ and $RA=\mathbb{Z}^+$ in both Rayleigh and Ricean cases.
\end{IEEEproof}

\vspace{0.25cm}

\begin{cor}
Consider a Nakagami fading coherent channel as in \eqref{eq: Additive Gaussian Noise channel} driven by an $\infty$-PSK, $\infty$-PAM, $\infty$-QAM or standard complex Gaussian continuous input, where \eqref{eq: Nakagami fading model} holds. Then, in the regime of high-$snr$ the average MMSE behaves as follows:
\begin{enumerate}
	\item
If $x$ is distributed according to $\infty$-PSK then
\begin{equation*}
\overline{mmse}\left(snr\right)\sim\frac{1}{2snr}+O\left(\frac{1}{snr^R}\right), \qquad snr\rightarrow +\infty, \qquad \forall R\in]1,\min\{\mu+1,2\}[
\end{equation*}
	\item
If $x$ is distributed according to $\infty$-PAM then
\begin{equation*}
\overline{mmse}\left(snr\right)\sim\frac{1}{2snr}+O\left(\frac{1}{snr^R}\right), \qquad snr\rightarrow +\infty, \qquad \forall R\in\left]1,\frac{3}{2}\right[
\end{equation*}
	\item
If $x$ is distributed according to $\infty$-QAM then
\begin{equation*}
\overline{mmse}\left(snr\right)\sim\frac{1}{snr}+O\left(\frac{1}{snr^R}\right), \qquad snr\rightarrow +\infty, \qquad \forall R\in\left]1,\frac{3}{2}\right[
\end{equation*}
	\item
If $x$ is distributed according to standard complex Gaussian it follows that
\begin{enumerate}
	\item
If $\mu\in\left[\frac{1}{2},+\infty\right[\cap\mathbb{Z}^+$ then:
\begin{align*}
\overline{mmse}\left(snr\right)&\sim\frac{1}{snr}\sum_{m=0}^{\mu-1}\frac{(-1)^m\frac{1}{\Gamma\left(\mu\right)}\left(\frac{w}{\mu}\right)^{-m}\Gamma\left(\mu-m\right)}{snr^m}+\nonumber\\
&\quad +
\frac{1}{snr^{1+\mu}}\sum_{m=0}^{+\infty}-\frac{1}{snr^m}
\sum_{j=0}^1
\frac{\left(-\log{\left(snr\right)}\right)^j}{j!\left(1-j\right)!}\times\nonumber\\
&\quad\quad\times\left(\frac{d}{dz}\right)^{\left(1-j\right)}\left\{\left(z-m-1-\mu\right)^{2}\frac{\pi \frac{1}{\Gamma\left(\mu\right)}\left(\frac{w}{\mu}\right)^{1-z}\Gamma\left(\mu+1-z\right)}{\sin{\left(\pi z\right)}}\right\}\Bigg|_{z=m+1+\mu}\\
&\quad,\qquad snr\rightarrow +\infty
\end{align*}
	\item
If $\mu\in\left[\frac{1}{2},+\infty\right[\setminus\mathbb{Z}^+$ then:
\begin{align*}
\overline{mmse}\left(snr\right)&\sim\frac{1}{snr}\sum_{m=0}^{+\infty}\frac{(-1)^m\frac{1}{\Gamma\left(\mu\right)}\left(\frac{w}{\mu}\right)^{-m}\Gamma\left(\mu-m\right)}{snr^m}+\\
&\quad+
\frac{1}{snr^{1+\mu}}\sum_{m=0}^{+\infty}\frac{1}{snr^m}\frac{\mu^{\mu}}{\Gamma\left(\mu\right)w^{\mu}}\frac{\left(-1\right)^m\mu^m}{m!w^m}\frac{\pi}{\sin{\left(\pi\left(1+\mu+m\right)\right)}}\\
&\quad,\qquad snr\rightarrow +\infty
\end{align*}
\end{enumerate}

\end{enumerate}
\label{cor: Additive Gaussian Noise average mmse h Continuous Inputs Nakagami}
\end{cor}

\begin{IEEEproof}
The conditions required for this specialization have already been established in Appendix \ref{section: Proof of Corollary Additive Gaussian Noise average mmse h Nakagami}. Note also that
\begin{align*}
R&=
\begin{cases}
\left\{0,1,2,\ldots,\mu-1\right\} &\text{if }\mu\in\left[\frac{1}{2},+\infty\right[\cap\mathbb{Z}^+,\\
\mathbb{Z}_0^+ &\text{if }\mu\in\left[\frac{1}{2},+\infty\right[\setminus\mathbb{Z}^+,\\
\end{cases}\\
A&=
\begin{cases}
\emptyset &\text{if }\mu\in\left[\frac{1}{2},+\infty\right[\cap\mathbb{Z}^+,\\
\mathbb{Z}_0^+ &\text{if }\mu\in\left[\frac{1}{2},+\infty\right[\setminus\mathbb{Z}^+,\\
\end{cases}\\
RA&=
\begin{cases}
\left\{\mu,\mu+1,\mu+2,\ldots\right\} &\text{if }\mu\in\left[\frac{1}{2},+\infty\right[\cap\mathbb{Z}^+,\\
\emptyset &\text{if }\mu\in\left[\frac{1}{2},+\infty\right[\setminus\mathbb{Z}^+.\\
\end{cases}
\end{align*}
in the Nakagami fading case.
\end{IEEEproof}

\vspace{0.25cm}

It is evident that the high-$snr$ behavior of the average MMSE for the most common fading coherent channel models is rather distinct for discrete and continuous inputs. For systems driven by discrete inputs the leading term of the high-$snr$ asymptotic expansion of the average MMSE is $O\left(\frac{1}{snr^2}\right)$ for Rayleigh and Ricean fading channels and $O\left(\frac{1}{snr^{1+\mu}}\right)$ for Nakagami fading channels; in contrast, for systems driven by $\infty$-PSK, $\infty$-PAM, $\infty$-QAM or standard complex Gaussian inputs the leading term in the asymptotic expansion of the average MMSE is $O\left(\frac{1}{snr}\right)$ for the four fading models. This is consistent with the fact that the high-$snr$ average mutual information for systems with continuous inputs is higher than for systems with discrete inputs, i.e., it grows without bound for continuous inputs but it is bounded by the input cardinality for discrete inputs.

The high-$snr$ asymptotic expansions of the average MMSE embodied in Corollaries \ref{cor: Additive Gaussian Noise average mmse h Continuous Inputs Rayleigh and Rice} and \ref{cor: Additive Gaussian Noise average mmse h Continuous Inputs Nakagami} do not lead directly to a high-$snr$ asymptotic expansion of the average mutual information, because it is not possible to explore a suitable integral representation of the connection between average MMSE and average mutual information in \eqref{eq: average mi-average mmse relation}.

\section{Low-snr Regime}
\label{section: Low-snr Regime}
We now consider the construction of low-$snr$ asymptotic expansions of the average MMSE and the average mutual information in a fading coherent channel driven by a range of inputs. The element of novelty, in view of the fact that examples of low-$snr$ asymptotic expansions of a series of estimation- and information-theoretic quantities are plentiful (see e.g.~\cite{Verdu02},~\cite{Second-Order_Asymptotics_Of_Mutual_Information_Prelov_Verdu}), relates to the use of Mellin transform expansions techniques to study the behavior of the quantities in such an asymptotic regime. This distinct unconventional approach, as opposed to the common method that involves in some suitable integral representation of the quantities the substitution of the integrand or part of the integrand by a series and its integration term by term, also illustrates that with Mellin transform expansions techniques one can often derive with little effort the asymptotic expansions as $snr\rightarrow 0^+$ from asymptotic expansions as $snr\rightarrow +\infty$ and vice versa thereby coupling the regimes~\cite{asymptotic_expansion_of_integrals_Bleistein_and_Handelsman}.

We also proceed by obtaining a low-$snr$ asymptotic expansion of the average MMSE and then -- via the connection between average MMSE and average mutual information in \eqref{eq: average mi-average mmse relation} -- a low-$snr$ asymptotic expansion of the average mutual information.

\subsection{Discrete and Continuous Inputs}
We consider arbitrary discrete inputs with finite support and unit-power $\infty$-PSK, $\infty$-PAM, $\infty$-QAM or standard complex Gaussian continuous inputs\footnote{The expansions are also applicable to finite-power inputs, i.e., $E_{x}\{|x|^2\}<+\infty$, such that the \emph{canonical} MMSE obeys $mmse(snr)=O\left(\frac{1}{snr}\right), snr\rightarrow +\infty$.}. The following Theorem, which applies to relatively general fading models, provides the asymptotic expansion of the average MMSE in fading coherent channels driven by a range of inputs.

\begin{thm}
Consider the fading coherent channel in \eqref{eq: Additive Gaussian Noise channel} driven by inputs that conform to: i\emph{)} a discrete distribution with finite support; or ii\emph{)} a $\infty$-PSK, $\infty$-PAM, $\infty$-QAM or standard complex Gaussian continuous distribution. Define
\begin{align*}
h_1\left(t\right)&:=\frac{\sqrt{t}f_{|h|}\left(\sqrt{t}\right)}{2}
\end{align*}
which relates to the distribution of the fading process and
\begin{align*}
\alpha_1 &:=\inf\left\{\alpha_1^*: h_1\left(t\right)=O\left(t^{-\alpha_1^*}\right), t\rightarrow 0^+\right\}\\
\beta_1 &:=\sup\left\{\beta_1^*:   h_1\left(t\right)=O\left(t^{-\beta_1^*}\right), t\rightarrow +\infty\right\}\\
C_1&:=
\begin{cases}
]\alpha_1,\beta_1[\cap]-\infty,1[ &\text{if $x$ is discrete},\\
]\alpha_1,\beta_1[\cap]0,1[ &\text{if $x$ is continuous}.\\
\end{cases}
\end{align*}

If
\begin{equation}
h_1\left(t\right)=O\left(\exp{\left(-k_1t^{v_1}\right)}\right), \qquad t\rightarrow +\infty
\label{eq: asymptotic behavior in Additive Gaussian Noise average mmse general h Low-snr}
\end{equation}
where $\mathcal{R}\left(k_1\right)>0$ and $v_1>0$,
\begin{gather}
\alpha_1<\beta_1\label{eq: gamma less delta in Additive Gaussian Noise average mmse general h Low-snr Mellin transform}\\
C_1\neq\emptyset\label{eq: C not empty in Additive Gaussian Noise average mmse general h Low-snr Mellin transform}\\
\exists c_1\in C_1: \forall x\in[c_1,+\infty[, M\left[h_1;x+iy\right]=O\left(|y|^{-2}\right), \qquad |y|\rightarrow +\infty\label{eq: asymptotic behavior in Additive Gaussian Noise average mmse general h Low-snr Mellin transform}
\end{gather}
where $M\left[h_1;x+iy\right]$ is to be understood as the analytic continuation of $M\left[h_1;x+iy\right]$ from $\{x+iy: \alpha_1<x<\beta_1\}$ to $\{x+iy: x>\alpha_1\}$]~\cite{introduction_to_complex_analysis_second_edition_Priestley}, then
\begin{equation*}
\overline{mmse}\left(snr\right)\sim\sum_{m=0}^{+\infty}\frac{1}{m!}M[h_1;m+1]mmse^{(m)}\left(z\right)\Bigg|_{z=0^+}snr^m, \qquad snr\rightarrow 0^+
\end{equation*}
\label{thm: Additive Gaussian Noise average mmse general h Low-snr}
\end{thm}

\begin{IEEEproof}
See Appendix \ref{section: Proof of Theorem Additive Gaussian Noise average mmse general h Low-snr}.
\end{IEEEproof}

\vspace{0.25cm}

Theorem \ref{thm: Additive Gaussian Noise average mmse general h Low-snr} can also be immediately specialized to the most common fading distributions, including : \emph{i}) Rayleigh fading; \emph{ii}) Ricean fading; and \emph{iii}) Nakagami fading.

\begin{cor}
Consider a Rayleigh or Ricean fading coherent channel as in \eqref{eq: Additive Gaussian Noise channel} driven by an arbitrary discrete input with finite support or by $\infty$-PSK, $\infty$-PAM, $\infty$-QAM or standard complex Gaussian continuous inputs, where \eqref{eq: Rayleigh fading model} or \eqref{eq: Rice fading model} hold, respectively. Then, in the regime of low-$snr$ the average MMSE behaves as follows:
\begin{align*}
\overline{mmse}\left(snr\right)&\sim\exp{\left(-\frac{|\mu|^2}{2\sigma^2}\right)}\sum_{m=0}^{+\infty}\left(m+1\right)\left(2\sigma^2\right)^{m+1}{_1}F_1\left(2+m;1;\frac{|\mu|^2}{2\sigma^2}\right)mmse^{(m)}\left(z\right)\Bigg|_{z=0^+}snr^m, \qquad snr\rightarrow 0^+
\end{align*}
where ${_1}F_1\left(a;b;c\right)$ is the confluent hypergeometric series~\cite[Equation 13.2.2]{abramowitz_stegun}.
\label{cor: Additive Gaussian Noise average mmse h Rayleigh and Rice Low-snr}
\end{cor}

\begin{IEEEproof}
See Appendix \ref{section: Proof of Corollary Additive Gaussian Noise average mmse h Rayleigh and Rice Low-snr}.
\end{IEEEproof}

\vspace{0.25cm}

\begin{cor}
Consider a Nakagami fading coherent channel as in \eqref{eq: Additive Gaussian Noise channel} driven by an arbitrary discrete input with finite support or by $\infty$-PSK, $\infty$-PAM, $\infty$-QAM or standard complex Gaussian continuous inputs, where \eqref{eq: Nakagami fading model} holds. Then, in the regime of low-$snr$ the average MMSE behaves as follows:
\begin{equation*}
\overline{mmse}\left(snr\right)\sim\sum_{m=0}^{+\infty}\frac{\Gamma\left(\mu+m+1\right)}{\Gamma\left(\mu\right)\Gamma\left(m+1\right)}\left(\frac{w}{\mu}\right)^{m+1}mmse^{(m)}\left(z\right)\Bigg|_{z=0^+}snr^m, \qquad snr\rightarrow 0^+
\end{equation*}
\label{cor: Additive Gaussian Noise average mmse h Nakagami Low-snr}
\end{cor}

\begin{IEEEproof}
See Appendix \ref{section: Proof of Corollary Additive Gaussian Noise average mmse h Nakagami Low-snr}.
\end{IEEEproof}

\vspace{0.25cm}

The low-$snr$ asymptotic expansions of the average MMSE embodied in Corollaries \ref{cor: Additive Gaussian Noise average mmse h Rayleigh and Rice Low-snr} and \ref{cor: Additive Gaussian Noise average mmse h Nakagami Low-snr} also lead immediately to a low-$snr$ asymptotic expansion of the average mutual information, by leveraging the following integral representation of the connection between average MMSE and average mutual information in \eqref{eq: average mi-average mmse relation} given by:
\begin{equation}
\overline{I}\left(snr\right)=\int_0^{snr}\overline{mmse}\left(\epsilon\right)d\epsilon
\label{eq: average mi-average mmse relation integral representation low snr}.
\end{equation}

\begin{cor}
Consider a Rayleigh or Ricean fading coherent channel as in \eqref{eq: Additive Gaussian Noise channel} driven by an arbitrary discrete input with finite support or by $\infty$-PSK, $\infty$-PAM, $\infty$-QAM or standard complex Gaussian continuous inputs, where \eqref{eq: Rayleigh fading model} or \eqref{eq: Rice fading model} hold, respectively. Then, the average mutual information obeys the low-$snr$ expansion given by:
\begin{equation*}
\overline{I}\left(snr\right)\sim\exp{\left(-\frac{|\mu|^2}{2\sigma^2}\right)}\sum_{m=0}^{+\infty}\left(2\sigma^2\right)^{m+1}{_1}F_1\left(2+m;1;\frac{|\mu|^2}{2\sigma^2}\right)mmse^{(m)}\left(z\right)\Bigg|_{z=0^+}snr^{m+1}, \qquad snr\rightarrow 0^+
\end{equation*}
where ${_1}F_1\left(a;b;c\right)$ is the confluent hypergeometric series~\cite[Equation 13.2.2]{abramowitz_stegun}.
\label{cor: Additive Gaussian Noise average mi h Rayleigh and Rice Low-snr}
\end{cor}

\begin{cor}
Consider a Nakagami fading coherent channel as in \eqref{eq: Additive Gaussian Noise channel} driven by an arbitrary discrete input with finite support or by $\infty$-PSK, $\infty$-PAM, $\infty$-QAM or standard complex Gaussian continuous inputs, where \eqref{eq: Nakagami fading model} holds. Then, the average mutual information obeys the low-$snr$ expansion given by:
\begin{equation*}
\overline{I}\left(snr\right)\sim\sum_{m=0}^{+\infty}\frac{1}{m+1}\frac{\Gamma\left(\mu+m+1\right)}{\Gamma\left(\mu\right)\Gamma\left(m+1\right)}\left(\frac{w}{\mu}\right)^{m+1}mmse^{(m)}\left(z\right)\Bigg|_{z=0^+}snr^{m+1}, \qquad snr\rightarrow 0^+
\end{equation*}
\label{cor: Additive Gaussian Noise average mi h Nakagami Low-snr}
\end{cor}

\begin{IEEEproof}[Proof of Corollaries \ref{cor: Additive Gaussian Noise average mi h Rayleigh and Rice Low-snr} and \ref{cor: Additive Gaussian Noise average mi h Nakagami Low-snr}]
The Corollaries follow immediately from Corollaries \ref{cor: Additive Gaussian Noise average mmse h Rayleigh and Rice Low-snr} or \ref{cor: Additive Gaussian Noise average mmse h Nakagami Low-snr} and the relationship in \eqref{eq: average mi-average mmse relation integral representation low snr}, together with the fact that an order relation can be integrated with respect to the independent variable~\cite{asymptotic_expansion_of_integrals_Bleistein_and_Handelsman}.
\end{IEEEproof}

\vspace{0.25cm}

It is interesting to note the role that the \emph{canonical} MMSE plays in the definition of the asymptotic expansions of the average MMSE and the average mutual information as $snr\rightarrow 0^+$ and $snr\rightarrow +\infty$. The high-$snr$ behavior of the quantities (for discrete inputs) is dictated via the Mellin transform and the higher-order derivatives of the Mellin transform of the \emph{canonical} MMSE or, equivalently, the repeated integrals of the \emph{canonical} MMSE. In contrast, the low-$snr$ behavior of the quantities is dictated via the derivatives of the \emph{canonical} MMSE.

\section{Generalizations}
\label{section: Generalizations}

We now consider a more general frequency-flat fading channel, which for a single time instant, can be modeled as follows:
\begin{equation}
\boldsymbol{y}=\sqrt{snr}\boldsymbol{h}x+\boldsymbol{n}
\label{eq: Additive Gaussian Noise channel Vector}
\end{equation}
where $\boldsymbol{y}=[y_1,\ldots,y_k]^T \in \mathbb{C}^k$ represents the channel output, $x\in\mathbb{C}$ represents the channel input, $\boldsymbol{h}=[h_1,\ldots,h_k]^T$ is a complex random vector such that $E_{\boldsymbol{h}}\left\{\|\boldsymbol{h}\|^2\right\}<+\infty$, which represents the random channel fading gains between the input and the outputs of the channel, and $\boldsymbol{n}=[n_1,\ldots,n_k]^T \in \mathbb{C}^k$ is a circularly symmetric complex Gaussian random vector with zero mean and identity covariance matrix which represents standard noise. The scaling factor $snr\in\mathbb{R}^+$ also relates to the signal-to-noise ratio. We assume that $x$, $\boldsymbol{h}$ and $\boldsymbol{n}$ are independent random variables/vectors. We also assume that the receiver knows the exact realization of the channel gains but the transmitter knows only the distribution of the channel gains. Note that the model in \eqref{eq: Additive Gaussian Noise channel Vector} represents a generalization of the model in \eqref{eq: Additive Gaussian Noise channel}, that captures various communications scenarios such as the use of multiple antennas at the receiver.

We write the average MMSE associated with the model in \eqref{eq: Additive Gaussian Noise channel Vector} as follows:
\begin{equation}
\overline{mmse}\left(snr\right):=E_{\boldsymbol{h}}\left\{mmse_{\boldsymbol{h}}\left(snr\right)\right\}
\label{eq: definition average mmse_vector}
\end{equation}
where
\begin{equation*}
mmse_{\boldsymbol{h}_0}\left(snr\right):=E_{x,\boldsymbol{y}|\boldsymbol{h}}\left\{\left\|\boldsymbol{h}_0x-E_{x|\boldsymbol{y},\boldsymbol{h}}\left\{\boldsymbol{h}_0x|\boldsymbol{y},\boldsymbol{h}_0\right\}\right\|^2\big|\boldsymbol{h}=\boldsymbol{h}_0\right\}
\end{equation*}
represents the conditional MMSE given that the channel gains $\boldsymbol{h}=\boldsymbol{h}_0$ associated with the estimation of the noiseless output given the noisy output of the channel model in \eqref{eq: Additive Gaussian Noise channel Vector}. We also write the average mutual information associated with the model in \eqref{eq: Additive Gaussian Noise channel Vector} as follows:
\begin{equation}
\overline{I}\left(snr\right):=E_{\boldsymbol{h}}\left\{I_{\boldsymbol{h}}\left(snr\right)\right\}
\label{eq: definition average mi_vector}
\end{equation}
where
\begin{equation*}
I_{\boldsymbol{h}_0}\left(snr\right):=E_{x,\boldsymbol{y}|\boldsymbol{h}}\left\{\log{\left(\frac{f_{x,\boldsymbol{y}|\boldsymbol{h}}\left(x,\boldsymbol{y}|\boldsymbol{h}_0\right)}{f_{x|\boldsymbol{h}}\left(x|\boldsymbol{h}_0\right)f_{\boldsymbol{y}|\boldsymbol{h}}\left(\boldsymbol{y}|\boldsymbol{h}_0\right)}\right)}\Bigg|\boldsymbol{h}=\boldsymbol{h}_0\right\}
\end{equation*}
represents the conditional mutual information given that the channel gains $\boldsymbol{h}=\boldsymbol{h}_0$ between the input and the output of the channel model in \eqref{eq: Additive Gaussian Noise channel Vector}. We can also immediately write the average MMSE in \eqref{eq: definition average mmse_vector} in terms of the \emph{canonical} MMSE in \eqref{eq: relation between cammse and commse} as:
\begin{equation}
\overline{mmse}\left(snr\right)=E_{\|\boldsymbol{h}\|}\left\{\|\boldsymbol{h}\|^2mmse\left(snr\|\boldsymbol{h}\|^2\right)\right\}
\label{eq: average mmse-mmse relation_vector}
\end{equation}
and the average mutual information in \eqref{eq: definition average mi_vector} in terms of the \emph{canonical} mutual information in \eqref{eq: relation between cami and comi} as:
\begin{equation}
\overline{I}\left(snr\right)=E_{\|\boldsymbol{h}\|}\left\{I\left(snr\|\boldsymbol{h}\|^2\right)\right\}
\label{eq: average mi-mi relation_vector}
\end{equation}

The fact that the form of \eqref{eq: average mmse-mmse relation_vector} and \eqref{eq: average mi-mi relation_vector} is identical to the form of \eqref{eq: average mmse-mmse relation} and \eqref{eq: average mi-mi relation}, respectively, enables the use of the previous  machinery to characterize the asymptotic behavior, as $snr\rightarrow\infty$ or as $snr\rightarrow 0^+$, of the quantities.\footnote{It is important to note that, whilst the generalization of the techniques is immediate from single-input--single-output to single-input--multiple-output channel models, such generalization does not seem possible to multiple-input--single-output or multiple-input--multiple-output channel models.} 

\subsection{High-snr Regime}

\subsubsection{Discrete Inputs}
We now concentrate on the generalization of the characterizations of the asymptotic behavior as $snr \to +\infty$ of the average MMSE and the average mutual information for vector Rayleigh and Ricean fading coherent channels driven by arbitrary discrete inputs with finite support.

\begin{cor}
Consider a vector fading coherent channel as in \eqref{eq: Additive Gaussian Noise channel Vector} driven by an arbitrary discrete input with finite support, where
$\boldsymbol{h}\sim\mathcal{CN}\left(\boldsymbol{\mu},2\sigma^2\boldsymbol{I}_k\right)$ with $\boldsymbol{\mu}\in\mathbb{C}^{k\times 1}$, and $\sigma>0$. Then, in the regime of high-$snr$ the average MMSE obeys:
\begin{align*}
&{\overline{mmse}}\left(snr\right)\sim\\
&\sim\frac{\exp{\left(-\frac{\|\boldsymbol{\mu}\|^2}{2\sigma^2}\right)}}{snr^{k+1}}\sum_{m=0}^{+\infty}\Bigg(\sum_{n=0}^m\left(\frac{\left(-1\right)^{m-n}}{\left(m-n\right)!\left(2\sigma^2\right)^{m+n+k}}\frac{\|\boldsymbol{\mu}\|^{2n}}{n!\Gamma\left(n+k\right)}\right)M\left[{mmse};k+1+m\right]\Bigg)\frac{1}{snr^m},\qquad snr\rightarrow +\infty
\end{align*}
\label{cor: Additive Gaussian Noise average mmse h Rayleigh and Rice Vector Version}
\end{cor}

\begin{IEEEproof}
See Appendix \ref{section: Proof of Corollary Additive Gaussian Noise average mmse h Rayleigh and Rice Vector Version}.
\end{IEEEproof}

\vspace{0.25cm}

\begin{cor}
Consider a vector fading coherent channel as in \eqref{eq: Additive Gaussian Noise channel Vector} driven by an arbitrary discrete input with finite support, where
$\boldsymbol{h}\sim\mathcal{CN}\left(\boldsymbol{\mu},2\sigma^2\boldsymbol{I}_k\right)$ with $\boldsymbol{\mu}\in\mathbb{C}^{k\times 1}$, and $\sigma>0$. Then, in the regime of high-$snr$ the average mutual information obeys the expansion:
\begin{align*}
&{\overline{I}}\left(snr\right)\sim\log{\left(m\right)}-\\
&-\frac{\exp{\left(-\frac{\|\boldsymbol{\mu}\|^2}{2\sigma^2}\right)}}{snr^k}\sum_{m=0}^{+\infty}\Bigg(\sum_{n=0}^m\left(\frac{\left(-1\right)^{m-n}}{\left(m+k\right)\left(m-n\right)!\left(2\sigma^2\right)^{m+n+k}}\frac{\|\boldsymbol{\mu}\|^{2n}}{n!\Gamma\left(n+k\right)}\right)M\left[{mmse};k+1+m\right]\Bigg)\frac{1}{snr^m},\qquad snr\rightarrow +\infty
\end{align*}
\label{cor: Additive Gaussian Noise average i h Rayleigh and Ricean Vector Version}
\end{cor}

\begin{IEEEproof}
The expansion follows immediately from the expansion in Corollary \ref{cor: Additive Gaussian Noise average mmse h Rayleigh and Rice Vector Version} and the relationship in \eqref{eq: average mi-average mmse relation integral representation}, together with the fact that an order relation can be integrated with respect to the independent variable~\cite{asymptotic_expansion_of_integrals_Bleistein_and_Handelsman}.
\end{IEEEproof}

\vspace{0.25cm}

Note that the rate at which the average MMSE and the average mutual information tend to their infinite-$snr$ values in the scalar channel model in \eqref{eq: Additive Gaussian Noise channel} is equal to $2$ and $1$, respectively, whereas the rate at which such quantities tend to their infinite-$snr$ values in the vector channel model in \eqref{eq: Additive Gaussian Noise channel Vector} is equal to $k+1$ and $k$, respectively. This indicts, as expected, that the availability of multiple receive dimensions leads to a lower value for the average MMSE and a higher value for the average mutual information for a certain signal-to-noise ratio.

\subsubsection{Continuous Inputs}
We now focus on the generalization of the characterizations of the asymptotic behavior as $snr \to +\infty$ of the average MMSE and the average mutual information for vector Rayleigh and Ricean fading coherent channels driven by continuous inputs.

\begin{cor}
Consider a vector fading coherent channel as in \eqref{eq: Additive Gaussian Noise channel Vector} driven by an $\infty$-PSK, $\infty$-PAM, $\infty$-QAM or standard complex Gaussian continuous input, where
$\boldsymbol{h}\sim\mathcal{CN}\left(\boldsymbol{\mu},2\sigma^2\boldsymbol{I}_k\right)$ with $\boldsymbol{\mu}\in\mathbb{C}^{k\times 1}$, and $\sigma>0$. Then, in the regime of high-$snr$ the average MMSE obeys:
\begin{enumerate}
	\item
If $x$ is distributed according to $\infty$-PSK then
\begin{equation*}
{\overline{mmse}}\left(snr\right)\sim\frac{1}{2snr}+O\left(\frac{1}{snr^R}\right), \qquad snr\rightarrow +\infty, \qquad \forall R\in]1,2[
\end{equation*}
	\item
If $x$ is distributed according to $\infty$-PAM then
\begin{equation*}
{\overline{mmse}}\left(snr\right)\sim\frac{1}{2snr}+O\left(\frac{1}{snr^R}\right), \qquad snr\rightarrow +\infty, \qquad \forall R\in\left]1,\frac{3}{2}\right[
\end{equation*}
	\item
If $x$ is distributed according to $\infty$-QAM then
\begin{equation*}
{\overline{mmse}}\left(snr\right)\sim\frac{1}{snr}+O\left(\frac{1}{snr^R}\right), \qquad snr\rightarrow +\infty, \qquad \forall R\in\left]1,\frac{3}{2}\right[
\end{equation*}
	\item
If $x$ is distributed according to standard complex Gaussian then
\begin{align*}
&\overline{mmse}\left(snr\right)\sim\nonumber\\
&\sim\sum_{m=0}^{k-1}\frac{(-1)^m\exp{\left(-\frac{\|\boldsymbol{\mu}\|^2}{2\sigma^2}\right)}\left(2\sigma^2\right)^{-m}\frac{\Gamma\left(-m+k\right)}{\left(k-1\right)!}{_1}F_1\left(-m+k;k;\frac{\|\boldsymbol{\mu}\|^2}{2\sigma^2}\right)}{snr^{m+1}}+\nonumber\\
&\quad +
\frac{1}{snr^{k+1}}\sum_{m=0}^{+\infty}-\frac{1}{snr^{m}}
\sum_{j=0}^{1}
\frac{\left(-\log{\left(snr\right)}\right)^j}{j!\left(1-j\right)!}\times\nonumber\\
&\quad\quad\times\left(\frac{d}{dz}\right)^{\left(1-j\right)}\left\{\left(z-m-k-1\right)^{2}\frac{\pi \exp{\left(-\frac{\|\boldsymbol{\mu}\|^2}{2\sigma^2}\right)}\left(2\sigma^2\right)^{1-z}\frac{\Gamma\left(1-z+k\right)}{\left(k-1\right)!}{_1}F_1\left(1-z+k;k;\frac{\|\boldsymbol{\mu}\|^2}{2\sigma^2}\right)}{\sin{\left(\pi z\right)}}\right\}\Bigg|_{z=m+k+1}\\
&\quad,\qquad snr\rightarrow +\infty
\end{align*}
where ${_1}F_1\left(a;b;c\right)$ is the confluent hypergeometric series~\cite[Equation 13.2.2]{abramowitz_stegun}.
\end{enumerate}
\label{cor: Additive Gaussian Noise average mmse h Continuous Inputs Rayleigh and Rice Vector Version}
\end{cor}

\begin{IEEEproof}
The proof for the case $\boldsymbol{\mu}=\boldsymbol{0}$ follows the steps in Appendix \ref{subsection: Proof of Corollary Additive Gaussian Noise average mmse h Rayleigh and Rice case mu=0} with some minor modifications that have been reported in Appendix \ref{subsection: Proof of Corollary Additive Gaussian Noise average mmse h Rayleigh and Rice Vector Version case mu=0}. Similarly, the proof for the case $\boldsymbol{\mu}\neq\boldsymbol{0}$ follows the steps in Appendix \ref{subsection: Proof of Corollary Additive Gaussian Noise average mmse h Rayleigh and Rice case mu not 0} with some minor modifications that have been reported in Appendix \ref{subsection: Proof of Corollary Additive Gaussian Noise average mmse h Rayleigh and Rice Vector Version case mu not 0}. Note also that $R=\left\{0,1,\ldots,k-1\right\}$, $A=\emptyset$ and $RA=\mathbb{Z}_0^+\setminus\left\{0,1,\ldots,k-1\right\}$ in both vector Rayleigh and vector Ricean cases.
\end{IEEEproof}

\vspace{0.25cm}

Note that the rates at which the average MMSE and the average mutual information tend to their infinite-$snr$ values in the scalar channel model in \eqref{eq: Additive Gaussian Noise channel} and the vector channel model in \eqref{eq: Additive Gaussian Noise channel Vector} are identical. In fact, the leading first-order term in the expansions of the quantities is identical but subsequent higher-order terms in the expansions may differ.

\subsection{Low-snr Regime}

\subsubsection{Discrete and Continuous Inputs}
The generalization of the characterizations of the asymptotic behavior as $snr \to 0^+$ of the average MMSE and the average mutual information for vector Rayleigh and Ricean fading coherent channels is also immediate.

\begin{cor}
Consider a vector fading coherent channel as in \eqref{eq: Additive Gaussian Noise channel Vector} driven by an arbitrary discrete input with finite support or by $\infty$-PSK, $\infty$-PAM, $\infty$-QAM or standard complex Gaussian continuous inputs, where
$\boldsymbol{h}\sim\mathcal{CN}\left(\boldsymbol{\mu},2\sigma^2\boldsymbol{I}_k\right)$ with $\boldsymbol{\mu}\in\mathbb{C}^{k\times 1}$, and $\sigma>0$. Then, in the regime of low-$snr$ the average MMSE obeys:
\begin{align*}
&{\overline{mmse}}\left(snr\right)\sim\\
&\sim\exp{\left(-\frac{\|\boldsymbol{\mu}\|^2}{2\sigma^2}\right)}\sum_{m=0}^{+\infty}\frac{1}{m!}\left(2\sigma^2\right)^{m+1}\frac{\Gamma\left(m+1+k\right)}{\left(k-1\right)!}{_1}F_1\left(m+1+k;k;\frac{\|\boldsymbol{\mu}\|^2}{2\sigma^2}\right)mmse^{(m)}\left(z\right)\Bigg|_{z=0^+}snr^m,\qquad snr\rightarrow 0^+
\end{align*}
where ${_1}F_1\left(a;b;c\right)$ is the confluent hypergeometric series~\cite[Equation 13.2.2]{abramowitz_stegun}.
\label{cor: Additive Gaussian Noise average mmse h Rayleigh and Rice Low-snr Vector Version}
\end{cor}

\begin{IEEEproof}
The proof for the case $\boldsymbol{\mu}=\boldsymbol{0}$ follows the steps in Appendix \ref{subsection: Proof of Corollary Additive Gaussian Noise average mmse h Rayleigh and Rice Low-snr case mu=0} with some minor modifications that have been reported in Appendix \ref{subsection: Proof of Corollary Additive Gaussian Noise average mmse h Rayleigh and Rice Vector Version case mu=0}. Similarly, the proof for the case $\boldsymbol{\mu}\neq\boldsymbol{0}$ follows the steps in Appendix \ref{subsection: Proof of Corollary Additive Gaussian Noise average mmse h Rayleigh and Rice Low-snr case mu not 0} with some minor modifications that have been reported in Appendix \ref{subsection: Proof of Corollary Additive Gaussian Noise average mmse h Rayleigh and Rice Vector Version case mu not 0}.
\end{IEEEproof}

\vspace{0.25cm}

\begin{cor}
Consider a vector fading coherent channel as in \eqref{eq: Additive Gaussian Noise channel Vector} driven by an arbitrary discrete input with finite support or by $\infty$-PSK, $\infty$-PAM, $\infty$-QAM or standard complex Gaussian continuous inputs, where
$\boldsymbol{h}\sim\mathcal{CN}\left(\boldsymbol{\mu},2\sigma^2\boldsymbol{I}_k\right)$ with $\boldsymbol{\mu}\in\mathbb{C}^{k\times 1}$, and $\sigma>0$. Then, in the regime of low-$snr$ the average mutual information obeys the expansion:
\begin{align*}
&{\overline{I}}\left(snr\right)\sim\\
&\sim\exp{\left(-\frac{\|\boldsymbol{\mu}\|^2}{2\sigma^2}\right)}\sum_{m=0}^{+\infty}\frac{1}{\left(m+1\right)!}\left(2\sigma^2\right)^{m+1}\frac{\Gamma\left(m+1+k\right)}{\left(k-1\right)!}{_1}F_1\left(m+1+k;k;\frac{\|\boldsymbol{\mu}\|^2}{2\sigma^2}\right)mmse^{(m)}\left(z\right)\Bigg|_{z=0^+}snr^{m+1},\qquad snr\rightarrow 0^+
\end{align*}
where ${_1}F_1\left(a;b;c\right)$ is the confluent hypergeometric series~\cite[Equation 13.2.2]{abramowitz_stegun}.
\label{cor: Additive Gaussian Noise average mi h Rayleigh and Rice Low-snr Vector Version}
\end{cor}

\begin{IEEEproof}
The expansion follows immediately from the expansion in Corollary \ref{cor: Additive Gaussian Noise average mmse h Rayleigh and Rice Low-snr Vector Version} and the relationship in \eqref{eq: average mi-average mmse relation integral representation}, together with the fact that an order relation can be integrated with respect to the independent variable~\cite{asymptotic_expansion_of_integrals_Bleistein_and_Handelsman}.
\end{IEEEproof}

\vspace{0.25cm}

\section{Some MMSE Mellin Transform Results}
\label{section: Some MMSE Mellin Transform Results}
It has been established that the Mellin transform of the \emph{canonical} MMSE given by:
\begin{equation}
M\left[mmse;1+z\right]:=\int_0^{+\infty}t^zmmse\left(t\right)dt,
\label{eq: Mellin transform of the canonical mmse}
\end{equation}
plays an important role in the definition of the high-$snr$ behavior of the average MMSE and the average mutual information in fading coherent channels driven by a range of inputs. The objective now is to compute either analytically or numerically such a quantity for the most common input distributions, which when used in conjunction with the previous results, leads to concrete asymptotic expansions.

\subsection{BPSK and QPSK Inputs}

The Mellin transform of the \emph{canonical} MMSE associated with binary phase shift keying (BPSK) and quadrature phase shift keying (QPSK) inputs can be obtained analytically. These results capitalize on the following Theorem.

\begin{thm}
Consider the \emph{canonical} AWGN channel model in \eqref{eq: Canonical Additive Gaussian Noise channel}, where the input $x$ is uniformly distributed over $\left\{-\frac{d}{2},\frac{d}{2}\right\}$ with $d>0$. Then, $\forall z\in\mathbb{C}: \mathcal{R}\left(z\right)>0$,
\begin{align*}
&M\left[mmse;1+z\right]=\\
&=\frac{d^2}{2}\left(\left(\frac{2}{d}\right)^{2\left(1+z\right)}\frac{\Gamma\left(\frac{3}{2}+z\right)}{\sqrt{\pi}\left(1+z\right)}+2d^{-2\left(1+z\right)}\frac{\Gamma\left(2+2z\right)}{\Gamma\left(2+z\right)}\sum_{l=1}^{+\infty}\left(-1\right)^l\frac{{_2}F_1\left(1,\frac{1}{2};2+z;1-\frac{1}{\left(1+2l\right)^2}\right)}{1+2l}\right)
\end{align*}
where ${_2}F_1\left(a,b;c;d\right)$ is the Gauss hypergeometric series~\cite[Equation 15.2.1]{abramowitz_stegun}.
\label{thm: specialized Mellin transform of mmse for a small generalization of BPSK}
\end{thm}

\begin{IEEEproof}
See Appendix \ref{section: Proof of Theorem specialized Mellin transform of mmse for a small generalization of BPSK}.
\end{IEEEproof}

\vspace{0.25cm}

This Mellin transform can be immediately specialized for BPSK and QPSK inputs.

\begin{thm}
Consider the \emph{canonical} AWGN channel model in \eqref{eq: Canonical Additive Gaussian Noise channel} driven by a standard unit-power BPSK input. Then, $\forall z\in\mathbb{C}: \mathcal{R}\left(z\right)>0$,
\begin{equation*}
M\left[mmse;1+z\right]=2\Bigg(\frac{\Gamma\left(\frac{3}{2}+z\right)}{\sqrt{\pi}\left(1+z\right)}+2^{-1-2z}\frac{\Gamma\left(2+2z\right)}{\Gamma\left(2+z\right)}\sum_{l=1}^{+\infty}\left(-1\right)^l\frac{{_2}F_1\left(1,\frac{1}{2};2+z;1-\frac{1}{\left(1+2l\right)^2}\right)}{1+2l}\Bigg)
\end{equation*}
\label{thm: specialized Mellin transform of mmse for 2-PAM/BPSK}
\end{thm}

\begin{IEEEproof}
The result follows from Theorem \ref{thm: specialized Mellin transform of mmse for a small generalization of BPSK} and the fact that the input $x$ is uniformly distributed over $\left\{-1,1\right\}$.
\end{IEEEproof}

\vspace{0.25cm}

\begin{thm}
Consider the \emph{canonical} AWGN channel model in \eqref{eq: Canonical Additive Gaussian Noise channel} driven by a standard unit-power QPSK input. Then, $\forall z\in\mathbb{C}: \mathcal{R}\left(z\right)>0$,
\begin{equation*}
M\left[mmse;1+z\right]=2^{2+z}\Bigg(\frac{\Gamma\left(\frac{3}{2}+z\right)}{\sqrt{\pi}\left(1+z\right)}+2^{-1-2z}\frac{\Gamma\left(2+2z\right)}{\Gamma\left(2+z\right)}\sum_{l=1}^{+\infty}\left(-1\right)^l\frac{{_2}F_1\left(1,\frac{1}{2};2+z;1-\frac{1}{\left(1+2l\right)^2}\right)}{1+2l}\Bigg)
\end{equation*}
\label{cor: specialized Mellin transform of mmse for 4-QAM/QPSK}
\end{thm}

\begin{IEEEproof}
The result follows from Theorem \ref{thm: specialized Mellin transform of mmse for 2-PAM/BPSK} and the fact that the input $x$ is uniformly distributed over $\left\{-\frac{1}{\sqrt{2}}-j\frac{1}{\sqrt{2}},+\frac{1}{\sqrt{2}}-j\frac{1}{\sqrt{2}},+\frac{1}{\sqrt{2}}+j\frac{1}{\sqrt{2}},-\frac{1}{\sqrt{2}}+j\frac{1}{\sqrt{2}}\right\}$, which implies that
\begin{equation*}
M\left[mmse^{\text{QPSK}};1+z\right]=\int_0^{+\infty}t^zmmse^{\text{QPSK}}\left(t\right)dt=\int_0^{+\infty}t^zmmse^{\text{BPSK}}\left(\frac{t}{2}\right)dt=2^{1+z}M\left[mmse^{\text{BPSK}};1+z\right]
\end{equation*}
\end{IEEEproof}

\subsection{Other Inputs}
The Mellin transform of the \emph{canonical} MMSE associated with $4$-PAM, $16$-QAM, $8$-PAM and $64$-QAM inputs, for points $z\in\left\{\frac{1}{2}+\frac{w}{4}: w\in\{0,1,\ldots,20\}\right\}$ has been obtained by numerical evaluation of \eqref{eq: Mellin transform of the canonical mmse}. These results are summarized in Table \ref{table: Numerical approximation of Mellin transform of the canonical mmse}.

\begin{center}
\begin{table}[!t]
\renewcommand{\arraystretch}{1.1}
\caption{Numerical approximation of the Mellin transform of the \emph{canonical} MMSE \eqref{eq: Mellin transform of the canonical mmse} associated with $4$-PAM, $16$-QAM, $8$-PAM and $64$-QAM inputs (Entries with $-$ have not been computed)}
\label{table: Numerical approximation of Mellin transform of the canonical mmse}
\centering
    \begin{tabular}{c|c|c|c|c}
        \hline
  & \multicolumn{4}{|c}{Input} \\ \hline
$z$ & $4$-PAM & $16$-QAM & $8$-PAM & $64$-QAM \\ \hline
$\frac{1}{2}$ & $2.04943$ & $5.79667$ & $5.30675$ & $1.50097\times 10^{1}$ \\ \cline{1-5}
$\frac{1}{2}+\frac{1}{4}$ & $2.88309$ & $9.69751$ & $1.03121\times 10^{1}$ & $3.46857\times 10^{1}$ \\ \cline{1-5}
$1$ & $4.34356$ & $1.73742\times 10^{1}$ & $2.18091\times 10^{1}$ & $8.72366\times 10^{1}$ \\ \cline{1-5}
$\frac{1}{2}+\frac{3}{4}$ & $6.91253$ & $3.28817\times 10^{1}$ & $4.91577\times 10^{1}$ & $2.33835\times 10^{2}$ \\ \cline{1-5}
$\frac{3}{2}$ & $1.15073\times 10^{1}$ & $6.50951\times 10^{1}$ & $1.16461\times 10^{2}$ & $6.588\times 10^{2}$ \\ \cline{1-5}
$\frac{3}{2}+\frac{1}{4}$ & $1.98962\times 10^{1}$ & $1.33845\times 10^{2}$ & $2.87314\times 10^{2}$ & $1.93281\times 10^{3}$ \\ \cline{1-5}
$2$ & $3.55419\times 10^{1}$ & $2.84336\times 10^{2}$ & $7.3338\times 10^{2}$ & $5.86704\times 10^{3}$ \\ \cline{1-5}
$\frac{3}{2}+\frac{3}{4}$ & $6.53372\times 10^{1}$ & $6.21596\times 10^{2}$ & $1.92794\times 10^{3}$ & $1.83418\times 10^{4}$ \\ \cline{1-5}
$\frac{5}{2}$ & $1.23221\times 10^{2}$ & $1.39409\times 10^{3}$ & $5.20186\times 10^{3}$ & $5.88523\times 10^{4}$ \\ \cline{1-5}
$\frac{5}{2}+\frac{1}{4}$ & $2.37821\times 10^{2}$ & $3.19972\times 10^{3}$ & $1.43672\times 10^{4}$ & $1.93302\times 10^{5}$ \\ \cline{1-5}
$3$ & $4.68794\times 10^{2}$ & $7.50071\times 10^{3}$ & $4.05343\times 10^{4}$ & $6.48549\times 10^{5}$ \\ \cline{1-5}
$\frac{5}{2}+\frac{3}{4}$ & $9.42243\times 10^{2}$ & $1.79284\times 10^{4}$ & $1.16616\times 10^{5}$ & $2.21889\times 10^{6}$ \\ \cline{1-5}
$\frac{7}{2}$ & $1.92833\times 10^{3}$ & $4.36331\times 10^{4}$ & $3.41629\times 10^{5}$ & $7.73017\times 10^{6}$ \\ \cline{1-5}
$\frac{7}{2}+\frac{1}{4}$ & $4.01341\times 10^{3}$ & $1.07996\times 10^{5}$ & $1.01784\times 10^{6}$ & $2.73886\times 10^{7}$ \\ \cline{1-5}
$4$ & $8.48601\times 10^{3}$ & $2.71552\times 10^{5}$ & $3.08083\times 10^{6}$ & $9.85865\times 10^{7}$ \\ \cline{1-5}
$\frac{7}{2}+\frac{3}{4}$ & $1.82117\times 10^{4}$ & $6.93041\times 10^{5}$ & - & - \\ \cline{1-5}
$\frac{9}{2}$ & $3.96371\times 10^{4}$ & $1.79377\times 10^{6}$ & - & - \\ \cline{1-5}
$\frac{9}{2}+\frac{1}{4}$ & $8.7425\times 10^{4}$ & $4.70498\times 10^{6}$ & - & - \\ \cline{1-5}
$5$ & $1.95284\times 10^{5}$ & $1.24982\times 10^{7}$ & - & - \\ \cline{1-5}
$\frac{9}{2}+\frac{3}{4}$ & $4.41507\times 10^{5}$ & $3.36027\times 10^{7}$ & - & - \\ \cline{1-5}
$\frac{11}{2}$ & $1.00974\times 10^{6}$ & $9.13912\times 10^{7}$ & - & - \\ \cline{1-5}
        \hline
    \end{tabular}
\end{table}
\end{center}

\section{Numerical Results}
\label{section: Numerical Results}
We illustrate the accuracy of the results by comparing the high-$snr$ asymptotic expansions of the quantities to the numerical approximation, in a range of scenarios.

\begin{figure}[H]
\begin{minipage}[b]{0.45\linewidth}
\centering
\includegraphics[scale=0.4]{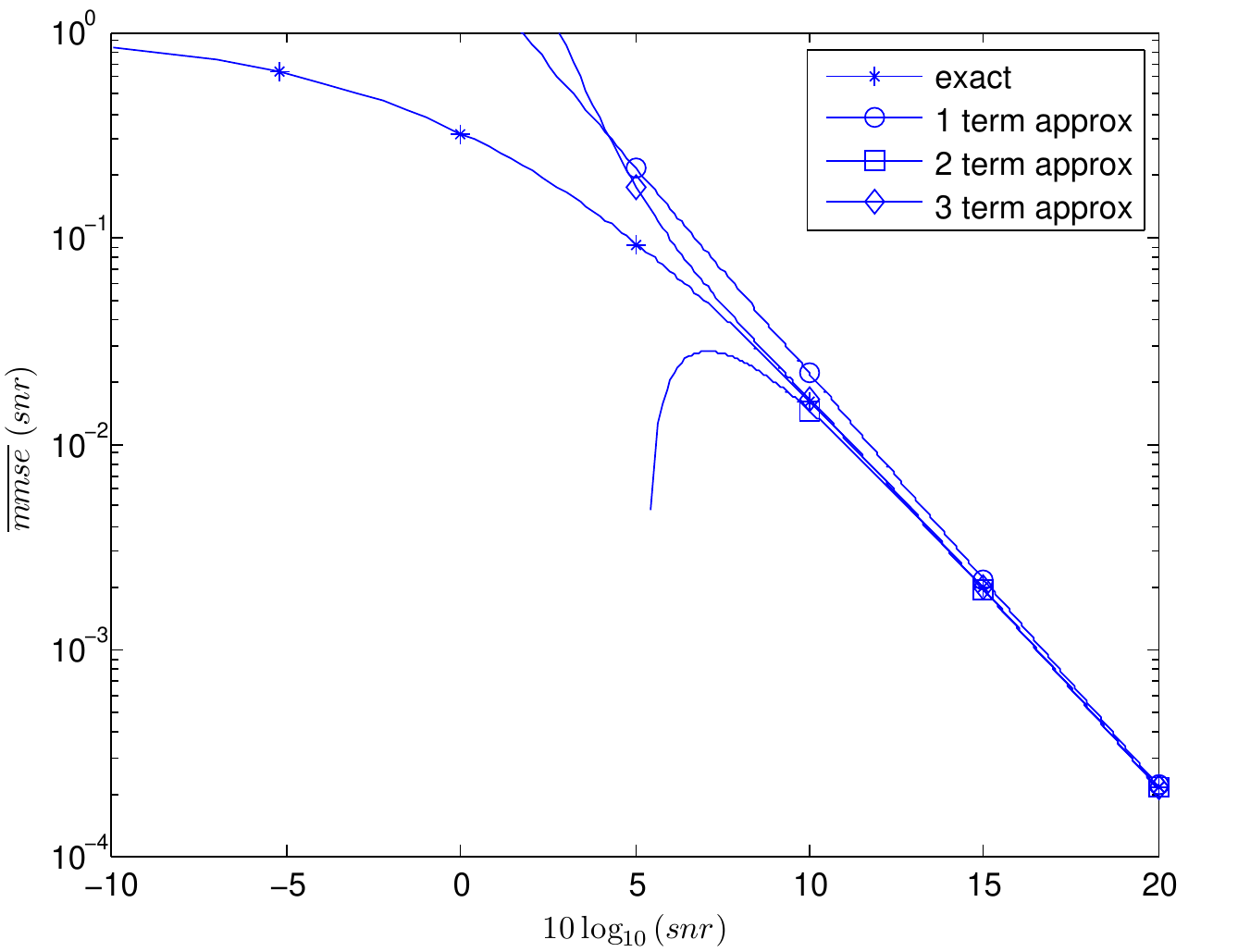}
\caption{Average MMSE in a Rayleigh fading coherent channel driven by a QPSK input $(\sigma=\frac{1}{\sqrt{2}})$.}
\label{figure: average_mmse_QPSK_Rayleigh_1overSqrt2}
\end{minipage}
\hspace{1.25cm}
\begin{minipage}[b]{0.45\linewidth}
\centering
\includegraphics[scale=0.4]{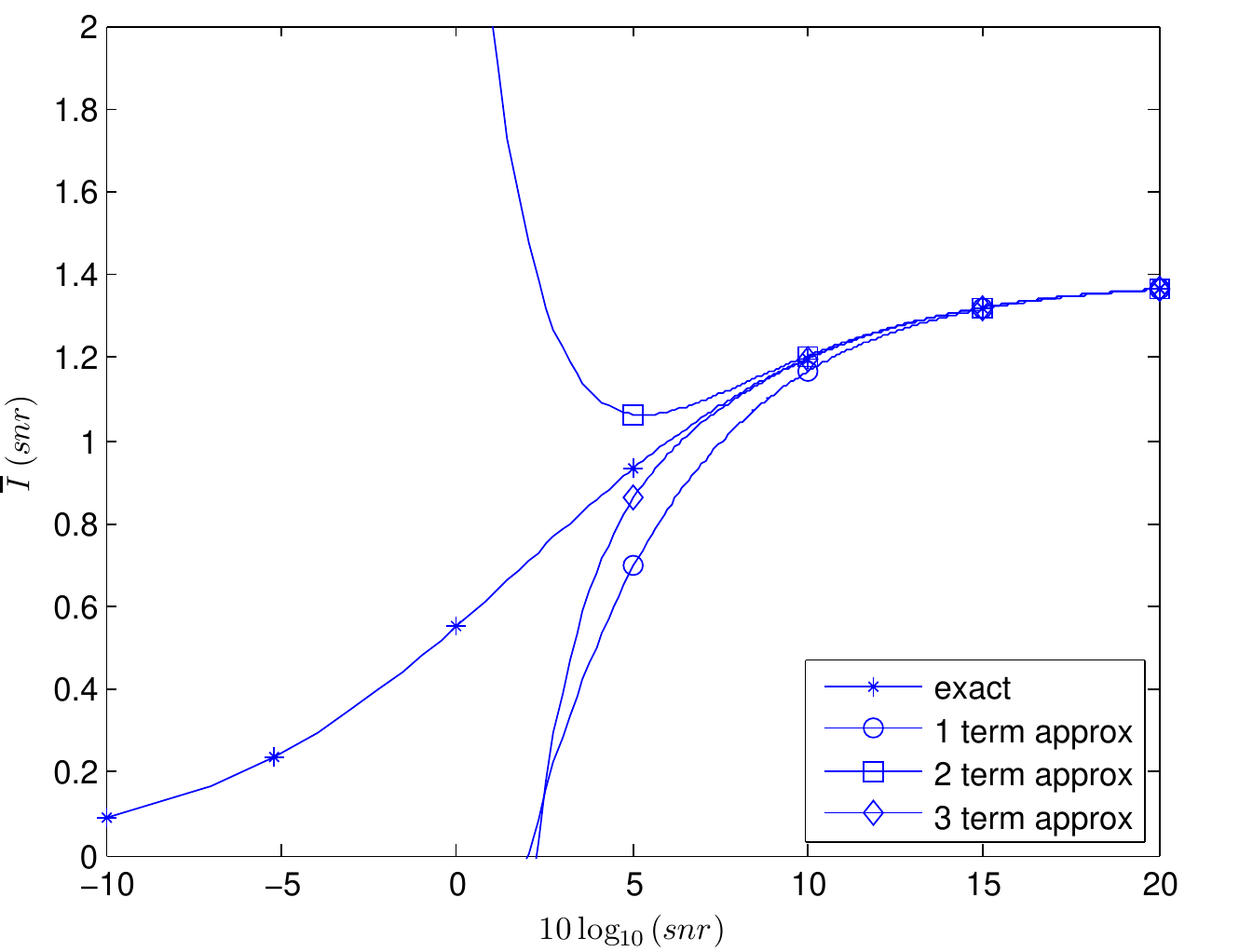}
\caption{Average mutual information in a Rayleigh fading coherent channel driven by a QPSK $(\sigma=\frac{1}{\sqrt{2}})$.}
\label{figure: average_mi_QPSK_Rayleigh_1overSqrt2}
\end{minipage}
\end{figure}

\begin{figure}[H]
\begin{minipage}[b]{0.45\linewidth}
\centering
\includegraphics[scale=0.4]{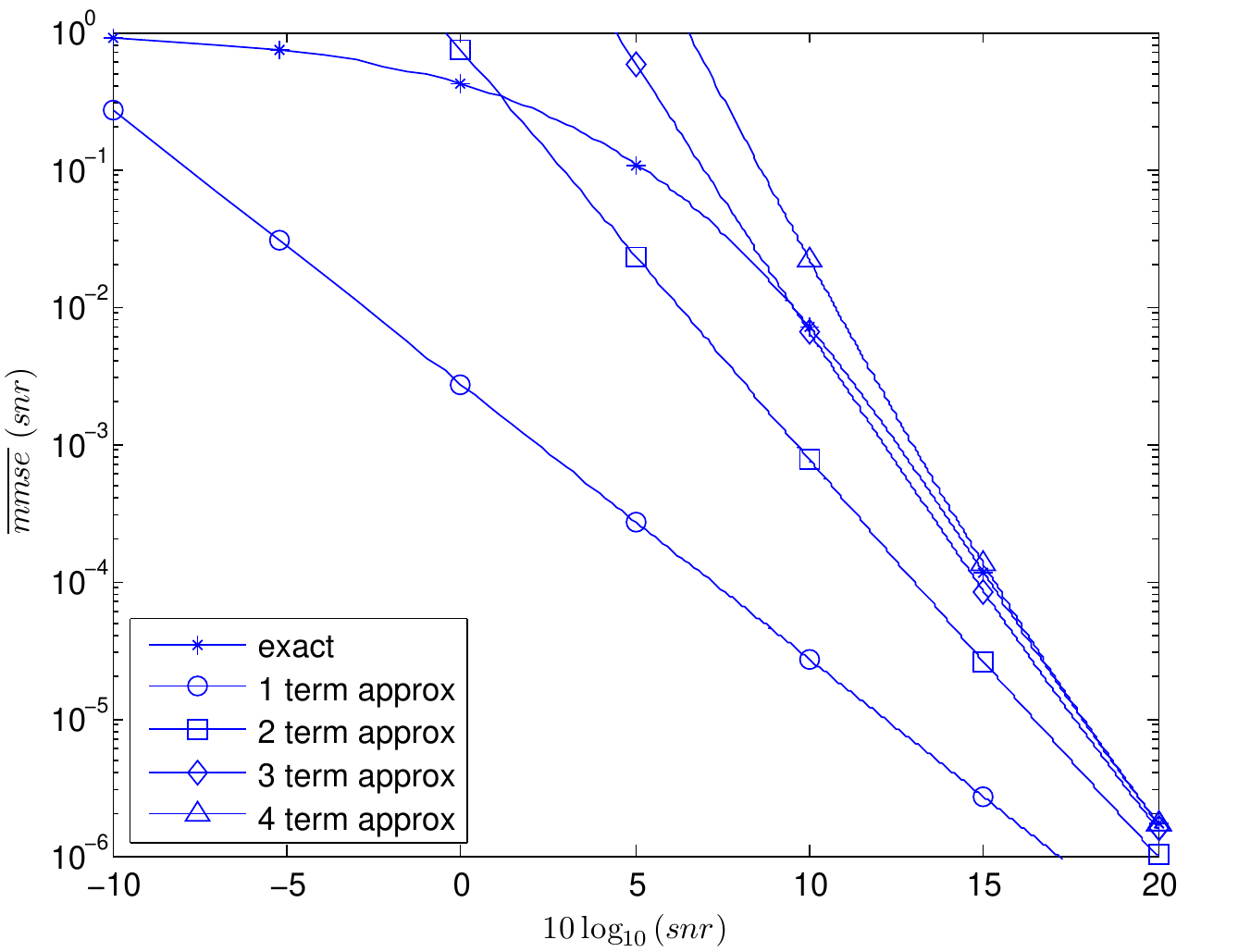}
\caption{Average MMSE in a Ricean fading coherent channel driven by a QPSK input $(|\mu|=\sqrt{\frac{9}{10}}~\text{and}~\sigma=\frac{1}{2\sqrt{5}})$.}
\label{figure: average_mmse_QPSK_Rice_v_Sqrt(9over10)_sigma_1over(2Sqrt5)}
\end{minipage}
\hspace{1.25cm}
\begin{minipage}[b]{0.45\linewidth}
\centering
\includegraphics[scale=0.4]{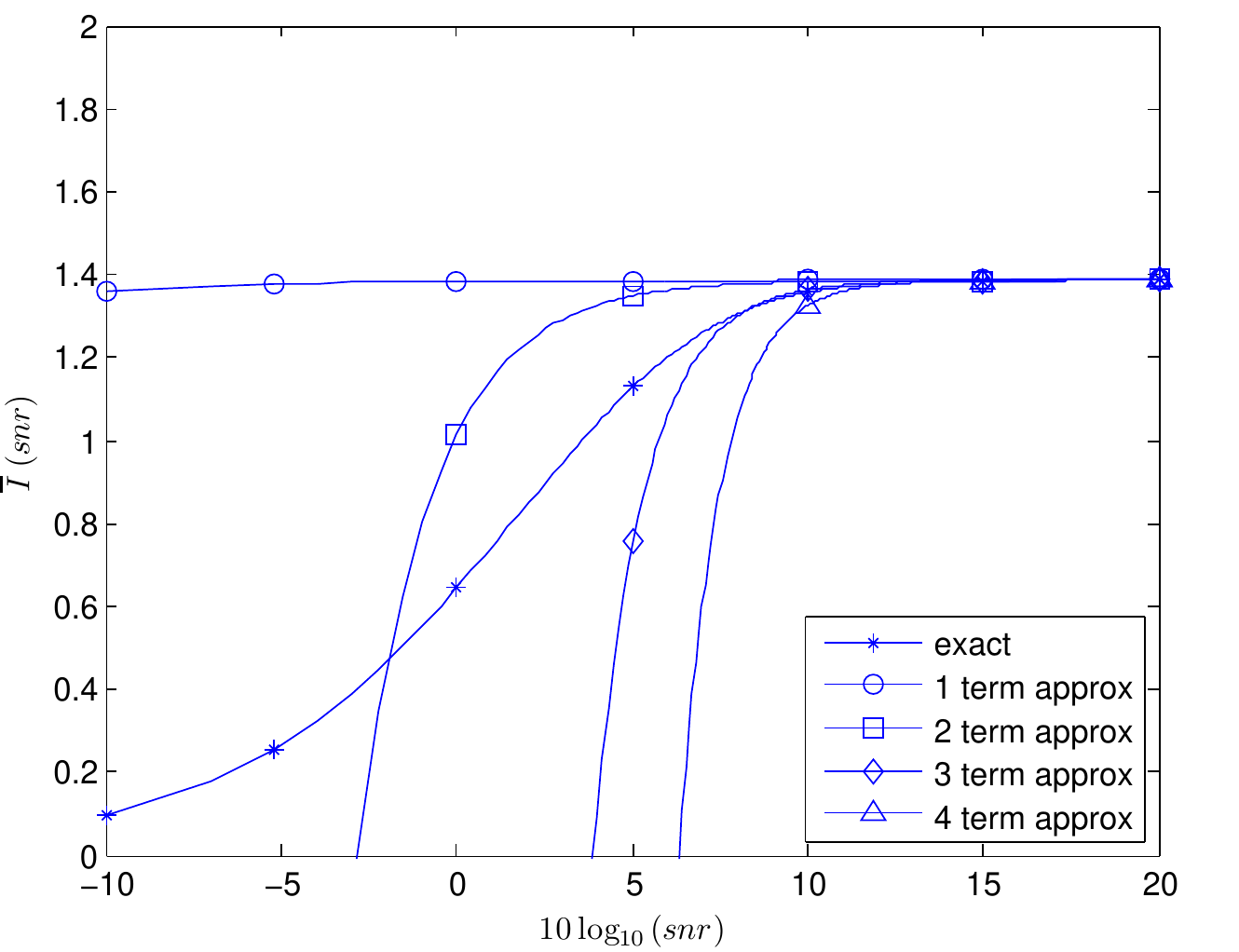}
\caption{Average mutual information in a Ricean fading coherent channel driven by a QPSK input $(|\mu|=\sqrt{\frac{9}{10}}~\text{and}~\sigma=\frac{1}{2\sqrt{5}})$.}
\label{figure: average_mi_QPSK_Rice_v_Sqrt(9over10)_sigma_1over(2Sqrt5)}
\end{minipage}
\end{figure}

\begin{figure}[H]
\begin{minipage}[b]{0.45\linewidth}
\centering
\includegraphics[scale=0.4]{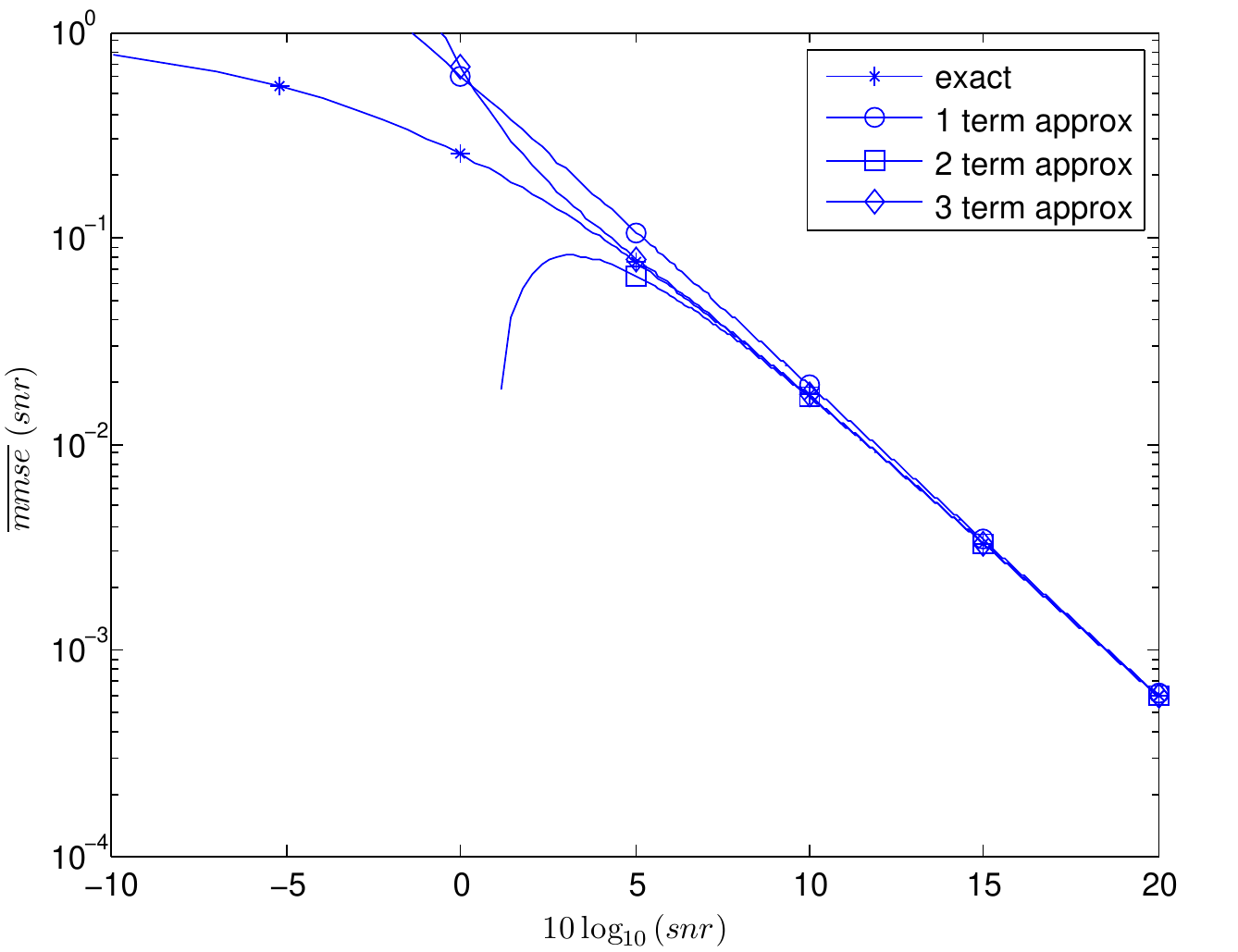}
\caption{Average MMSE in a Nakagami fading coherent channel driven by a QPSK input $(\mu =\frac{1}{2}~\text{and}~w=1)$.}
\label{figure: average_mmse_QPSK_Nakagami_mu_1over2_w_1}
\end{minipage}
\hspace{1.25cm}
\begin{minipage}[b]{0.45\linewidth}
\centering
\includegraphics[scale=0.4]{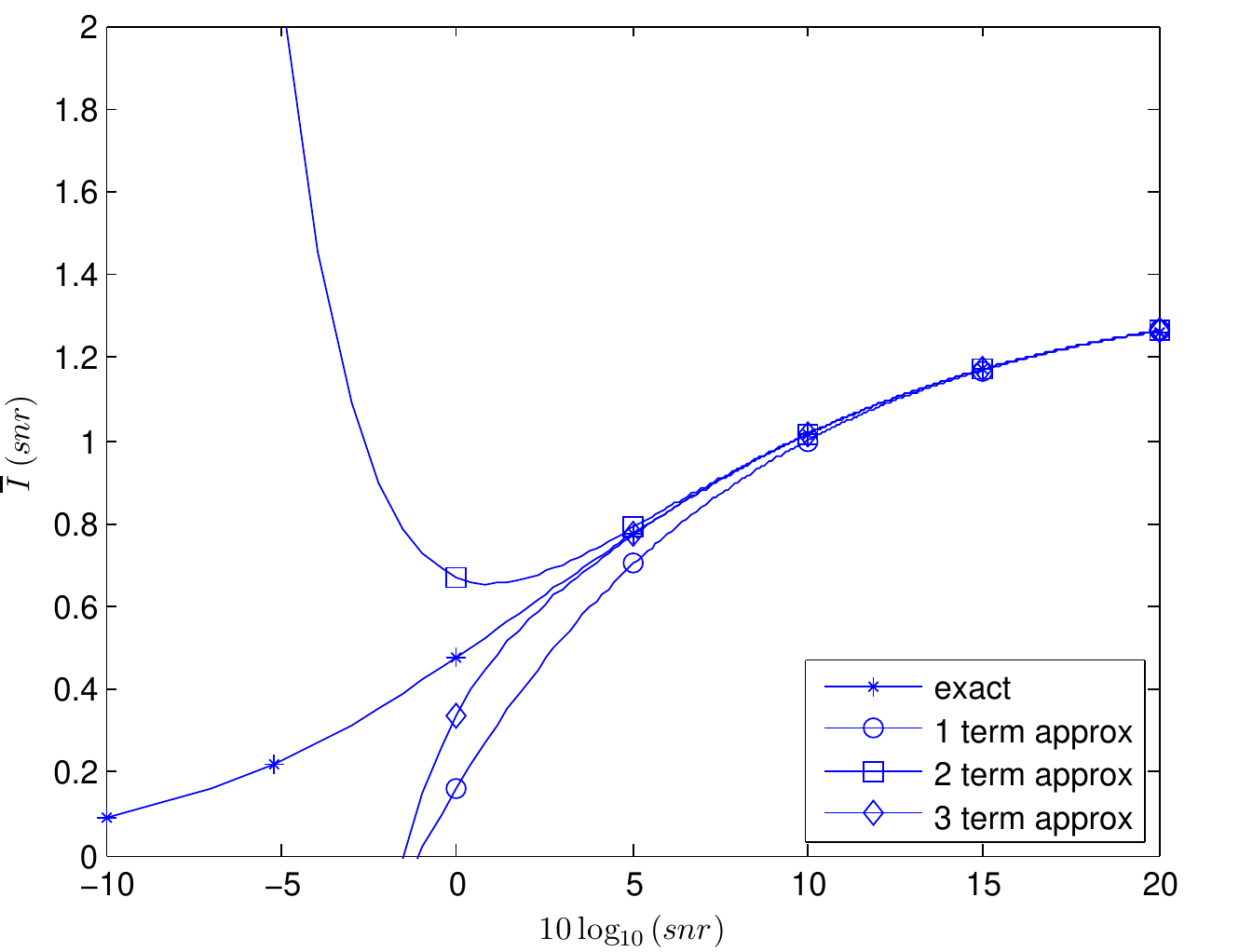}
\caption{Average mutual information in a Nakagami fading coherent channel driven by a QPSK input $(\mu =\frac{1}{2}~\text{and}~w=1)$.}
\label{figure: average_mi_QPSK_Nakagami_mu_1over2_w_1}
\end{minipage}
\end{figure}

Figures \ref{figure: average_mmse_QPSK_Rayleigh_1overSqrt2} -- \ref{figure: average_mi_QPSK_Nakagami_mu_1over2_w_1} consider the average MMSE and the average mutual information in Rayleigh, Ricean and Nakagami fading coherent channels driven by QPSK inputs. We observe that the expansions capture very well the high-$snr$ behavior of the quantities. In particular, one concludes that it is possible to approximate the behavior of the quantities over a wider $snr$ range by using several terms in the asymptotic expansions. We also observe that a single term expansion is sufficient to approximate well the high-$snr$ behavior of the average MMSE and the average mutual information in channels subject to Rayleigh and Nakagami fading. However, expansions with a higher number of terms are necessary to approximate the high-$snr$ behavior of the average MMSE and the average mutual information in channels subject to Ricean fading. This phenomenon, which is specially pronounced in the regime $|\mu| \gg \sigma$, is due to the fact that in such a scenario the behavior of the fading channel approaches the behavior of an AWGN channel, where the average MMSE and the average mutual information tend to their infinite-$snr$ values at rates greater than $2$ and $1$, respectively (see also \eqref{rate_mmse} and \eqref{rate_i}). This can only be captured by incorporating more than a single term in the asymptotic expansions.

\begin{figure}[H]
\begin{minipage}[b]{0.33\linewidth}
\centering
\includegraphics[scale=0.4]{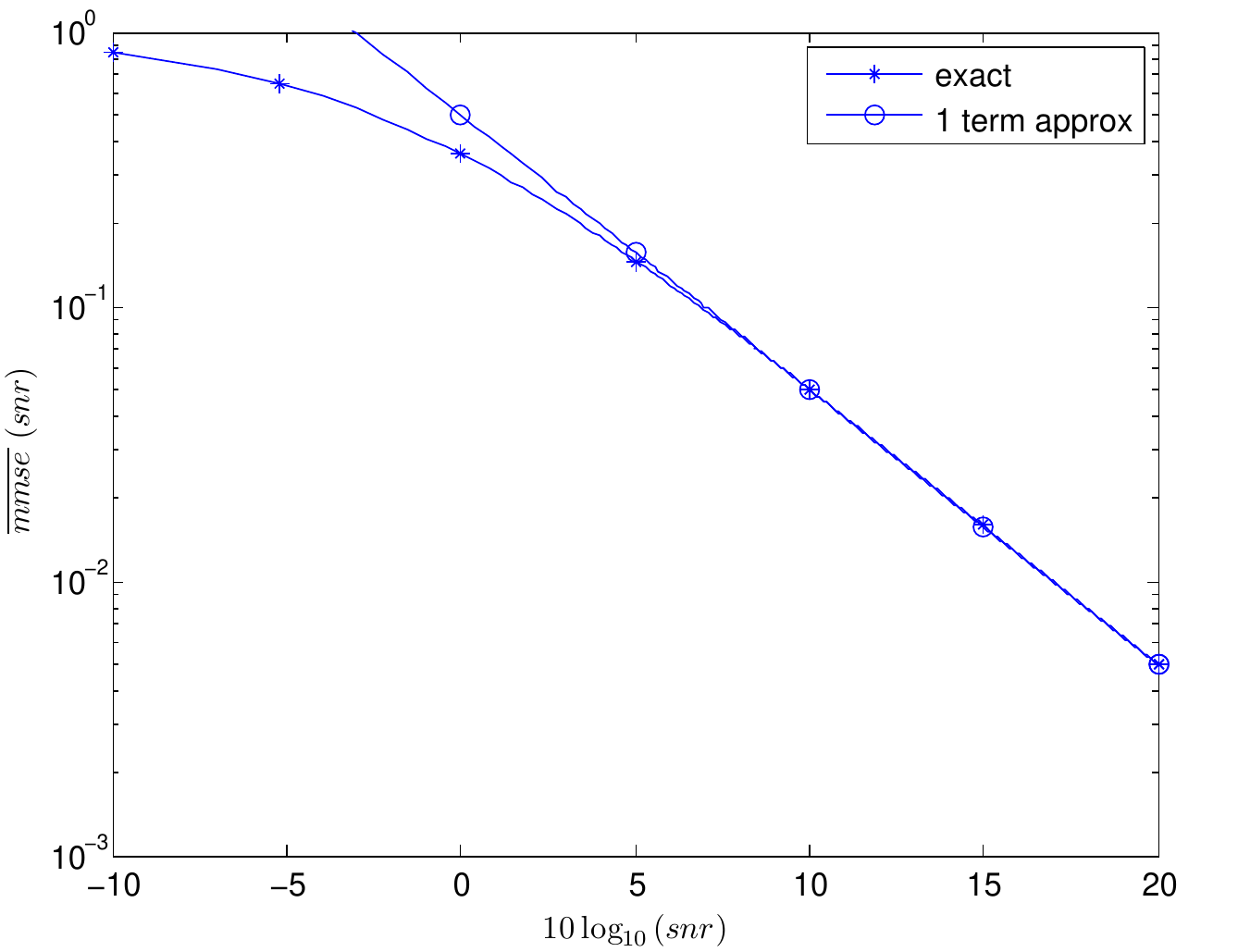}
\caption{Average MMSE in a Rayleigh fading coherent channel driven by $\infty$-PSK input $(\sigma=\frac{1}{\sqrt{2}})$.}
\label{figure: average_mmse_Infinity-PSK_Rayleigh_1overSqrt2}
\end{minipage}
\hspace{0.25cm}
\begin{minipage}[b]{0.33\linewidth}
\centering
\includegraphics[scale=0.4]{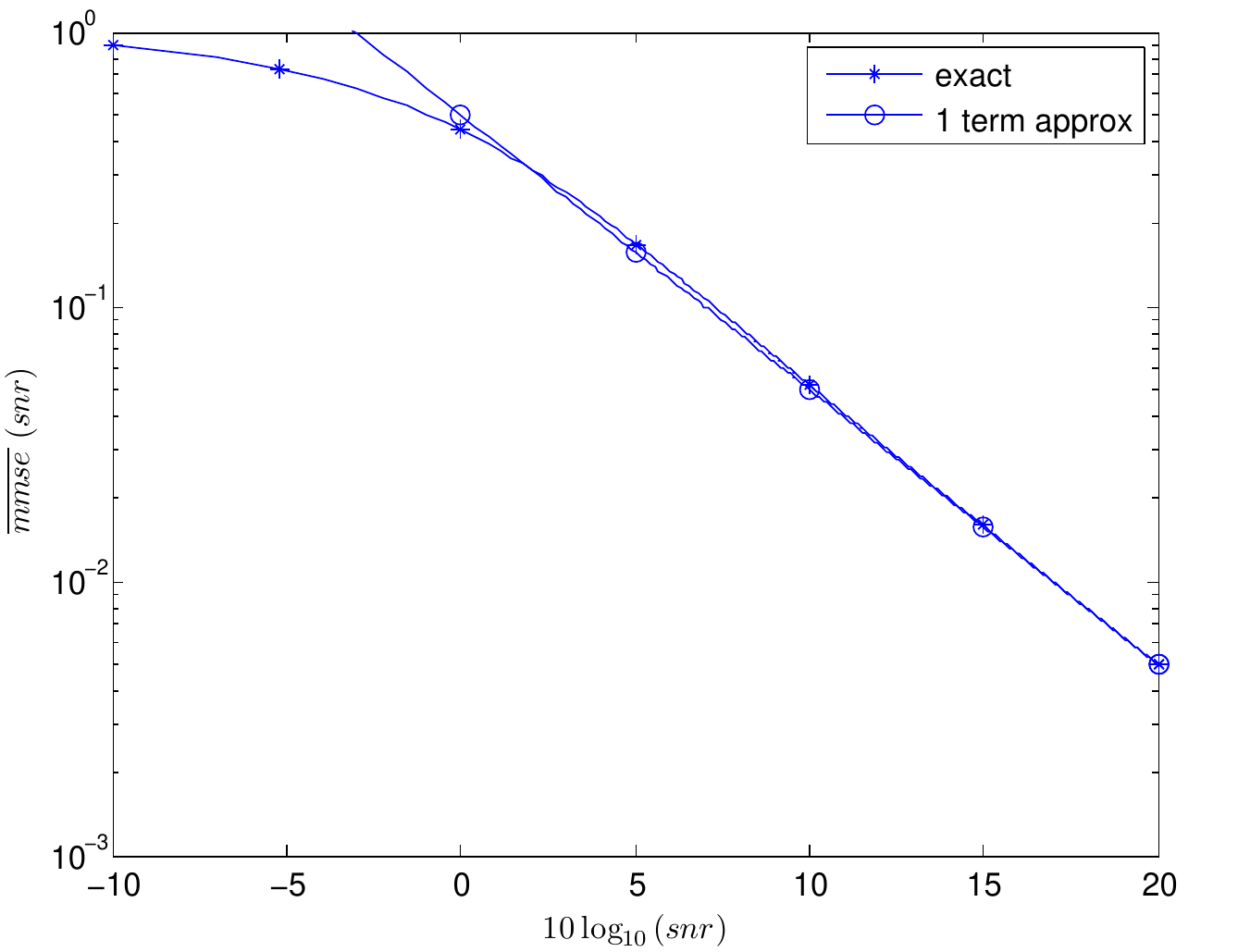}
\caption{Average MMSE in a Ricean fading coherent channel driven by $\infty$-PSK input $(|\mu|=\sqrt{\frac{9}{10}}~\text{and}~\sigma=\frac{1}{2\sqrt{5}})$.}
\label{figure: average_mmse_Infinity-PSK_Rice_v_Sqrt(9over10)_sigma_1over(2Sqrt5)}
\end{minipage}
\hspace{0.25cm}
\begin{minipage}[b]{0.33\linewidth}
\centering
\includegraphics[scale=0.4]{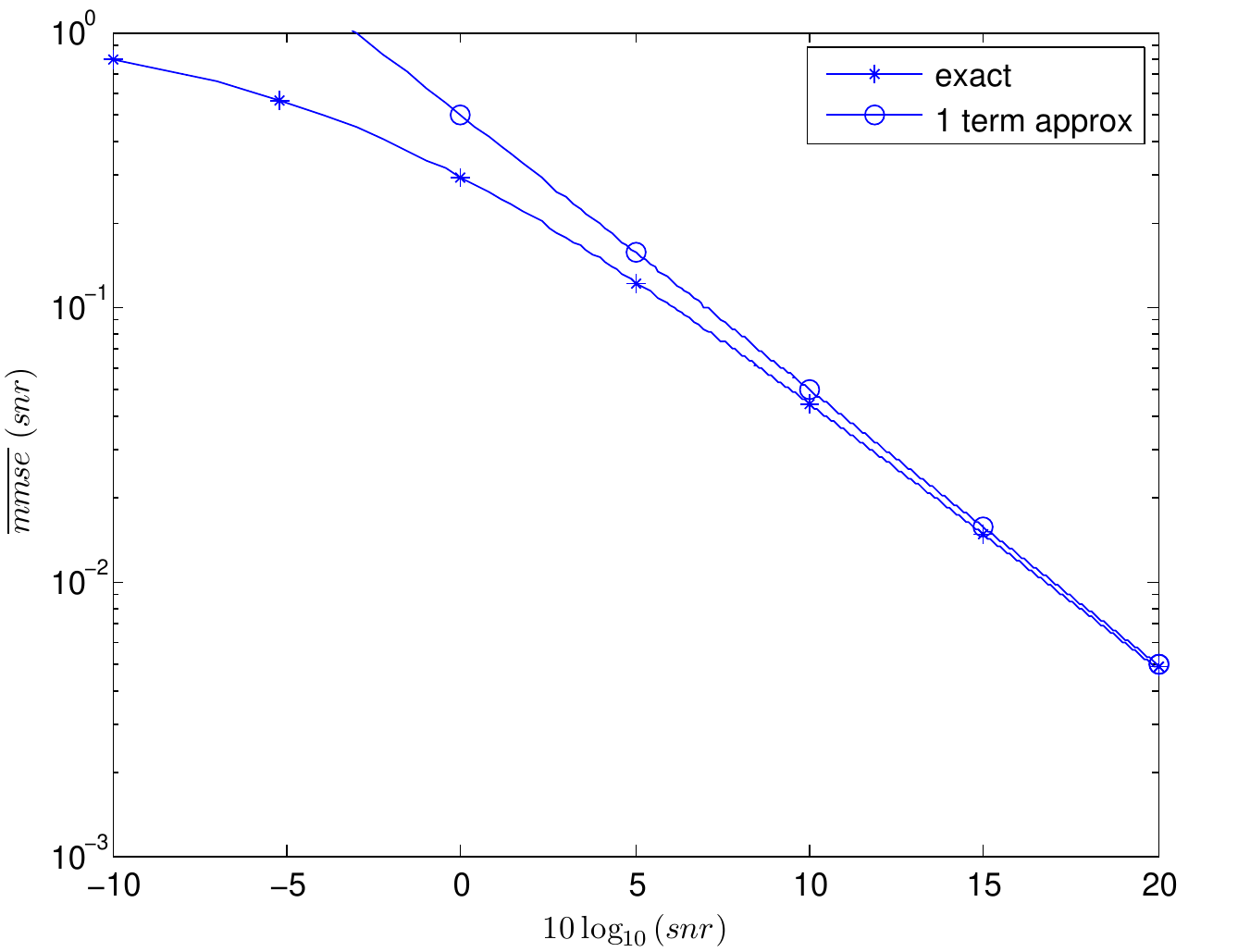}
\caption{Average MMSE in a Nakagami fading coherent channel driven by $\infty$-PSK $(\mu =\frac{1}{2}~\text{and}~w=1)$.}
\label{figure: average_mmse_Infinity-PSK_Nakagami_mu_1over2_w_1}
\end{minipage}
\end{figure}

Figures \ref{figure: average_mmse_Infinity-PSK_Rayleigh_1overSqrt2} -- \ref{figure: average_mmse_Infinity-PSK_Nakagami_mu_1over2_w_1} consider the average MMSE in Rayleigh, Ricean and Nakagami fading coherent channels driven by $\infty$-PSK inputs. We also observe that the single-term expansions capture well the high-$snr$ behavior of the quantities.

\section{Practical Applications}
\label{section: Applications: Optimal Power Allocation in a Bank of Parallel Independent Fading Coherent Channels driven by Arbitrary Inputs}
We conclude by considering a problem of optimal power allocation in a bank of $k$ parallel independent fading coherent channels driven by arbitrary discrete inputs, in order to showcase the application of the results. The channel model is given by:
\begin{equation}
y_i=\sqrt{snr}h_i\sqrt{p_i}x_i+n_i, \quad i = 1,\ldots,k
\label{eq: i-th Additive Gaussian Noise channel}
\end{equation}
where $y_i\in\mathbb{C}$ represents the $i$-th sub-channel output, $x_i\in\mathbb{C}$ represents the $i$-th sub-channel input, $h_i$ is a complex scalar random variable with support $\mathbb{C}$ or $\mathbb{C}\setminus\{0\}$ such that $E_{h_i}\left\{|h_i|^2\right\}<+\infty$ which represents the random channel fading gain between the input and the output of the $i$-th sub-channel, and $n_i\in\mathbb{C}$ is a circularly symmetric complex scalar Gaussian random variable with zero mean and unit variance which represents standard noise. The scaling factor $p_i \in\mathbb{R}^+_0$ represents the power injected into sub-channel $i$. The scaling factor $snr\in\mathbb{R}^+$ relates to the signal-to-noise ratio. We assume that $x_i, i = 1,\ldots,k$, $h_i, i = 1,\ldots,k$ and $n_i, i = 1,\ldots,k$ are independent random variables. We also assume that the receiver knows the exact realization of the sub-channel gains but the transmitter knows only the distribution of the sub-channel gains. This channel model is applicable to a OFDM and multi-user OFDM communications system \cite{Lozano06optimumpower,Lozano_optimumpower}.

We denote the average MMSE and the \emph{canonical} MMSE of sub-channel $i$ in the model in \eqref{eq: i-th Additive Gaussian Noise channel} as ${\overline{mmse}}_i\left(\cdot\right)$ and ${mmse}_i\left(\cdot\right)$, respectively. We also denote the average mutual information and the \emph{canonical} mutual information of sub-channel $i$ in the model in \eqref{eq: i-th Additive Gaussian Noise channel} as ${\overline{I}_i}\left(\cdot\right)$ and ${I}_i\left(\cdot\right)$, respectively.

The objective is to determine the power allocation policy that maximizes the constrained capacity given by:
\begin{equation*}
\overline{I}\left(snr;p_1,\ldots,p_k\right) = \sum_{i=1}^k{\overline{I}}_i\left(snr\cdot p_i\right)
\end{equation*}
subject to a total power constraint:
\begin{equation*}
\sum_{i=1}^{k} p_i \leq P
\end{equation*}
and $p_i \geq 0, i = 1,\ldots,k$. The following Theorem, which is based on the asymptotic expansions put forth in the previous sections, defines the optimal power allocation policy in the asymptotic regime of high $snr$ for Rayleigh and Ricean fading models.

\begin{thm}
Consider a bank of $k$ parallel independent Rayleigh or Ricean fading coherent channels as in~\eqref{eq: i-th Additive Gaussian Noise channel} driven by arbitrary discrete inputs with finite support, where $h_i \sim \mathcal{CN}\left(\mu_i,2\sigma_i^2\right)$ with $\mu_i=0$ or $\mu_i\neq0$, respectively, and $\sigma_i > 0$, $i = 1,\ldots,k$. Then, in the regime of high $snr$ the optimal power allocation policy obeys:
\begin{equation*}
p_i^* = \sqrt{\exp{\left(-\frac{|\mu_i|^2}{2\sigma_i^2}\right)} \cdot \frac{M\left[{mmse}_i;2\right]}{2\sigma_i^2} \cdot \frac{1}{\lambda snr}}+O\left(\frac{1}{snr}\right), \qquad snr\rightarrow +\infty, \qquad i = 1,\ldots,k
\end{equation*}
where $\lambda$ is such that $\sum_{i=1}^{k} p_i^* = P$.
\label{thm: rayleigh and ricean applications}
\end{thm}

\begin{IEEEproof}
See Appendix \ref{section: Optimal Power Allocation Rayleigh and Rice}.
\end{IEEEproof}

\vspace{0.25cm}

\begin{figure}[H]
\begin{minipage}[b]{0.45\linewidth}
\centering
\includegraphics[scale=0.4]{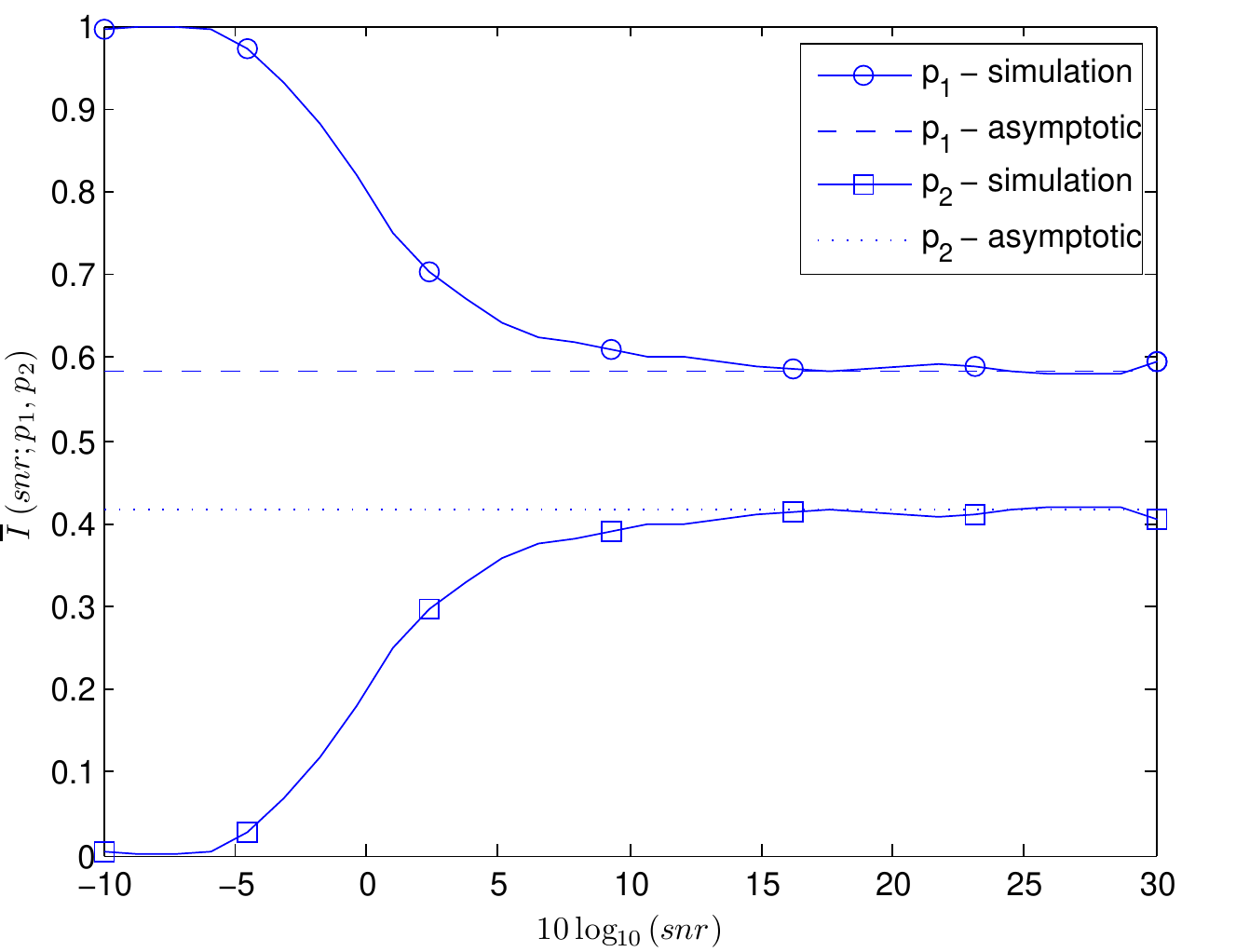}
\caption{Optimal and asymptotic power allocation in a bank of two parallel independent Rayleigh fading channels ($2 \sigma_1^2 = 4$ and $2 \sigma_2^2=1$). First sub-channel driven by equiprobable unit-power 16-QAM and second sub-channel driven by equiprobable unit-power QPSK.}
\label{figure: Rayleigh}
\end{minipage}
\hspace{1.25cm}
\begin{minipage}[b]{0.45\linewidth}
\centering
\includegraphics[scale=0.4]{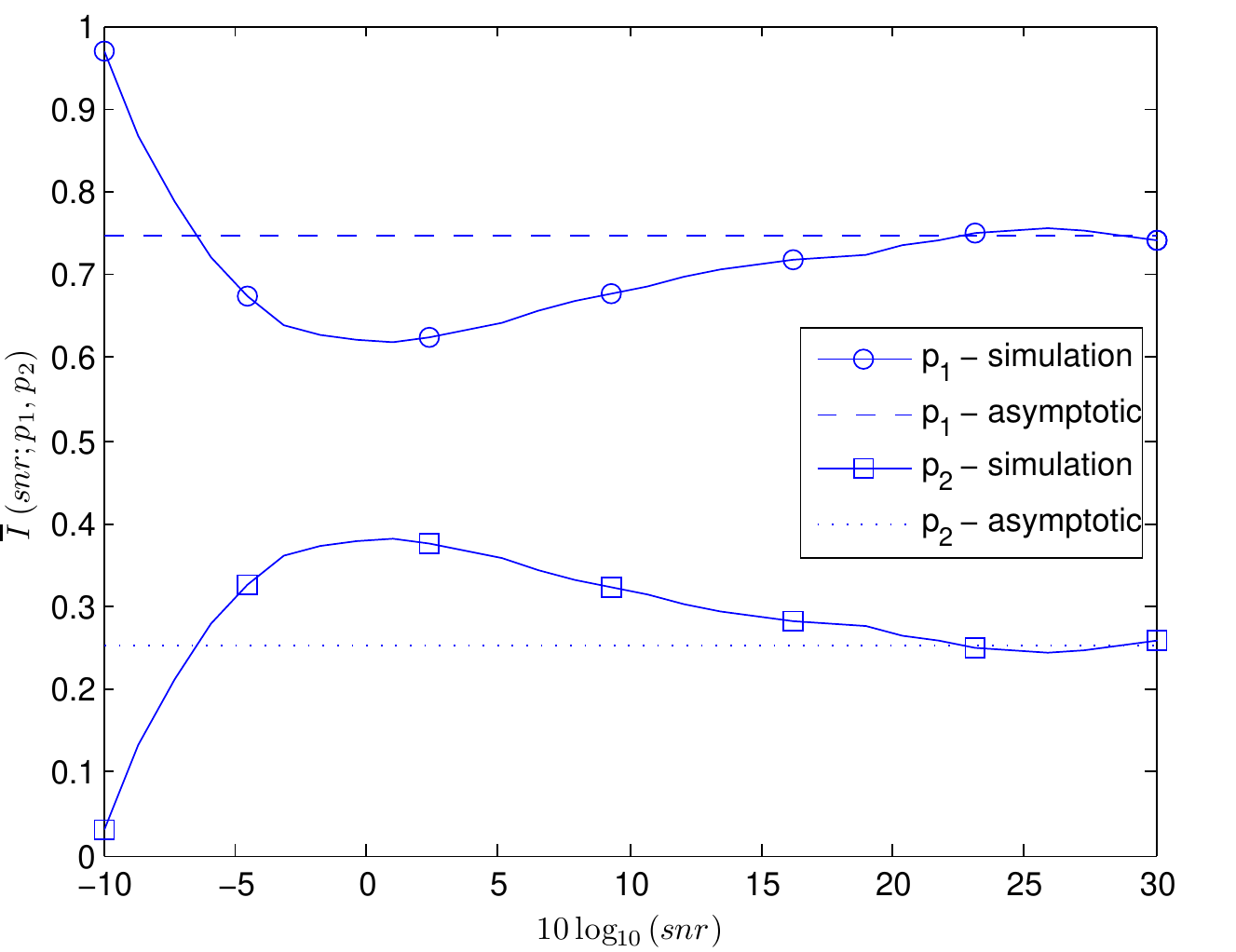}
\caption{Optimal and asymptotic power allocation in a bank of two parallel independent Ricean fading channels ($\mu_1 = 1+i$, $2 \sigma_1^2 = 4$, $\mu_2 = 1+i$ and $2 \sigma_2^2=1$). First sub-channel driven by equiprobable unit-power 16-QAM and second sub-channel driven by equiprobable unit-power QPSK.}
\label{figure: Ricean}
\end{minipage}
\end{figure}

Theorem \ref{thm: rayleigh and ricean applications} reveals the impact of the nature of the fading distribution and the input distribution on the high-$snr$ optimal power allocation policy. In Rayleigh fading channels, given equal sub-channel inputs, it can be seen that the higher the average sub-channel strength (i.e., the higher $2\sigma_i^2$) then the lower the allocated power. In Ricean fading channels, it can also be seen that the presence of line-of-sight components affects dramatically the power allocation policy. It is interesting to note that, as expected, the nature of the inputs affects the optimal power allocation policy via the Mellin transform of the \emph{canonical} MMSE. It is also interesting to note that the power allocation policies embodied in Theorem \ref{thm: rayleigh and ricean applications} in fact represent a generalization of the power allocation policy put forth in~\cite{Lozano_optimumpower}, in the sense that -- in the single-user setting -- it applies to Ricean fading in addition to Rayleigh fading and to scenarios where the different input signals conform to different discrete constellations. Figures \ref{figure: Rayleigh} and \ref{figure: Ricean} confirm that the optimal power allocation rapidly converges to the high-$snr$ power allocation uncovered by Theorem \ref{thm: rayleigh and ricean applications} for a bank of two parallel independent fading coherent channels.

\section{Conclusions}
\label{section: Conclusions}
Motivated by the need to understand the behavior of the constrained capacity of fading channels, we have unveiled asymptotic expansions of key estimation- and information-theoretic measures in scalar and vector fading coherent channels, where the receiver knows the exact fading channel state but the transmitter knows only the fading channel distribution, driven by a range of inputs. In particular, we have constructed low-$snr$ and -- at the heart of the novelty of the contribution -- high-$snr$ asymptotic expansions for the average minimum mean-squared error and the average mutual information for coherent channels subject to Rayleigh fading, Ricean fading or Nakagami fading and driven by arbitrary discrete inputs (with finite support) or by $\infty$-PSK, $\infty$-PAM, $\infty$-QAM, and standard complex Gaussian continuous inputs. The most relevant element for the construction of the asymptotic expansions is the realization that the integral representation of the measures can be seen as an $h$-transform of a kernel with a monotonic argument. This paves the way to the use of a range of expansion of integrals techniques, most notably, Mellin transform methods, that yield the asymptotic expansions for the average minimum mean-squared error and -- via the now well-known I-MMSE relationship -- for the average mutual information.


We have also considered as a case study a standard power allocation problem over a bank of parallel independent fading coherent channels driven by arbitrary discrete inputs, a scenario representative of orthogonal frequency division multiplexing communications systems. In particular,  we have illustrated how to determine the power allocation policy that maximizes the constrained capacity of the bank of parallel independent fading channels in key asymptotic regimes.

\appendices
\section{Proof of Theorem \ref{thm: Additive Gaussian Noise average mmse general h}}
\label{section: Proof of Theorem Additive Gaussian Noise average mmse general h}
Let
\begin{align*}
f\left(t\right)&:=\frac{\sqrt{t}f_{|h|}\left(\sqrt{t}\right)}{2}\\
h\left(t\right)&:=mmse\left(t\right)
\end{align*}
Then
\begin{equation}
\overline{mmse}\left(snr\right)=E_{|h|}\left\{|h|^2mmse\left(snr|h|^2\right)\right\}=\int_0^{+\infty}f\left(u\right)h\left(snru\right)du
\label{eq: Proof of Theorem Additive Gaussian Noise average mmse general h Part 1}
\end{equation}

We note that \eqref{eq: Proof of Theorem Additive Gaussian Noise average mmse general h Part 1} is an $h$-transform with Kernel of Monotonic Argument, so that we can capitalize on the method of Mellin transforms~\cite[Section 4.4]{asymptotic_expansion_of_integrals_Bleistein_and_Handelsman} to obtain the asymptotic expansion of $\overline{mmse}\left(snr\right)$ as $snr \to +\infty$ via~\cite[Theorem 4.4]{asymptotic_expansion_of_integrals_Bleistein_and_Handelsman}. The application of~\cite[Theorem 4.4]{asymptotic_expansion_of_integrals_Bleistein_and_Handelsman} requires that:
\begin{enumerate}
	\item\label{item: requirement 1}
Both $h\left(\cdot\right)$ and $f\left(\cdot\right)$ are locally integrable functions on $\mathbb{R}^+$;
	\item\label{item: requirement 2}
The following holds true
\begin{enumerate}
	\item\label{item: requirement 2.1}
The function $h\left(\cdot\right)$ decays as
\begin{equation}
h\left(t\right)\sim\exp{\left(-kt^v\right)}\sum_{m=0}^{+\infty}\sum_{n=0}^{N\left(m\right)}c_{mn}t^{-r_m}\left(\log{\left(t\right)}\right)^n, \qquad t\rightarrow +\infty
\label{eq: user obtain h}
\end{equation}
where $\mathcal{R}\left(k\right)\geq 0$, $v>0$, $c_{mn}\in\mathbb{C}$, $\mathcal{R}\left(r_m\right)\uparrow +\infty$ and $N\left(m\right)$ is finite for each $m$.

We note that in the case $k\neq 0$ a small modification of the proof of~\cite[Lemma 4.3.1]{asymptotic_expansion_of_integrals_Bleistein_and_Handelsman} also yields the conclusions in~\cite[Theorem 4.4]{asymptotic_expansion_of_integrals_Bleistein_and_Handelsman} if instead of \eqref{eq: user obtain h} we have
\begin{equation*}
h\left(t\right)=O\left(\exp{\left(-kt^v\right)}\right), \qquad t\rightarrow +\infty
\end{equation*}
where $\mathcal{R}\left(k\right)>0$ and $v>0$.

We also note that in the case $k=0$, if it suffices to obtain an asymptotic expansion with a finite number of terms then the proof of~\cite[Lemma 4.3.3]{asymptotic_expansion_of_integrals_Bleistein_and_Handelsman} also reveals that we obtain the same conclusions in~\cite[Theorem 4.4]{asymptotic_expansion_of_integrals_Bleistein_and_Handelsman} if instead of \eqref{eq: user obtain h} we have
\begin{equation}
h\left(t\right)=\sum_{m=0}^{M_h}\sum_{n=0}^{N\left(m\right)}c_{mn}t^{-r_m}\left(\log{\left(t\right)}\right)^n+O\left(t^{-r_{M_h+1}}\left(\log{\left(t\right)}\right)^{N\left(M_h+1\right)}\right), \qquad t\rightarrow +\infty
\label{eq: user obtain h finite number of terms}
\end{equation}
where $M_h<+\infty$ and $N\left(m\right)$ is finite for each $m$.
	\item\label{item: requirement 2.2}
The function $f\left(\cdot\right)$ decays as
\begin{equation}
f\left(t\right)\sim\exp{\left(-qt^{-\mu}\right)}\sum_{m=0}^{+\infty}\sum_{n=0}^{\overline{N}\left(m\right)}p_{mn}t^{a_m}\left(\log{\left(t\right)}\right)^n, \qquad t\rightarrow 0^+
\label{eq: user obtain f}
\end{equation}
where $\mathcal{R}\left(q\right)\geq 0$, $\mu>0$, $p_{mn}\in\mathbb{C}$, $\mathcal{R}\left(a_m\right)\uparrow +\infty$ and $\overline{N}\left(m\right)$ is finite for each $m$.

We note that in the case $q\neq 0$ a small modification of the proof of~\cite[Lemma 4.3.4]{asymptotic_expansion_of_integrals_Bleistein_and_Handelsman} also yields the conclusions in~\cite[Theorem 4.4]{asymptotic_expansion_of_integrals_Bleistein_and_Handelsman} if instead of \eqref{eq: user obtain f} we have
\begin{equation*}
f\left(t\right)=O\left(\exp{\left(-qt^{-\mu}\right)}\right), \qquad t\rightarrow 0^+
\end{equation*}
where $\mathcal{R}\left(q\right)>0$ and $\mu>0$.

We also note that in the case $q=0$, if it suffices to obtain an asymptotic expansion with a finite number of terms then the proof of~\cite[Lemma 4.3.6]{asymptotic_expansion_of_integrals_Bleistein_and_Handelsman} also reveals that we obtain the same conclusions in~\cite[Theorem 4.4]{asymptotic_expansion_of_integrals_Bleistein_and_Handelsman} if instead of \eqref{eq: user obtain f} we have
\begin{equation}
f\left(t\right)=\sum_{m=0}^{M_f}\sum_{n=0}^{\overline{N}\left(m\right)}p_{mn}t^{a_m}\left(\log{\left(t\right)}\right)^n+O\left(t^{a_{M_f+1}}\left(\log{\left(t\right)}\right)^{\overline{N}\left(M_f+1\right)}\right), \qquad t\rightarrow 0^+
\label{eq: user obtain f finite number of terms}
\end{equation}
where $M_f<+\infty$ and $\overline{N}\left(m\right)$ is finite for each $m$.
\end{enumerate}
	\item\label{item: requirement 3}
Let
\begin{align*}
\alpha &:=\inf\left\{\alpha^*: h\left(t\right)=O\left(t^{-\alpha^*}\right), t\rightarrow 0^+\right\}\\
\beta &:=\sup\left\{\beta^*: h\left(t\right)=O\left(t^{-\beta^*}\right), t\rightarrow +\infty\right\}\\
\gamma &:=\inf\left\{\gamma^*: f\left(t\right)=O\left(t^{-\gamma^*}\right), t\rightarrow 0^+\right\}\\
\delta &:=\sup\left\{\delta^*: f\left(t\right)=O\left(t^{-\delta^*}\right), t\rightarrow +\infty\right\}
\end{align*}
We require that $\alpha<\beta$ and $\gamma<\delta$, so that $M\left[h;z\right]$ and $M\left[f;z\right]$ are holomorphic in the strips $\alpha<\mathcal{R}\left(z\right)<\beta$ and $\gamma<\mathcal{R}\left(z\right)<\delta$, respectively,~\cite[p.~106]{asymptotic_expansion_of_integrals_Bleistein_and_Handelsman}.
	\item\label{item: requirement 4}
Let
\begin{equation*}
C:=\left\{z\in\mathbb{C}: \left(\alpha<\mathcal{R}\left(z\right)<\beta\right)\wedge\left(1-\delta<\mathcal{R}\left(z\right)<1-\gamma\right)\right\}
\end{equation*}
We require that
\begin{equation*}
C\neq\emptyset
\end{equation*}
so that
\begin{equation*}
G\left(z\right):=M\left[h;z\right]M\left[f;1-z\right]
\end{equation*}
is holomorphic in $C$ (which is then continued to the right as a meromorphic function at worst~\cite[p.~118]{asymptotic_expansion_of_integrals_Bleistein_and_Handelsman});
	\item\label{item: requirement 5}
The following holds true
\begin{equation}
\exists c\in C\cap\mathbb{R}: \Bigg(\int_0^{+\infty}f\left(t\right)h\left(t\right)dt=\frac{1}{2\pi i}\int_{c-i\infty}^{c+i\infty}G\left(z\right)dz \wedge \forall x\in[c,+\infty[, \lim_{|y|\rightarrow +\infty}G\left(x+iy\right)=0\Bigg)
\label{eq: a c in C}
\end{equation}
	\item\label{item: requirement 6}
The following holds true
\begin{enumerate}
	\item\label{item: requirement 6.1}
If $k\neq 0\wedge q\neq 0$, then we require that there exists a real sequence $u_n$ such that $u_n\uparrow +\infty$ and $\forall n\in\mathbb{Z}^+, \int_{-\infty}^{+\infty}\left|G\left(u_n+iy\right)\right|dy<+\infty$;
	\item\label{item: requirement 6.2}
If $k\neq 0\wedge q=0$, let $U:=\{\mathcal{R}\left(a_m\right)+1: m\in\mathbb{Z}^+\}$ and let $u_n$ be the real sequence such that $u_n\uparrow +\infty$ and $U=\{u_n: n\in\mathbb{Z}^+\}$. Then we require that $\forall n\in\mathbb{Z}^+, \exists x\in]u_n,u_{n+1}[: \int_{-\infty}^{+\infty}\left|G\left(x+iy\right)\right|dy<+\infty$;
	\item\label{item: requirement 6.3}
If $k=0\wedge q\neq 0$, let $V:=\{\mathcal{R}\left(r_m\right): m\in\mathbb{Z}^+\}$ and let $u_n$ be the real sequence such that $u_n\uparrow +\infty$ and $V=\{u_n: n\in\mathbb{Z}^+\}$. Then we require that $\forall n\in\mathbb{Z}^+, \exists x\in]u_n,u_{n+1}[: \int_{-\infty}^{+\infty}\left|G\left(x+iy\right)\right|dy<+\infty$;
	\item\label{item: requirement 6.4}
If $k=0\wedge q=0$, let $W:=\{\mathcal{R}\left(a_m\right)+1: m\in\mathbb{Z}^+\}\cup\{\mathcal{R}\left(r_m\right): m\in\mathbb{Z}^+\}$ and let $u_n$ be the real sequence such that $u_n\uparrow +\infty$ and $W=\{u_n: n\in\mathbb{Z}^+\}$. Then we require that $\forall n\in\mathbb{Z}^+, \exists x\in]u_n,u_{n+1}[: \int_{-\infty}^{+\infty}\left|G\left(x+iy\right)\right|dy<+\infty$;
\end{enumerate}

We note that in the case \eqref{eq: user obtain h finite number of terms} and/or \eqref{eq: user obtain f finite number of terms} the real sequence $u_n$ must be replaced by the finite list $\{u_n: u_n<\min\{\mathcal{R}\left(r_{M_h+1}\right),\mathcal{R}\left(a_{M_f+1}\right)+1\}\}$.
\end{enumerate}

The asymptotic expansion of \eqref{eq: Proof of Theorem Additive Gaussian Noise average mmse general h Part 1} as $snr\rightarrow +\infty$ is then given by~\cite[Theorem 4.4]{asymptotic_expansion_of_integrals_Bleistein_and_Handelsman}:
\begin{equation}
\int_0^{+\infty}f\left(u\right)h\left(snru\right)du=-\sum_{c<\mathcal{R}\left(z\right)}\res\{snr^{-z}G\left(z\right)\}
\label{eq: sum of residues}
\end{equation}
(where $c$ is as in \eqref{eq: a c in C} and does not need to be unique, and $\res\{snr^{-z}G\left(z\right)\}$ denotes the residue of the meromorphic function $snr^{-\cdot}G\left(\cdot\right)$ at $z$~\cite{introduction_to_complex_analysis_second_edition_Priestley}) which can be written more explicitly by using $N\left(m\right)$, $c_{mn}$, $r_m$, $\overline{N}\left(m\right)$, $p_{mn}$ and $a_m$ as well as by identifying the appropriate scenario, i.e., $k\neq 0\wedge q\neq 0$, $k\neq 0\wedge q=0$, $k=0\wedge q\neq 0$ or $k=0\wedge q=0$. We note that in the case \eqref{eq: user obtain h finite number of terms} and/or \eqref{eq: user obtain f finite number of terms} the sum in \eqref{eq: sum of residues} will be taken with respect to $c<\mathcal{R}\left(z\right)<\min\{\mathcal{R}\left(r_{M_h+1}\right),\mathcal{R}\left(a_{M_f+1}\right)+1\}$ instead of with respect to $c<\mathcal{R}\left(z\right)$.

\begin{IEEEproof}[Proof of Theorem \ref{thm: Additive Gaussian Noise average mmse general h}]
We now establish the requirements \ref{item: requirement 1}-\ref{item: requirement 6} for the application of the Mellin transform method.

The function $f\left(\cdot\right)$ is locally integrable on $\mathbb{R}^+$ because of the hypothesis $E_{h}\left\{|h|^2\right\}<+\infty$ which ensures that
\begin{equation*}
a<b\Rightarrow\int_a^bf\left(t\right)dt=\int_a^b\frac{\sqrt{t}f_{|h|}\left(\sqrt{t}\right)}{2}dt=\int_{\sqrt{a}}^{\sqrt{b}}u^2f_{|h|}\left(u\right)du\leq\int_0^{+\infty}u^2f_{|h|}\left(u\right)du=E_{h}\left\{|h|^2\right\}<+\infty
\end{equation*}

The Mellin transform $M\left[f;z\right]$ converges absolutely and is holomorphic in the strip $\gamma<\mathcal{R}\left(z\right)<\delta$ because of hypothesis \eqref{eq: gamma less delta in Additive Gaussian Noise average mmse general h Mellin transform}~\cite[p.~106]{asymptotic_expansion_of_integrals_Bleistein_and_Handelsman}.

The function $h\left(\cdot\right)$ is locally integrable on $\mathbb{R}^+$ and the Mellin transform $M\left[h;z\right]$ converges absolutely and is holomorphic in the strip $\mathcal{R}\left(z\right)>0$ because of
\begin{gather}
h\left(0\right)<+\infty\label{eq: finite mmse}\\
\forall t\in\mathbb{R}_0^+, h\left(t\right)>0\label{eq: positive mmse}\\
\forall t\in\mathbb{R}_0^+, h'\left(t\right)<0\label{eq: decreasing decay mmse}\\
h\left(t\right)=O\left(\exp{\left(-\frac{d^2}{4}t\right)}\right), \qquad t\rightarrow +\infty\label{eq: exponential decay mmse}
\end{gather}
where $d>0$ denotes the minimum distance between the elements of the support of the input distribution~\cite{Estimation_in_Gaussian_Noise_Properties_of_the_Minimum_Mean-Square_Error_Guo_Wu_Shamai_and_Verdu}~\cite[Theorem 4]{Lozano06optimumpower}. Hence, requirements \ref{item: requirement 1} and \ref{item: requirement 3} are satisfied.

Requirement \ref{item: requirement 2} is ensured by hypothesis \eqref{eq: asymptotic behavior in Additive Gaussian Noise average mmse general h} and \eqref{eq: exponential decay mmse}.

Requirement \ref{item: requirement 4} is ensured by hypothesis \eqref{eq: C not empty in Additive Gaussian Noise average mmse general h Mellin transform}.

Combining
\begin{equation*}
M\left[f;1-c-iy\right]\in L^1\left(-\infty<y<+\infty\right)
\end{equation*}
(which is true due to hypothesis \eqref{eq: asymptotic behavior in Additive Gaussian Noise average mmse general h Mellin transform} and the fact that $M\left[f;1-c-iy\right]$ is holomorphic in the line $]c-i\infty,c+i\infty[$) with
\begin{equation*}
t^{c-1}h\left(t\right)\in L^1\left(0\leq t<+\infty\right)
\end{equation*}
(which is true due to \eqref{eq: finite mmse}, \eqref{eq: positive mmse}, \eqref{eq: decreasing decay mmse}, \eqref{eq: exponential decay mmse} and $\emptyset\neq C\subseteq\mathbb{R}^+$) it is clear~\cite[p.~108]{asymptotic_expansion_of_integrals_Bleistein_and_Handelsman} that
\begin{equation*}
\int_0^{+\infty}f\left(t\right)h\left(t\right)dt=\frac{1}{2\pi i}\int_{c-i\infty}^{c+i\infty}G\left(z\right)dz
\end{equation*}

Combining hypothesis \eqref{eq: asymptotic behavior in Additive Gaussian Noise average mmse general h Mellin transform} with
\begin{equation*}
\forall x\in\mathbb{R}^+, M\left[h;x+iy\right]=o\left(1\right), \qquad |y|\rightarrow +\infty
\end{equation*}
(which is true because $M\left[h;z\right]$ is holomorphic in the strip $\mathcal{R}\left(z\right)>0$) yields
\begin{equation*}
\forall x\in[c,+\infty[, G\left(x+iy\right)=O\left(|y|^{-2}\right), \qquad |y|\rightarrow +\infty
\end{equation*}
which in turn implies
\begin{equation*}
\forall x\in[c,+\infty[, \lim_{|y|\rightarrow +\infty}G\left(x+iy\right)=0
\end{equation*}
as well as (note that \eqref{eq: exponential decay mmse} implies that $k\neq 0$)
\begin{itemize}
	\item
If $k\neq 0\wedge q\neq 0$, there exists a real sequence $u_n$ such that $u_n\uparrow +\infty$ and -- because $\forall n\in\mathbb{Z}^+, G\left(u_n+iy\right)$ is holomorphic in the line $]u_n-i\infty,u_n+i\infty[$ -- $\forall n\in\mathbb{Z}^+, \int_{-\infty}^{+\infty}\left|G\left(u_n+iy\right)\right|dy<+\infty$,
	\item
If $k\neq 0\wedge q=0$, let $U:=\{\mathcal{R}\left(a_m\right)+1: m\in\mathbb{Z}^+\}$ and let $u_n$ be the real sequence such that $u_n\uparrow +\infty$ and $U=\{u_n: n\in\mathbb{Z}^+\}$. Since $\forall x\in]u_n,u_{n+1}[, G\left(x+iy\right)$ is holomorphic in the line $]x-i\infty,x+i\infty[$, we have that $\forall n\in\mathbb{Z}^+, \exists x\in]u_n,u_{n+1}[: \int_{-\infty}^{+\infty}\left|G\left(x+iy\right)\right|dy<+\infty$.
\end{itemize}

Hence, requirements \ref{item: requirement 5} and \ref{item: requirement 6} are satisfied.

These established requirements lead immediately -- via \eqref{eq: sum of residues} -- to the expansions:
\begin{itemize}
	\item
If $q=0$ then
\begin{align*}
&\overline{mmse}\left(snr\right)\sim\nonumber\\
&\sim\sum_{m=0}^{+\infty}snr^{-1-a_m}\sum_{n=0}^{\overline{N}\left(m\right)}p_{mn}\sum_{j=0}^n\binom{n}{j}\left(-\log{\left(snr\right)}\right)^jM^{\left(n-j\right)}\left[mmse;z\right]\Bigg|_{z=1+a_m}, \qquad snr\rightarrow +\infty
\end{align*}
	\item
If $q\neq 0$ then
\begin{equation*}
\forall R\in\mathbb{R}^+, \overline{mmse}\left(snr\right)=o\left(snr^{-R}\right), \qquad snr\rightarrow +\infty
\end{equation*}
\end{itemize}
\end{IEEEproof}

\section{Proof of Corollary \ref{cor: Additive Gaussian Noise average mmse h Rayleigh and Rice}}
\label{section: Proof of Corollary Additive Gaussian Noise average mmse h Rayleigh and Rice}
\subsection{Case \texorpdfstring{$\mu=0$}{mu=0}}
\label{subsection: Proof of Corollary Additive Gaussian Noise average mmse h Rayleigh and Rice case mu=0}
Since, by Taylor's Theorem~\cite{introduction_to_complex_analysis_second_edition_Priestley},
\begin{equation*}
f\left(t\right):=\frac{\sqrt{t}f_{|h|}\left(\sqrt{t}\right)}{2}=\frac{t}{2\sigma^2}\exp{\left(-\frac{t}{2\sigma^2}\right)}\sim\sum_{m=0}^{+\infty}\frac{\left(-1\right)^m}{m!\left(2\sigma^2\right)^{m+1}}t^{m+1}, \qquad t\rightarrow 0^+
\end{equation*}
we have that requirement \eqref{eq: asymptotic behavior in Additive Gaussian Noise average mmse general h} holds with
\begin{equation*}
q=0,\qquad\overline{N}\left(m\right)=0,\qquad p_{m0}=\frac{\left(-1\right)^m}{m!\left(2\sigma^2\right)^{m+1}},\qquad a_m=m+1
\end{equation*}

Since, the Mellin transform of $f\left(\cdot\right)$
\begin{equation*}
M[f;z]=\int_0^{+\infty}t^{z-1}\frac{t}{2\sigma^2}\exp{\left(-\frac{t}{2\sigma^2}\right)}dt=\left(2\sigma^2\right)^z\Gamma\left(z+1\right)<+\infty
\end{equation*}
converges absolutely and is holomorphic in the strip $\mathcal{R}\left(z\right)>-1$~\cite[Equation 5.2.1]{abramowitz_stegun}, we have that $\gamma=-1$ and $\delta =+\infty$ which satisfy $\gamma<\delta$ and $]0,+\infty[\cap]1-\delta,1-\gamma[=]0,+\infty[\cap]-\infty,2[=]0,2[\neq\emptyset$, i.e., requirements \eqref{eq: gamma less delta in Additive Gaussian Noise average mmse general h Mellin transform} and \eqref{eq: C not empty in Additive Gaussian Noise average mmse general h Mellin transform} are satisfied. Also, since~\cite[p.~138]{asymptotic_expansion_of_integrals_Bleistein_and_Handelsman}
\begin{equation*}
\forall x\in\mathbb{R}, \Gamma\left(x+iy\right)=O\left(\exp{\left(-\left(\frac{\pi}{2}-\epsilon\right)|y|\right)}\right), \qquad |y|\rightarrow +\infty
\end{equation*}
where $\epsilon$ is any small positive real number, we also have that
\begin{equation*}
\forall x\in[\frac{1}{2},+\infty[, M\left[f;1-x-iy\right]=O\left(|y|^{-2}\right), \qquad |y|\rightarrow +\infty
\end{equation*}
i.e., we also have requirement \eqref{eq: asymptotic behavior in Additive Gaussian Noise average mmse general h Mellin transform}.

The result now follows from Theorem \ref{thm: Additive Gaussian Noise average mmse general h}.

\subsection{Case \texorpdfstring{$\mu\neq0$}{mu not 0}}
\label{subsection: Proof of Corollary Additive Gaussian Noise average mmse h Rayleigh and Rice case mu not 0}
Since
\begin{align}
f\left(t\right):&=\frac{\sqrt{t}f_{|h|}\left(\sqrt{t}\right)}{2}\nonumber\\
&=\frac{t}{2\sigma^2}\exp{\left(-\frac{|\mu|^2}{2\sigma^2}\right)}\exp{\left(-\frac{t}{2\sigma^2}\right)}I_0\left(\frac{\sqrt{t}|\mu|}{\sigma^2}\right)\nonumber\\
&=\frac{t}{2\sigma^2}\exp{\left(-\frac{|\mu|^2}{2\sigma^2}\right)}\left(\sum_{n=0}^{+\infty}\frac{\left(-1\right)^n}{n!\left(2\sigma^2\right)^n}t^n\right)\left(\sum_{m=0}^{+\infty}\frac{|\mu|^{2m}}{\left(m!\right)^2\left(2\sigma^2\right)^{2m}}t^m\right)\nonumber\\
&=\frac{t}{2\sigma^2}\exp{\left(-\frac{|\mu|^2}{2\sigma^2}\right)}\sum_{k=0}^{+\infty}\sum_{l=0}^k\left(\frac{\left(-1\right)^{k-l}}{\left(k-l\right)!\left(2\sigma^2\right)^{k-l}}\frac{|\mu|^{2l}}{\left(l!\right)^2\left(2\sigma^2\right)^{2l}}\right)t^k\nonumber\\
&=\exp{\left(-\frac{|\mu|^2}{2\sigma^2}\right)}\sum_{k=0}^{+\infty}\sum_{l=0}^k\left(\frac{\left(-1\right)^{k-l}}{\left(k-l\right)!\left(2\sigma^2\right)^{k+l+1}}\frac{|\mu|^{2l}}{\left(l!\right)^2}\right)t^{k+1}
\label{eq: Taylor expansion of f related to Rice fading}
\end{align}
where the second equality is due to the definition of the modified Bessel function of the first kind~\cite[Equation 10.25.2]{abramowitz_stegun}, we have that requirement \eqref{eq: asymptotic behavior in Additive Gaussian Noise average mmse general h} holds with
\begin{equation*}
q=0,\qquad\overline{N}\left(m\right)=0,\qquad p_{k0}=\exp{\left(-\frac{|\mu|^2}{2\sigma^2}\right)}\sum_{l=0}^k\left(\frac{\left(-1\right)^{k-l}}{\left(k-l\right)!\left(2\sigma^2\right)^{k+l+1}}\frac{|\mu|^{2l}}{\left(l!\right)^2}\right),\qquad a_k=k+1
\end{equation*}

Since, the Mellin transform of $f\left(\cdot\right)$
\begin{align*}
M[f;z]&=\int_0^{+\infty}t^{z-1}\frac{t}{2\sigma^2}\exp{\left(-\frac{|\mu|^2}{2\sigma^2}\right)}\exp{\left(-\frac{t}{2\sigma^2}\right)}I_0\left(\frac{\sqrt{t}|\mu|}{\sigma^2}\right)dt\nonumber\\
&=\exp{\left(-\frac{|\mu|^2}{2\sigma^2}\right)}\left(2\sigma^2\right)^z\int_0^{+\infty}u^z\exp{\left(-u\right)}I_0\left(2\sqrt{u\frac{|\mu|^2}{2\sigma^2}}\right)du\nonumber\\
&=\exp{\left(-\frac{|\mu|^2}{2\sigma^2}\right)}\left(2\sigma^2\right)^z\int_0^{+\infty}u^z\exp{\left(-u\right)}\sum_{n=0}^{+\infty}\frac{1}{\left(n!\right)^2}u^n\left(\frac{|\mu|^2}{2\sigma^2}\right)^ndu\nonumber\\
&=\exp{\left(-\frac{|\mu|^2}{2\sigma^2}\right)}\left(2\sigma^2\right)^z\sum_{n=0}^{+\infty}\int_0^{+\infty}u^{z+n}\exp{\left(-u\right)}du\frac{1}{\left(n!\right)^2}\left(\frac{|\mu|^2}{2\sigma^2}\right)^n\nonumber\\
&=\exp{\left(-\frac{|\mu|^2}{2\sigma^2}\right)}\left(2\sigma^2\right)^z\Gamma\left(z+1\right)\frac{\Gamma\left(1\right)}{\Gamma\left(z+1\right)}\sum_{n=0}^{+\infty}\frac{\Gamma\left(z+1+n\right)}{\Gamma\left(1+n\right)}\frac{\left(\frac{|\mu|^2}{2\sigma^2}\right)^n}{n!}\nonumber\\
&=\exp{\left(-\frac{|\mu|^2}{2\sigma^2}\right)}\left(2\sigma^2\right)^z\Gamma\left(z+1\right){_1}F_1\left(z+1;1;\frac{|\mu|^2}{2\sigma^2}\right)\nonumber\\
&<+\infty
\end{align*}
where the third equality is due to the definition of the modified Bessel function of the first kind~\cite[Equation 10.25.2]{abramowitz_stegun}, the fourth equality is due to Fubini Theorem~\cite[Theorem 6.5]{Introduction_to_Measure_and_Probability_Kingman_and_Taylor}, the fifth equality is due to $\mathcal{R}\left(z\right)>-1$ which implies $\forall n\in\mathbb{Z}_0^+, \mathcal{R}\left(z\right)+n>-1$ and the sixth equality is due to the definition of the Confluent hypergeometric series~\cite[Equation 13.2.2]{abramowitz_stegun}, converges absolutely and is holomorphic in the strip $\mathcal{R}\left(z\right)>-1$, we have that $\gamma=-1$ and $\delta =+\infty$ which satisfy $\gamma<\delta$ and $]0,+\infty[\cap]1-\delta,1-\gamma[=]0,+\infty[\cap]-\infty,2[=]0,2[\neq\emptyset$, i.e., we satisfy requirements \eqref{eq: gamma less delta in Additive Gaussian Noise average mmse general h Mellin transform} and \eqref{eq: C not empty in Additive Gaussian Noise average mmse general h Mellin transform}.

It is now important to examine the asymptotic behavior of the Mellin transform of $f\left(\cdot\right)$. We can conclude from \eqref{eq: Taylor expansion of f related to Rice fading} that $f\left(\cdot\right)$ is infinitely continuously differentiable on $\mathbb{R}$. We can thus also conclude -- in view of the fact that the power series \eqref{eq: Taylor expansion of f related to Rice fading} has infinite radius of convergence -- that $f^{\left(j\right)}\left(\cdot\right)$, $j=1,2,\ldots$ is given by term-by-term differentiation of \eqref{eq: Taylor expansion of f related to Rice fading}~\cite[p.~74]{introduction_to_complex_analysis_second_edition_Priestley}. This leads to the fact that
\begin{equation*}
g\left(t,x,p\right):=\left(t\left(\frac{d}{dt}\right)\right)^p\left(t^xf\left(t\right)\right)
\end{equation*}
is a finite sum where the terms are given by a constant, times a power of $t$ (namely, $t^y$ with $y\in\mathbb{R}$), times $\exp{\left(-\frac{t}{2\sigma^2}\right)}$ and times $I_0\left(\frac{\sqrt{t}|\mu|}{\sigma^2}\right)$ and/or a derivative of $I_0\left(t\right)$ evaluated at $t=\frac{\sqrt{t}|\mu|}{\sigma^2}$, and together with~\cite[Equation 10.29.5]{abramowitz_stegun}
\begin{equation*}
I_v^{\left(k\right)}\left(z\right)=\frac{1}{2^k}\left(I_{v-k}\left(z\right)+\binom{k}{1}I_{v-k+2}\left(z\right)+\binom{k}{2}I_{v-k+4}\left(z\right)+\cdots +I_{v+k}\left(z\right)\right)
\end{equation*}
and~\cite[Equation 10.40.1]{abramowitz_stegun}
\begin{equation*}
I_v\left(z\right)\sim\frac{\exp{\left(z\right)}}{\left(2\pi z\right)^{\frac{1}{2}}}\sum_{k=0}^{+\infty}\left(-1\right)^k\frac{a_k\left(v\right)}{z^k}, \qquad z\rightarrow +\infty
\end{equation*}
where~\cite[Equation 10.17.1]{abramowitz_stegun}
\begin{equation*}
a_0\left(v\right)=1 \qquad k\geq 1\Rightarrow a_k\left(v\right)=\frac{\left(4v^2-1^2\right)\left(4v^2-3^2\right)\ldots\left(4v^2-\left(2k-1\right)^2\right)}{k!8^k}
\end{equation*}
to the fact that
\begin{equation*}
g\left(t,x,p\right)=O\left(\exp{\left(-k\left(p\right)t\right)}\right), \qquad t\rightarrow +\infty
\label{eq: limit g related to Rice fading}
\end{equation*}
where $\forall p\in\mathbb{Z}_0^+, k\left(p\right)>0$.

We have now established that $f\left(t\right)$ is infinitely continuously differentiable on $\mathbb{R}^+$, that
\begin{equation*}
f\left(t\right)\sim\sum_{m=0}^{+\infty}p_{m0}t^{a_m}, \qquad t\rightarrow 0^+
\end{equation*}
where $\mathcal{R}\left(a_m\right)\uparrow +\infty$, that the asymptotic expansion of $f^{\left(j\right)}\left(t\right), j=1,2,\ldots$ as $t\rightarrow 0^+$ is obtained from the asymptotic expansion of $f\left(\cdot\right)$ by successively differentiating term-by-term, and that $g\left(t,x,p\right)$ vanishes as $t\rightarrow +\infty$ for $p=0,1,\ldots$ and $x>-\mathcal{R}\left(a_0\right)$. This implies~\cite[Corollary 6.2.3]{asymptotic_expansion_of_integrals_Bleistein_and_Handelsman} that
\begin{equation*}
\forall R\in\mathbb{R}^+, \forall x\in\mathbb{R}, M[f;x+iy]=O\left(|y|^{-R}\right), \qquad |y|\rightarrow +\infty
\end{equation*}
where $M\left[f;x+iy\right]$ is to be understood as the analytic continuation of $M\left[f;x+iy\right]$ from $\{x+iy: 1-\delta<x<1-\gamma\}$ to the entire $z$ plane~\cite[p.~181]{introduction_to_complex_analysis_second_edition_Priestley} and hence that
\begin{equation*}
\forall x\in[\frac{1}{2},+\infty[, M\left[f;1-x-iy\right]=O\left(|y|^{-2}\right), \qquad |y|\rightarrow +\infty
\end{equation*}
i.e., requirement \eqref{eq: asymptotic behavior in Additive Gaussian Noise average mmse general h Mellin transform}.

The result now follows from Theorem \ref{thm: Additive Gaussian Noise average mmse general h}.

\section{Proof of Corollary \ref{cor: Additive Gaussian Noise average mmse h Nakagami}}
\label{section: Proof of Corollary Additive Gaussian Noise average mmse h Nakagami}
Since, by Taylor's Theorem~\cite{introduction_to_complex_analysis_second_edition_Priestley},
\begin{equation*}
f\left(t\right):=\frac{\sqrt{t}f_{|h|}\left(\sqrt{t}\right)}{2}=\frac{\mu^{\mu}}{\Gamma\left(\mu\right)w^{\mu}}t^{\mu}\exp{\left(-\frac{\mu}{w}t\right)}\sim\sum_{m=0}^{+\infty}\frac{\mu^{\mu}}{\Gamma\left(\mu\right)w^{\mu}}\frac{\left(-1\right)^m\mu^m}{m!w^m}t^{m+\mu}, \qquad t\rightarrow 0^+
\end{equation*}
we have that requirement \eqref{eq: asymptotic behavior in Additive Gaussian Noise average mmse general h} holds with
\begin{equation*}
q=0,\qquad\overline{N}\left(m\right)=0,\qquad p_{m0}=\frac{\mu^{\mu}}{\Gamma\left(\mu\right)w^{\mu}}\frac{\left(-1\right)^m\mu^m}{m!w^m},\qquad a_m=m+\mu
\end{equation*}

Since, the Mellin transform of $f\left(\cdot\right)$
\begin{equation*}
M[f;z]=\int_0^{+\infty}t^{z-1}\frac{\mu^{\mu}}{\Gamma\left(\mu\right)w^{\mu}}t^{\mu}\exp{\left(-\frac{\mu}{w}t\right)}dt=\frac{1}{\Gamma\left(\mu\right)}\left(\frac{w}{\mu}\right)^z\Gamma\left(z+\mu\right)<+\infty
\end{equation*}
converges absolutely and is holomorphic in the strip $\mathcal{R}\left(z\right)>-\mu$~\cite[Equation 5.2.1]{abramowitz_stegun}, we have that $\gamma=-\mu$ and $\delta =+\infty$ which satisfy (in view of the fact that $\mu\geq\frac{1}{2}$) $\gamma<\delta$ and $]0,+\infty[\cap]1-\delta,1-\gamma[=]0,+\infty[\cap]-\infty,\mu+1[\supseteq]0,+\infty[\cap]-\infty,\frac{3}{2}[=]0,\frac{3}{2}[\neq\emptyset$, i.e., requirements \eqref{eq: gamma less delta in Additive Gaussian Noise average mmse general h Mellin transform} and \eqref{eq: C not empty in Additive Gaussian Noise average mmse general h Mellin transform} are satisfied. Also, since~\cite[p.~138]{asymptotic_expansion_of_integrals_Bleistein_and_Handelsman}
\begin{equation*}
\forall x\in\mathbb{R}, \Gamma\left(x+iy\right)=O\left(\exp{\left(-\left(\frac{\pi}{2}-\epsilon\right)|y|\right)}\right), \qquad |y|\rightarrow +\infty
\end{equation*}
where $\epsilon$ is any small positive real number, we also have that
\begin{equation*}
\forall x\in[\frac{1}{2},+\infty[, M\left[f;1-x-iy\right]=O\left(|y|^{-2}\right), \qquad |y|\rightarrow +\infty
\end{equation*}
i.e., we also have requirement \eqref{eq: asymptotic behavior in Additive Gaussian Noise average mmse general h Mellin transform}.

The result now follows from Theorem \ref{thm: Additive Gaussian Noise average mmse general h}.

\section{Proof of Theorem \ref{thm: Additive Gaussian Noise average mmse general h via integration by parts}}
\label{section: Proof of Theorem Additive Gaussian Noise average mmse general h via integration by parts}
Let
\begin{align*}
f\left(t\right)&:=\frac{\sqrt{t}f_{|h|}\left(\sqrt{t}\right)}{2}\\
h\left(t\right)&:=mmse\left(t\right)
\end{align*}
Let also $n,M\in\mathbb{Z}_0^+$ and $x\in\mathbb{R}_0^+$. We have that
\begin{align}
{h}^{\left(-n-1\right)}\left(x\right)&=\int_{+\infty}^x\int_{+\infty}^{t_1}\cdots\int_{+\infty}^{t_{n-1}}\int_{+\infty}^{t_{n}}h\left(t_{n+1}\right)dt_{n+1}dt_n\cdots dt_2dt_1\nonumber\\
&=\frac{\left(-1\right)^{n+1}}{n!}\int_x^{+\infty}h\left(t\right)\left(t-x\right)^{n}dt\nonumber\\
&=\frac{\left(-1\right)^{n+1}}{n!}\int_0^{+\infty}h\left(t+x\right)t^{n}dt
\label{eq: Proof of Theorem Additive Gaussian Noise average mmse general h via integration by parts fundamental step}
\end{align}
where in the first equality we use \eqref{eq: [m+1]th repeated integral} and in the second equality we use~\cite[Equation 1.4.31]{abramowitz_stegun} and hence that
\begin{align}
\left|{h}^{\left(-n-1\right)}\left(x\right)\right|&=\left|\frac{\left(-1\right)^{n+1}}{n!}\int_0^{+\infty}h\left(t+x\right)t^{n}dt\right|\nonumber\\
&=\frac{1}{n!}\int_0^{+\infty}h\left(t+x\right)t^{n}dt\nonumber\\
&\leq\frac{1}{n!}\int_0^{+\infty}h\left(t\right)t^{n}dt\nonumber\\
&=\frac{1}{n!}M\left[h;n+1\right]\nonumber\\
&<+\infty
\label{eq: Proof of Theorem Additive Gaussian Noise average mmse general h via integration by parts fundamental step part 2}
\end{align}
where the second equality and the first inequality are due to \eqref{eq: finite mmse}, \eqref{eq: positive mmse} and \eqref{eq: decreasing decay mmse} and the second inequality has been justified in Appendix \ref{section: Proof of Theorem Additive Gaussian Noise average mmse general h}. We also have that
\begin{align}
\lim_{x\rightarrow 0^+}{h}^{\left(-n-1\right)}\left(x\right)&=\frac{\left(-1\right)^{n+1}}{n!}\lim_{x\rightarrow 0^+}\int_0^{+\infty}h\left(t+x\right)t^{n}dt\nonumber\\
&=\frac{\left(-1\right)^{n+1}}{n!}\int_0^{+\infty}\lim_{x\rightarrow 0^+}\left(h\left(t+x\right)t^{n}\right)dt\nonumber\\
&=\frac{\left(-1\right)^{n+1}}{n!}\int_0^{+\infty}h\left(t\right)t^{n}dt\nonumber\\
&={h}^{\left(-n-1\right)}\left(0\right)
\label{eq: Proof of Theorem Additive Gaussian Noise average mmse general h via integration by parts fundamental step number 2}
\end{align}
where the first equality is due to \eqref{eq: Proof of Theorem Additive Gaussian Noise average mmse general h via integration by parts fundamental step} and the second equality is due to \eqref{eq: Proof of Theorem Additive Gaussian Noise average mmse general h via integration by parts fundamental step part 2} and to Lebesgue Dominated Convergence Theorem~\cite[Theorem 5.8]{Introduction_to_Measure_and_Probability_Kingman_and_Taylor}. One can now use \eqref{eq: Proof of Theorem Additive Gaussian Noise average mmse general h via integration by parts fundamental step part 2} and \eqref{eq: Proof of Theorem Additive Gaussian Noise average mmse general h via integration by parts fundamental step number 2} with the set of hypotheses \eqref{eq: differentiability in Additive Gaussian Noise average mmse general h via integration by parts}, \eqref{eq: existence and finiteness of limits in Additive Gaussian Noise average mmse general h via integration by parts} and \eqref{eq: asymptotic behavior in Additive Gaussian Noise average mmse general h via integration by parts} to conclude that
\begin{equation*}
\forall snr\in\mathbb{R}^+, \exists\lim_{t\rightarrow +\infty}f^{\left(n\right)}\left(t\right)h^{\left(-n-1\right)}\left(snrt\right)=0
\end{equation*}
\begin{equation*}
\forall snr\in\mathbb{R}^+, \exists\lim_{t\rightarrow 0^+}f^{\left(n\right)}\left(t\right)h^{\left(-n-1\right)}\left(snrt\right)<+\infty
\end{equation*}
\begin{equation*}
\exists\int_0^{+\infty}f^{\left(n+1\right)}\left(t\right)h^{\left(-n-1\right)}\left(snrt\right)dt=O\left(1\right), \qquad snr\rightarrow +\infty
\end{equation*}
This leads immediately -- via integration by parts -- to the asymptotic expansion given by
\begin{align*}
&\overline{mmse}\left(snr\right)=\\
&=E_{|h|}\left\{|h|^2mmse\left(snr|h|^2\right)\right\}\\
&=\int_0^{+\infty}f\left(u\right)h\left(snru\right)du\\
&=\sum_{m=0}^{M}\frac{\left(-1\right)^{m}}{snr^{m+1}}\left[f^{\left(m\right)}\left(t\right)h^{\left(-m-1\right)}\left(snrt\right)\right]_0^{+\infty}+\frac{\left(-1\right)^{M+1}}{snr^{M+1}}\int_0^{+\infty}f^{\left(M+1\right)}\left(t\right)h^{\left(-M-1\right)}\left(snrt\right)dt\\
&=\sum_{m=0}^{M}\frac{\left(-1\right)^{m}}{snr^{m+1}}\left[f^{\left(m\right)}\left(t\right)h^{\left(-m-1\right)}\left(snrt\right)\right]_0^{+\infty}+\\
&\quad\quad\quad\quad+\frac{\left(-1\right)^{M+1}}{snr^{M+2}}\left(\left[f^{\left(M+1\right)}\left(t\right)h^{\left(-M-2\right)}\left(snrt\right)\right]_0^{+\infty}-\int_0^{+\infty}f^{\left(M+2\right)}\left(t\right)h^{\left(-M-2\right)}\left(snrt\right)dt\right)\\
&=\sum_{m=0}^{M}\frac{\left(-1\right)^{m+1}}{snr^{m+1}}f^{\left(m\right)}\left(0\right)h^{\left(-m-1\right)}\left(0\right)+\\
&\quad\quad\quad\quad+\frac{\left(-1\right)^{M+2}}{snr^{M+2}}\left(f^{\left(M+1\right)}\left(0\right)h^{\left(-M-2\right)}\left(0\right)+\int_0^{+\infty}f^{\left(M+2\right)}\left(t\right)h^{\left(-M-2\right)}\left(snrt\right)dt\right)\\
&=\sum_{m=0}^{M}\frac{\left(-1\right)^{m+1}}{snr^{m+1}}f^{\left(m\right)}\left(0\right)h^{\left(-m-1\right)}\left(0\right)+O\left(\frac{1}{snr^{M+2}}\right), \qquad snr\rightarrow +\infty\\
&\sim\sum_{m=0}^{+\infty}\frac{\left(-1\right)^{m+1}}{snr^{m+1}}f^{\left(m\right)}\left(0\right){h}^{\left(-m-1\right)}\left(0\right), \qquad snr\rightarrow +\infty
\end{align*}

\section{Proof of Theorem \ref{thm: Additive Gaussian Noise average mmse general h Continuous Inputs}}
\label{section: Proof of Theorem Additive Gaussian Noise average mmse general h Continuous Inputs}
Let
\begin{align*}
f\left(t\right)&:=\frac{\sqrt{t}f_{|h|}\left(\sqrt{t}\right)}{2}\\
h\left(t\right)&:=mmse\left(t\right)
\end{align*}
Then
\begin{equation}
\overline{mmse}\left(snr\right)=E_{|h|}\left\{|h|^2mmse\left(snr|h|^2\right)\right\}=\int_0^{+\infty}f\left(u\right)h\left(snru\right)du
\label{eq: Proof of Theorem Additive Gaussian Noise average mmse general h Continuous Inputs Part 1}
\end{equation}

We note that \eqref{eq: Proof of Theorem Additive Gaussian Noise average mmse general h Continuous Inputs Part 1} is also an $h$-transform with Kernel of Monotonic Argument, so that we can capitalize on the method of Mellin transforms~\cite[Section 4.4]{asymptotic_expansion_of_integrals_Bleistein_and_Handelsman} to obtain the asymptotic expansion of $\overline{mmse}\left(snr\right)$ as $snr \to +\infty$ via~\cite[Theorem 4.4]{asymptotic_expansion_of_integrals_Bleistein_and_Handelsman}.

We now establish the requirements for the application of the Mellin transforms method.

\subsection{Case: \texorpdfstring{$x\sim\infty$-PSK, $x\sim\infty$-PAM or $x\sim\infty$-QAM}{infty -PSK, -PAM or -QAM}}
We can establish that $f\left(\cdot\right)$ is locally integrable on $\mathbb{R}^+$ and that $M\left[f;z\right]$ converges absolutely and is holomorphic in the strip $\gamma<\mathcal{R}\left(z\right)<\delta$ by capitalizing on $E_{h}\left\{|h|^2\right\}<+\infty$ and hypothesis \eqref{eq: gamma less delta in Additive Gaussian Noise average mmse general h Mellin transform Continuous Inputs}, respectively, as we did in Appendix \ref{section: Proof of Theorem Additive Gaussian Noise average mmse general h}. We now extend $M\left[f;z\right]$ to the left of the strip: consider the two cases $q=0$ and $q\neq 0$:
\begin{enumerate}
	\item
Case $q=0$: In this case we have~\cite[Lemma 4.3.6]{asymptotic_expansion_of_integrals_Bleistein_and_Handelsman} that $M\left[f;1-z\right]$ can be analytically continued as a meromorphic function from $1-\delta<\mathcal{R}\left(z\right)<1-\gamma$ into $\mathcal{R}\left(z\right)>1-\delta$ with poles at $z=a_m+1$ for every $m\in\mathbb{Z}_0^+$.
	\item
Case $q\neq 0$: In this case we have~\cite[Lemma 4.3.4]{asymptotic_expansion_of_integrals_Bleistein_and_Handelsman} that $M\left[f;1-z\right]$ can be analytically continued as a holomorphic function from $1-\delta<\mathcal{R}\left(z\right)<1-\gamma$ into $\mathcal{R}\left(z\right)>1-\delta$.
\end{enumerate}

We can also establish that $h\left(t\right)$ is locally integrable on $\mathbb{R}^+$ because $h\left(\cdot\right)$ is continuous on $\mathbb{R}_0^+$ and that $M\left[h;z\right]$ converges absolutely and is holomorphic in the strip $0<\mathcal{R}\left(z\right)<1$ for $\infty$-PSK, $\infty$-PAM and $\infty$-QAM because of \eqref{eq: finite mmse}, \eqref{eq: positive mmse} and \eqref{eq: decreasing decay mmse}, and of \eqref{eq: decay mmse Infinite PSK intro}, \eqref{eq: decay mmse Infinite PAM intro} and \eqref{eq: decay mmse Infinite QAM intro}, i.e., of
\begin{equation}
h\left(t\right)=\zeta t^{-r_0}+O\left(t^{-r_1}\right), \qquad t\rightarrow +\infty
\label{eq: decay mmse Infinite PSK, PAM and QAM}
\end{equation}
where
\begin{align*}
\zeta:=
\begin{cases}
\frac{1}{2}&\Leftarrow\infty\text{-PSK}\\
\frac{1}{2}&\Leftarrow\infty\text{-PAM}\\
1&\Leftarrow\infty\text{-QAM}\\
\end{cases}
& &
r_0:=1
& &
r_1:=
\begin{cases}
2&\Leftarrow\infty\text{-PSK}\\
\frac{3}{2}&\Leftarrow\infty\text{-PAM}\\
\frac{3}{2}&\Leftarrow\infty\text{-QAM}\\
\end{cases}
\end{align*}
We now extend $M\left[h;z\right]$ to the right of the strip: indeed, it can be~\cite[Lemma 4.3.3]{asymptotic_expansion_of_integrals_Bleistein_and_Handelsman} analytically continued as a meromorphic function into $0<\mathcal{R}\left(z\right)<r_1$ with a pole at $z=r_0$.

Note that hypothesis \eqref{eq: asymptotic behavior in Additive Gaussian Noise average mmse general h Continuous Inputs} and \eqref{eq: decay mmse Infinite PSK, PAM and QAM} ensure a ``correct'' type of decay.

Note also that hypothesis \eqref{eq: C not empty in Additive Gaussian Noise average mmse general h Mellin transform Continuous Inputs} ensures that the function $G\left(\cdot\right):=M\left[h;\cdot\right]M\left[f;1-\cdot\right]$ is holomorphic in $C:=]0,1[\cap]1-\delta,1-\gamma[\neq\emptyset$. Note also that $G\left(\cdot\right)$ can be analytically continued: consider the two cases $q=0$ and $q\neq 0$:
\begin{enumerate}
	\item
Case $q=0$: The function $G\left(z\right)$ can be analytically continued as a meromorphic function from $C$ into $\max\{0,1-\delta\}<\mathcal{R}\left(z\right)<\min\left\{\mathcal{R}\left(a_0\right)+1,r_1\right\}$ because $r_0=1$ and by hypothesis \eqref{eq: real part of a_0 greater than zero in Additive Gaussian Noise average mmse general h Mellin transform Continuous Inputs} and $\mathcal{R}\left(a_m\right)\uparrow +\infty$ implies that $\mathcal{R}\left(a_m\right)+1>r_0$.
	\item
Case $q\neq 0$: The function $G\left(z\right)$ can be analytically continued as a meromorphic function from $C$ into $\max\{0,1-\delta\}<\mathcal{R}\left(z\right)<r_1$.
\end{enumerate}

Combining
\begin{equation*}
M\left[f;1-c-iy\right]\in L^1\left(-\infty<y<+\infty\right)
\end{equation*}
(which is true due to hypothesis \eqref{eq: asymptotic behavior in Additive Gaussian Noise average mmse general h Mellin transform Continuous Inputs} and the fact that $M\left[f;1-c-iy\right]$ is holomorphic in the line $]c-i\infty,c+i\infty[$) with
\begin{equation*}
t^{c-1}h\left(t\right)\in L^1\left(0\leq t<+\infty\right)
\end{equation*}
(which is true due to \eqref{eq: finite mmse}, \eqref{eq: positive mmse}, \eqref{eq: decreasing decay mmse}, \eqref{eq: decay mmse Infinite PSK, PAM and QAM} and $\emptyset\neq C\subseteq]0,1[$) it is clear that~\cite[p.~108]{asymptotic_expansion_of_integrals_Bleistein_and_Handelsman}
\begin{equation*}
\int_0^{+\infty}f\left(t\right)h\left(t\right)dt=\frac{1}{2\pi i}\int_{c-i\infty}^{c+i\infty}G\left(z\right)dz
\end{equation*}

Combining hypothesis \eqref{eq: asymptotic behavior in Additive Gaussian Noise average mmse general h Mellin transform Continuous Inputs} with the fact that
\begin{enumerate}
	\item
Case $q=0$:
\begin{equation*}
\forall x\in[c,\min\left\{\mathcal{R}\left(a_0\right)+1,r_1\right\}[, M\left[h;x+iy\right]=o\left(1\right), \qquad |y|\rightarrow +\infty
\end{equation*}
we conclude -- due to $\emptyset\neq C\subseteq]0,1[$ -- that
\begin{equation*}
\forall x\in[c,\min\left\{\mathcal{R}\left(a_0\right)+1,r_1\right\}[, G\left(x+iy\right)=O\left(|y|^{-2}\right), \qquad |y|\rightarrow +\infty
\end{equation*}
which in turn implies that
\begin{equation*}
\forall x\in[c,\min\left\{\mathcal{R}\left(a_0\right)+1,r_1\right\}[, \lim_{|y|\rightarrow +\infty}G\left(x+iy\right)=0
\end{equation*}
and that
\begin{equation*}
\forall x\in]r_0,\min\left\{\mathcal{R}\left(a_0\right)+1,r_1\right\}[, \int_{-\infty}^{+\infty}\left|G\left(x+iy\right)\right|dy<+\infty
\end{equation*}
The fact that the requirements for the application of the Mellin transform method are met leads to the expansions~\cite[Theorem 4.4]{asymptotic_expansion_of_integrals_Bleistein_and_Handelsman}:
\begin{equation*}
 \overline{mmse}\left(snr\right)\sim\frac{\zeta M\left[f;0\right]}{snr}+O\left(\frac{1}{snr^R}\right), \qquad snr\rightarrow +\infty, \qquad \forall R\in]1,\min\left\{\mathcal{R}\left(a_0\right)+1,r_1\right\}[
\end{equation*}
	\item
Case $q\neq 0$:
\begin{equation*}
\forall x\in[c,r_1[, M\left[h;x+iy\right]=o\left(1\right), \qquad |y|\rightarrow +\infty
\end{equation*}
we conclude -- due to $\emptyset\neq C\subseteq]0,1[$ -- that
\begin{equation*}
\forall x\in[c,r_1[, G\left(x+iy\right)=O\left(|y|^{-2}\right), \qquad |y|\rightarrow +\infty
\end{equation*}
which in turn implies that
\begin{equation*}
\forall x\in[c,r_1[, \lim_{|y|\rightarrow +\infty}G\left(x+iy\right)=0
\end{equation*}
and that
\begin{equation*}
\forall x\in]r_0,r_1[, \int_{-\infty}^{+\infty}\left|G\left(x+iy\right)\right|dy<+\infty
\end{equation*}
The fact that the requirements for the application of the Mellin transform method are met leads to the expansions~\cite[Theorem 4.4]{asymptotic_expansion_of_integrals_Bleistein_and_Handelsman}:
\begin{equation*}
\overline{mmse}\left(snr\right)\sim\frac{\zeta M\left[f;0\right]}{snr}+O\left(\frac{1}{snr^R}\right), \qquad snr\rightarrow +\infty, \qquad \forall R\in]1,r_1[
\end{equation*}
\end{enumerate}

\subsection{Case: \texorpdfstring{$x\sim\mathcal{CN}\left(0,1\right)$}{standard complex Gaussian}}
This proof follows the steps of the previous proof. The difference is that we use \eqref{eq: decay mmse Infinite Gaussian intro}
\begin{equation}
h(t)=\frac{1}{1+t}=\frac{1}{t}\frac{1}{\frac{1}{t}+1}\sim\sum_{m=0}^{+\infty}(-1)^mt^{-\left(m+1\right)}, \qquad t\rightarrow +\infty
\label{eq: canonical mmse Gaussian input}
\end{equation}
and~\cite[p.~123]{asymptotic_expansion_of_integrals_Bleistein_and_Handelsman}
\begin{equation}
M\left[h;z\right]=\frac{\pi}{\sin{\left(\pi z\right)}}
\label{eq: canonical mmse Gaussian input Mellin transform}
\end{equation}
Hence, the Mellin transform $M\left[h;z\right]$ converges absolutely and is holomorphic in the strip $0<\mathcal{R}\left(z\right)<1$, and can be analytically continued from $0<\mathcal{R}\left(z\right)<1$ to the entire complex plane via \eqref{eq: canonical mmse Gaussian input Mellin transform}.

We note that, in this case, we have capitalized not only on~\cite[Theorem 4.4]{asymptotic_expansion_of_integrals_Bleistein_and_Handelsman}, but also on~\cite[Exercise 4.16]{asymptotic_expansion_of_integrals_Bleistein_and_Handelsman}.

\section{Proof of Theorem \ref{thm: Additive Gaussian Noise average mmse general h Low-snr}}
\label{section: Proof of Theorem Additive Gaussian Noise average mmse general h Low-snr}
Let
\begin{align*}
f_1\left(t\right)&:=mmse\left(t\right)\\
h_1\left(t\right)&:=\frac{\sqrt{t}f_{|h|}\left(\sqrt{t}\right)}{2}\\
\lambda &:=\frac{1}{snr}
\end{align*}
Then
\begin{gather}
\overline{mmse}\left(snr\right)=E_{|h|}\left\{|h|^2mmse\left(snr|h|^2\right)\right\}=\int_0^{+\infty}h_1\left(u\right)f_1\left(snru\right)du=\lambda\int_0^{+\infty}f_1\left(u\right)h_1\left(\lambda u\right)du
\label{eq: Proof of Theorem Additive Gaussian Noise average mmse general h Low-snr}\\
snr\rightarrow 0^+\Rightarrow\lambda\rightarrow +\infty\nonumber
\end{gather}
We also note that \eqref{eq: Proof of Theorem Additive Gaussian Noise average mmse general h Low-snr} is an $h$-transform with Kernel of Monotonic Argument, so that we can also capitalize on the method of Mellin transforms~\cite[Section 4.4]{asymptotic_expansion_of_integrals_Bleistein_and_Handelsman} to obtain the asymptotic expansion of $\overline{mmse}\left(snr\right)$ as $snr \to 0^+$ or $\lambda \to +\infty$.

We now establish the requirements for the application of the Mellin transforms method.

We showed in Appendix \ref{section: Proof of Theorem Additive Gaussian Noise average mmse general h} (for discrete inputs) and in Appendix \ref{section: Proof of Theorem Additive Gaussian Noise average mmse general h Continuous Inputs} (for the continuous inputs) that $h_1\left(t\right)$ and $f_1\left(t\right)$ are locally integrable functions on $\mathbb{R}^+$. We also showed in Appendix \ref{section: Proof of Theorem Additive Gaussian Noise average mmse general h} and in Appendix \ref{section: Proof of Theorem Additive Gaussian Noise average mmse general h Continuous Inputs} that $M\left[f_1;z\right]$ converges absolutely and is holomorphic in the strip $\mathcal{R}\left(z\right)>0$ (for discrete inputs) and in the strip $0<\mathcal{R}\left(z\right)<1$ (for the continuous inputs) and hence that $M\left[f_1;1-z\right]$ converges absolutely and is holomorphic in the strip $\mathcal{R}\left(z\right)<1$ (for discrete inputs) and in the strip $0<\mathcal{R}\left(z\right)<1$ (for the continuous inputs). The Mellin transform $M\left[h_1;z\right]$ also converges absolutely and is holomorphic in the strip $\alpha_1<\mathcal{R}\left(z\right)<\beta_1$ because of hypothesis \eqref{eq: gamma less delta in Additive Gaussian Noise average mmse general h Low-snr Mellin transform}~\cite[p.~106]{asymptotic_expansion_of_integrals_Bleistein_and_Handelsman}.

Note hypothesis \eqref{eq: asymptotic behavior in Additive Gaussian Noise average mmse general h Low-snr} and
\begin{align}
f_1\left(t\right)&=\sum_{m=0}^{+\infty}\frac{1}{m!}mmse^{(m)}\left(z\right)\Bigg|_{z=0^+}t^m, \qquad t\rightarrow 0^+\nonumber\\
&\sim\sum_{m=0}^{+\infty}p_mt^{a_m}, \qquad t\rightarrow 0^+
\label{eq: low snr Taylor series f1}
\end{align}
(which is true because the fact that the input has finite moments implies that the function $f_1\left(t\right)$ is infinitely right differentiable at $t=0$~\cite[Proposition 7]{Estimation_in_Gaussian_Noise_Properties_of_the_Minimum_Mean-Square_Error_Guo_Wu_Shamai_and_Verdu} which enables a straightforward application of Taylor's Theorem~\cite{introduction_to_complex_analysis_second_edition_Priestley}) which ensures a ``correct'' type of decay.

Note hypothesis \eqref{eq: C not empty in Additive Gaussian Noise average mmse general h Low-snr Mellin transform} which ensures that the function $G_1\left(\cdot\right):=M\left[h_1;\cdot\right]M\left[f_1;1-\cdot\right]$ is holomorphic in $C_1\neq\emptyset$.

Combining
\begin{equation*}
M\left[h_1;c_1+iy\right]\in L^1\left(-\infty<y<+\infty\right)
\end{equation*}
(which is true due to hypothesis \eqref{eq: asymptotic behavior in Additive Gaussian Noise average mmse general h Low-snr Mellin transform} and the fact that $M\left[h_1;c_1+iy\right]$ is holomorphic in the line $]c_1-i\infty,c_1+i\infty[$) with
\begin{equation*}
t^{-c_1}f_1\left(t\right)\in L^1\left(0\leq t<+\infty\right)
\end{equation*}
(which is true due to $\emptyset\neq C_1\subseteq]-\infty,1[$ (for discrete inputs) or $\emptyset\neq C_1\subseteq]0,1[$ (for the continuous inputs), \eqref{eq: finite mmse}, \eqref{eq: positive mmse}, \eqref{eq: decreasing decay mmse} and \eqref{eq: exponential decay mmse} (for discrete inputs) or \eqref{eq: decay mmse Infinite PSK, PAM and QAM} and \eqref{eq: canonical mmse Gaussian input} (for the continuous inputs)) it is clear that~\cite[p.~108]{asymptotic_expansion_of_integrals_Bleistein_and_Handelsman}
\begin{equation*}
\int_0^{+\infty}f_1\left(t\right)h_1\left(t\right)dt=\frac{1}{2\pi i}\int_{c_1-i\infty}^{c_1+i\infty}G_1\left(z\right)dz
\end{equation*}

Combining hypothesis \eqref{eq: asymptotic behavior in Additive Gaussian Noise average mmse general h Low-snr Mellin transform} with
\begin{equation*}
M\left[f_1;1-x-iy\right]=o\left(1\right), \qquad |y|\rightarrow +\infty
\end{equation*}
(which holds $\forall x\in\mathbb{R}$ (for discrete inputs) or $\forall x\in\mathbb{R}^+$ (for the continuous inputs)~\cite[Lemma 4.3.6]{asymptotic_expansion_of_integrals_Bleistein_and_Handelsman}) where $M\left[f_1;1-x-iy\right]$ is to be understood as the analytic continuation of $M\left[f_1;1-x-iy\right]$ from $\{x+iy: x<1\}$ (for discrete inputs) or from $\{x+iy: 0<x<1\}$ (for the continuous inputs) to the entire $z$ plane (for discrete inputs) or to the strip $\{x+iy: x>0\}$ (for the continuous inputs)~\cite[p.~181]{introduction_to_complex_analysis_second_edition_Priestley} yields
\begin{equation*}
\forall x\in[c_1,+\infty[, G\left(x+iy\right)=O\left(|y|^{-2}\right), \qquad |y|\rightarrow +\infty
\end{equation*}
which in turn implies
\begin{equation*}
\forall x\in[c_1,+\infty[, \lim_{|y|\rightarrow +\infty}G\left(x+iy\right)=0
\end{equation*}
as well as (note that \eqref{eq: asymptotic behavior in Additive Gaussian Noise average mmse general h Low-snr} implies that $k_1\neq 0$ and that \eqref{eq: low snr Taylor series f1} implies that $q_1=0$) if $U:=\{\mathcal{R}\left(a_m\right)+1: m\in\mathbb{Z}^+\}$ and $u_n$ is the real sequence such that $u_n\uparrow +\infty$ and $U=\{u_n: n\in\mathbb{Z}^+\}$ then, since $\forall x\in]u_n,u_{n+1}[, G\left(x+iy\right)$ is holomorphic in the line $]x-i\infty,x+i\infty[$, we have that $\forall n\in\mathbb{Z}^+, \exists x\in]u_n,u_{n+1}[: \int_{-\infty}^{+\infty}\left|G\left(x+iy\right)\right|dy<+\infty$.

This leads immediately to the expansions~\cite[Theorem 4.4., Case II]{asymptotic_expansion_of_integrals_Bleistein_and_Handelsman}:
\begin{align*}
\overline{mmse}\left(snr\right)&\sim\sum_{m=0}^{+\infty}p_mM[h_1;a_m+1]\lambda^{-a_m}, \qquad \lambda\rightarrow +\infty\\
&\sim\sum_{m=0}^{+\infty}\frac{1}{m!}M[h_1;m+1]mmse^{(m)}\left(z\right)\Bigg|_{z=0^+}snr^m, \qquad snr\rightarrow 0^+
\end{align*}

\section{Proof of Corollary \ref{cor: Additive Gaussian Noise average mmse h Rayleigh and Rice Low-snr}}
\label{section: Proof of Corollary Additive Gaussian Noise average mmse h Rayleigh and Rice Low-snr}
\subsection{Case \texorpdfstring{$\mu=0$}{mu=0}}
\label{subsection: Proof of Corollary Additive Gaussian Noise average mmse h Rayleigh and Rice Low-snr case mu=0}
Since
\begin{equation*}
h_1\left(t\right):=\frac{\sqrt{t}f_{|h|}\left(\sqrt{t}\right)}{2}=\frac{t}{2\sigma^2}\exp{\left(-\frac{t}{2\sigma^2}\right)}=O\left(\exp{\left(-k_1t^{v_1}\right)}\right), \qquad t\rightarrow +\infty
\end{equation*}
where $\mathcal{R}\left(k_1\right)>0$ and $v_1>0$, we have that requirement \eqref{eq: asymptotic behavior in Additive Gaussian Noise average mmse general h Low-snr} holds.

Since, we showed in Appendix \ref{subsection: Proof of Corollary Additive Gaussian Noise average mmse h Rayleigh and Rice case mu=0} that the Mellin transform of $h_1\left(\cdot\right)$, which is given by
\begin{equation*}
M[h_1;z]=\left(2\sigma^2\right)^z\Gamma\left(z+1\right),
\end{equation*}
converges absolutely and is holomorphic in the strip $\mathcal{R}\left(z\right)>-1$, we have that $\alpha_1=-1$ and $\beta_1 =+\infty$, which satisfy $\alpha_1<\beta_1$, and
\begin{equation*}
\begin{cases}
]\alpha_1,\beta_1[\cap]-\infty,1[=]-1,+\infty[\cap]-\infty,1[=]-1,1[\neq\emptyset &\text{if $x$ is discrete},\\
]\alpha_1,\beta_1[\cap]0,1[=]-1,+\infty[\cap]0,1[=]0,1[\neq\emptyset &\text{if $x$ is continuous}.\\
\end{cases}
\end{equation*}
i.e., we satisfy requirements \eqref{eq: gamma less delta in Additive Gaussian Noise average mmse general h Low-snr Mellin transform} and \eqref{eq: C not empty in Additive Gaussian Noise average mmse general h Low-snr Mellin transform}.

We also showed in Appendix \ref{subsection: Proof of Corollary Additive Gaussian Noise average mmse h Rayleigh and Rice case mu=0} that
\begin{equation*}
\forall x\in[\frac{1}{2},+\infty[, M\left[h_1;x+iy\right]=O\left(|y|^{-2}\right), \qquad |y|\rightarrow +\infty
\end{equation*}
i.e., we satisfy requirement \eqref{eq: asymptotic behavior in Additive Gaussian Noise average mmse general h Low-snr Mellin transform}.

The result now follows from Theorem \ref{thm: Additive Gaussian Noise average mmse general h Low-snr}.

\subsection{Case \texorpdfstring{$\mu\neq0$}{mu not 0}}
\label{subsection: Proof of Corollary Additive Gaussian Noise average mmse h Rayleigh and Rice Low-snr case mu not 0}
Since
\begin{align*}
h_1\left(t\right):&=\frac{\sqrt{t}f_{|h|}\left(\sqrt{t}\right)}{2}\\
&=\frac{t}{2\sigma^2}\exp{\left(-\frac{|\mu|^2}{2\sigma^2}\right)}\exp{\left(-\frac{t}{2\sigma^2}\right)}I_0\left(\frac{\sqrt{t}v}{\sigma^2}\right)\\
&\sim\frac{t}{2\sigma^2}\exp{\left(-\frac{|\mu|^2}{2\sigma^2}\right)}\exp{\left(-\frac{t}{2\sigma^2}\right)}\frac{\exp{\left(\frac{\sqrt{t}v}{\sigma^2}\right)}}{\left(2\pi\frac{\sqrt{t}v}{\sigma^2}\right)^{\frac{1}{2}}}\sum_{k=0}^{+\infty}\frac{\frac{\prod_{i=1}^k\left(2k-1\right)^2}{k!8^k}}{\left(\frac{\sqrt{t}v}{\sigma^2}\right)^k}, \qquad t\rightarrow +\infty\\
&=O\left(\exp{\left(-k_1t^{v_1}\right)}\right), \qquad t\rightarrow +\infty
\end{align*}
where the asymptotic expansion is due to~\cite[Equation 10.40.1]{abramowitz_stegun}, $\mathcal{R}\left(k_1\right)>0$ and $v_1>0$, we have that requirement \eqref{eq: asymptotic behavior in Additive Gaussian Noise average mmse general h Low-snr} holds.

Since, we showed in Appendix \ref{subsection: Proof of Corollary Additive Gaussian Noise average mmse h Rayleigh and Rice case mu not 0} that the Mellin transform of $h_1\left(\cdot\right)$, which is given by
\begin{equation*}
M[h_1;z]=\exp{\left(-\frac{|\mu|^2}{2\sigma^2}\right)}\left(2\sigma^2\right)^z\Gamma\left(z+1\right){_1}F_1\left(z+1;1;\frac{v^2}{2\sigma^2}\right),
\end{equation*}
converges absolutely and is holomorphic in the strip $\mathcal{R}\left(z\right)>-1$, we have that $\alpha_1=-1$ and $\beta_1 =+\infty$, which satisfy $\alpha_1<\beta_1$, and
\begin{equation*}
\begin{cases}
]\alpha_1,\beta_1[\cap]-\infty,1[=]-1,+\infty[\cap]-\infty,1[=]-1,1[\neq\emptyset &\text{if $x$ is discrete},\\
]\alpha_1,\beta_1[\cap]0,1[=]-1,+\infty[\cap]0,1[=]0,1[\neq\emptyset &\text{if $x$ is continuous}.\\
\end{cases}
\end{equation*}
i.e., we satisfy requirements \eqref{eq: gamma less delta in Additive Gaussian Noise average mmse general h Low-snr Mellin transform} and \eqref{eq: C not empty in Additive Gaussian Noise average mmse general h Low-snr Mellin transform}.

We also showed in Appendix \ref{subsection: Proof of Corollary Additive Gaussian Noise average mmse h Rayleigh and Rice case mu not 0} that
\begin{equation*}
\forall x\in[\frac{1}{2},+\infty[, M\left[h_1;x+iy\right]=O\left(|y|^{-2}\right), \qquad |y|\rightarrow +\infty
\end{equation*}
i.e., we satisfy requirement \eqref{eq: asymptotic behavior in Additive Gaussian Noise average mmse general h Low-snr Mellin transform}.

The result now follows from Theorem \ref{thm: Additive Gaussian Noise average mmse general h Low-snr}.

\section{Proof of Corollary \ref{cor: Additive Gaussian Noise average mmse h Nakagami Low-snr}}
\label{section: Proof of Corollary Additive Gaussian Noise average mmse h Nakagami Low-snr}
Since
\begin{equation*}
h_1\left(t\right):=\frac{\sqrt{t}f_{|h|}\left(\sqrt{t}\right)}{2}=\frac{\mu^{\mu}}{\Gamma\left(\mu\right)w^{\mu}}t^{\mu}\exp{\left(-\frac{\mu}{w}t\right)}=O\left(\exp{\left(-k_1t^{v_1}\right)}\right), \qquad t\rightarrow +\infty
\end{equation*}
where $\mathcal{R}\left(k_1\right)>0$ and $v_1>0$, we have that requirement \eqref{eq: asymptotic behavior in Additive Gaussian Noise average mmse general h Low-snr} holds.

Since, we showed in Appendix \ref{section: Proof of Corollary Additive Gaussian Noise average mmse h Nakagami} that the Mellin transform of $h_1\left(\cdot\right)$, which is given by
\begin{equation*}
M[h_1;z]=\frac{1}{\Gamma\left(\mu\right)}\left(\frac{w}{\mu}\right)^z\Gamma\left(z+\mu\right),
\end{equation*}
converges absolutely and is holomorphic in the strip $\mathcal{R}\left(z\right)>-\mu$, we have that $\alpha_1=-\mu$ and $\beta_1 =+\infty$ which satisfy $\alpha_1<\beta_1$ because $\mu\geq\frac{1}{2}$ and
\begin{equation*}
\begin{cases}
]\alpha_1,\beta_1[\cap]-\infty,1[=]-\mu,+\infty[\cap]-\infty,1[\supseteq]-\frac{1}{2},+\infty[\cap]-\infty,1[=]-\frac{1}{2},1[\neq\emptyset &\text{if $x$ is discrete},\\
]\alpha_1,\beta_1[\cap]0,1[=]-\mu,+\infty[\cap]0,1[\supseteq]-\frac{1}{2},+\infty[\cap]0,1[=]0,1[\neq\emptyset &\text{if $x$ is continuous}.\\
\end{cases}
\end{equation*}
i.e., we satisfy requirements \eqref{eq: gamma less delta in Additive Gaussian Noise average mmse general h Low-snr Mellin transform} and \eqref{eq: C not empty in Additive Gaussian Noise average mmse general h Low-snr Mellin transform}.

We also showed in Appendix \ref{section: Proof of Corollary Additive Gaussian Noise average mmse h Nakagami} that
\begin{equation*}
\forall x\in[\frac{1}{2},+\infty[, M\left[h_1;x+iy\right]=O\left(|y|^{-2}\right), \qquad |y|\rightarrow +\infty
\end{equation*}
i.e., we satisfy requirement \eqref{eq: asymptotic behavior in Additive Gaussian Noise average mmse general h Low-snr Mellin transform}.

The result now also follows from Theorem \ref{thm: Additive Gaussian Noise average mmse general h Low-snr}.

\section{Proof of Corollary \ref{cor: Additive Gaussian Noise average mmse h Rayleigh and Rice Vector Version}}
\label{section: Proof of Corollary Additive Gaussian Noise average mmse h Rayleigh and Rice Vector Version}
\subsection{Case \texorpdfstring{$\boldsymbol{\mu}=\boldsymbol{0}$}{mu=0}}
\label{subsection: Proof of Corollary Additive Gaussian Noise average mmse h Rayleigh and Rice Vector Version case mu=0}
The proof follows the steps of Appendix \ref{subsection: Proof of Corollary Additive Gaussian Noise average mmse h Rayleigh and Rice case mu=0} taking into account that now:
\begin{equation*}
f_{\|\boldsymbol{h}\|}\left(t\right)=\frac{2t^{2k-1}\exp{\left(-\frac{t^2}{2\sigma^2}\right)}}{\left(2\sigma^2\right)^k\left(k-1\right)!}
\end{equation*}

By Taylor's Theorem~\cite{introduction_to_complex_analysis_second_edition_Priestley} we have that
\begin{equation*}
f\left(t\right)\sim\sum_{m=0}^{+\infty}\frac{\left(-1\right)^m}{\left(k-1\right)!m!\left(2\sigma^2\right)^{m+k}}t^{m+k}, \qquad t\rightarrow 0^+
\end{equation*}

We also have that
\begin{equation*}
M[f;z]=\frac{\left(2\sigma^2\right)^z\Gamma\left(z+k\right)}{\left(k-1\right)!}<+\infty
\end{equation*}
converges absolutely and is holomorphic in the strip $\mathcal{R}\left(z\right)>-k$~\cite[Equation 5.2.1]{abramowitz_stegun}.

\subsection{Case \texorpdfstring{$\boldsymbol{\mu}\neq\boldsymbol{0}$}{mu not 0}}
\label{subsection: Proof of Corollary Additive Gaussian Noise average mmse h Rayleigh and Rice Vector Version case mu not 0}
The proof follows the steps of Appendix \ref{subsection: Proof of Corollary Additive Gaussian Noise average mmse h Rayleigh and Rice case mu not 0} taking into account that now:
\begin{equation*}
f_{\|\boldsymbol{h}\|}\left(t\right)=\frac{t^k}{\|\boldsymbol{\mu}\|^{k-1}\sigma^2}\exp{\left(-\frac{\|\boldsymbol{\mu}\|^2}{2\sigma^2}\right)}\exp{\left(-\frac{t^2}{2\sigma^2}\right)}I_{k-1}\left(\frac{t\|\boldsymbol{\mu}\|}{\sigma^2}\right)
\end{equation*}

By Taylor's Theorem~\cite{introduction_to_complex_analysis_second_edition_Priestley} and~\cite[Equation 10.25.2]{abramowitz_stegun} we have that
\begin{equation*}
f\left(t\right)\sim\exp{\left(-\frac{\|\boldsymbol{\mu}\|^2}{2\sigma^2}\right)}\sum_{a=0}^{+\infty}\sum_{b=0}^a\left(\frac{\left(-1\right)^{a-b}}{\left(a-b\right)!\left(2\sigma^2\right)^{a+b+k}}\frac{\|\boldsymbol{\mu}\|^{2b}}{b!\Gamma\left(b+k\right)}\right)t^{a+k}, \qquad t\rightarrow 0^+
\end{equation*}

We also have that
\begin{equation*}
M[f;z]=\exp{\left(-\frac{\|\boldsymbol{\mu}\|^2}{2\sigma^2}\right)}\left(2\sigma^2\right)^z\frac{\Gamma\left(z+k\right)}{\left(k-1\right)!}{_1}F_1\left(z+k;k;\frac{\|\boldsymbol{\mu}\|^2}{2\sigma^2}\right)<+\infty
\end{equation*}
converges absolutely and is holomorphic in the strip $\mathcal{R}\left(z\right)>-k$~\cite[Equation 5.2.1]{abramowitz_stegun}.

\section{Proof of Theorem \ref{thm: specialized Mellin transform of mmse for a small generalization of BPSK}}
\label{section: Proof of Theorem specialized Mellin transform of mmse for a small generalization of BPSK}
Note that
\begin{align*}
M\left[mmse;1+z\right]&=\int_0^{+\infty}t^zmmse\left(t\right)dt\\
&=\frac{d^2}{2}\int_0^{+\infty}t^z\sum_{l=0}^{+\infty}\left(-1\right)^l\exp{\left(l\left(l+1\right)td^2\right)}\erfc{\left(\left(2l+1\right)\sqrt{t}\frac{d}{2}\right)}dt\\
&=\frac{d^2}{2}\sum_{l=0}^{+\infty}\left(-1\right)^l\int_0^{+\infty}t^z\exp{\left(l\left(l+1\right)td^2\right)}\erfc{\left(\left(2l+1\right)\sqrt{t}\frac{d}{2}\right)}dt\\
&=\frac{d^2}{2}\Bigg(\left(\frac{2}{d}\right)^{2\left(1+z\right)}\frac{\Gamma\left(\frac{3}{2}+z\right)}{\sqrt{\pi}\left(1+z\right)}+2d^{-2\left(1+z\right)}\frac{\Gamma\left(2+2z\right)}{\Gamma\left(2+z\right)}\sum_{l=1}^{+\infty}\left(-1\right)^l\frac{{_2}F_1\left(1,\frac{1}{2};2+z;1-\frac{1}{\left(1+2l\right)^2}\right)}{1+2l}\Bigg)
\end{align*}
where the second equality is due to the characterization of the \emph{canonical} MMSE associated with BPSK in \cite{Lozano06optimumpower}, the third equality is due to uniform convergence and the fourth equality follows from algebraic manipulations.

\section{Proof of Theorem \ref{thm: rayleigh and ricean applications}}
\label{section: Optimal Power Allocation Rayleigh and Rice}


Consider the function:
\begin{align*}
p:\mathbb{R}^+&\rightarrow\mathbb{R}_0^+\times\cdots\times\mathbb{R}_0^+\nonumber\\
snr&\mapsto p\left(snr\right)=\left(p_1\left(snr\right),\ldots,p_k\left(snr\right)\right):=\arg\max_{\substack{
   \text{s.t.}\\
   \forall i\in\{1,\ldots,k\}, p_i\geq 0\\
   \sum_{i=1}^kp_i\leq P
   }}
\overline{I}\left(snr;p_1,\ldots,p_k\right)
\end{align*}

It is possible to establish, from the KKT conditions associated with this optimization problem, that
\begin{equation*}
\exists snr_0>0: \exists\epsilon>0: \forall i\in\{1,\ldots,k\}, \forall snr>snr_0, \left(\epsilon<p_i\left(snr\right)<\frac{1}{\epsilon}\wedge{\overline{mmse}_i}\left(snrp_i\left(snr\right)\right)snr=\lambda\left(snr\right)\right)
\end{equation*}

Therefore, due to the expansion embodied in Corollary \ref{cor: Additive Gaussian Noise average mmse h Rayleigh and Rice}, it follows, for $i = 1,\ldots,k$, that
\begin{equation*}
{\overline{mmse}_i}\left(snrp_i\left(snr\right)\right)=\frac{\tau_i}{snr^2p_i^2\left(snr\right)}+O\left(\frac{1}{snr^3}\right), \qquad snr\rightarrow +\infty
\end{equation*}
where
\begin{equation*}
\tau_i := \exp{\left(-\frac{|\mu_i|^2}{2\sigma_i^2}\right)}\frac{M\left[{mmse}_i;2\right]}{2\sigma_i^2}
\end{equation*}
which leads to
\begin{gather*}
\lambda\left(snr\right)snr=\frac{\tau_i}{p_i^2\left(snr\right)}+O\left(\frac{1}{snr}\right), \qquad snr\rightarrow +\infty\\
\frac{1}{\lambda\left(snr\right)snr}=O\left(1\right), \qquad snr\rightarrow +\infty
\end{gather*}
and
\begin{align*}
p_i\left(snr\right)&=\sqrt{\frac{\tau_i}{\lambda\left(snr\right)snr}}\frac{1}{\sqrt{1+\frac{1}{\lambda\left(snr\right)snr}O\left(\frac{1}{snr}\right)}}, \qquad snr\rightarrow +\infty\\
&=\sqrt{\frac{\tau_i}{\lambda\left(snr\right)snr}}\left(1+O\left(\frac{1}{snr}\right)\right), \qquad snr\rightarrow +\infty\\
&=\sqrt{\frac{\tau_i}{\lambda\left(snr\right)snr}}+O\left(\frac{1}{snr}\right), \qquad snr\rightarrow +\infty
\end{align*}


\bibliographystyle{IEEEtran}
\bibliography{IEEEabrv,References}

\end{document}